\documentclass[aip,pof,onecolumn,floatfix]{revtex4-2}
\usepackage{amsmath,amssymb,amsfonts}
\usepackage{graphicx}
\usepackage{dcolumn}
\usepackage{siunitx}
\usepackage{booktabs}
\usepackage{bm}
\usepackage{multirow}
\usepackage{longtable}
\usepackage{rotating}

\makeatletter
\providecommand{\@bibitemShut}{\@empty}
\providecommand{\bibitemNoStop}{\@empty}
\providecommand{\bibitemStop}{\@empty}
\def\selectlanguage#1{}
\makeatother

\usepackage{CJKutf8}
\newcommand{\cn}[1]{\begin{CJK*}{UTF8}{gbsn}#1\end{CJK*}}

\providecommand{\alttext}[1]{}

\begin{document}
\raggedbottom

\title{Tandem jets in crossflow: a one-parameter dilution law for the trajectory and the velocity decay}

\author{Xunchen Liu\,(\cn{刘训臣})}
\email{liuxunchen@sdust.edu.cn}
\affiliation{\vspace{-2.2pt}College of Mechanical and Electronic Engineering, Shandong University of Science and Technology, No.~579 Qianwan'gang Road, Qingdao 266590, Shandong, China}

\begin{abstract}
The front jet of a tandem pair in crossflow is not passive. This paper treats the mean-flow phenomenology and its scaling; the mechanism is left to the companion paper. In 40\,kHz time-resolved PIV measurements covering nine spacing--velocity-ratio conditions ($L=2d$, $3d$, $4d$; $r\approx3$, $5$, $7$) and a single-jet baseline, the front-jet $rd$-normalized trajectory lies below the single-jet reference in every condition (near-field segment up to 40\% shorter), while the shielded rear jet rises above the reference in eight of the nine and penetrates $1.4$--$2.0$ times deeper than the front jet. The mean pressure and reverse-flow statistics are inconsistent with the two classical explanations: the inter-jet region is pressurized rather than sub-atmospheric, the lateral pressure difference across the front jet is weakened rather than enhanced (within reconstruction uncertainty), and the reverse-flow region does not follow the ordering of the modification. In natural coordinates the centerline decay re-scales onto the single-jet law at the effective velocity ratio (seven of nine within 18\%), so bending and decay are two readouts of one dilution, which whole-trajectory matching condenses into the one-parameter law $r_{\mathrm{eff}}/r=r/(r+c)$, $c=3.31$, with the velocity ratio as the primary control and the spacing entering through the near-field capacity ($c_h=2.50$--$3.30$). The law has one fitted constant and needs only the single-jet reference family; leave-one-condition-out predictions reproduce each held-out condition within 15\%, so the procedure is portable to any tandem array in which the gap recirculation closes.
\end{abstract}


\maketitle

\section{\label{sec:intro}Introduction}

Two tandem jets in a crossflow pose a two-body problem that a single jet does not: the upstream (front) jet shields the downstream (rear) jet from the crossflow. The rear jet in turn blocks the flow in the gap and redirects part of it back toward the front jet. In a single jet in crossflow (JICF) the trajectory and mixing are controlled primarily by a single parameter, the jet-to-crossflow velocity ratio $r$~\citep{fric_vortical_1994}, and arrays of such jets are widely used, for example, in gas-turbine film cooling, combustion-chamber wall cooling and pollutant dispersion. The tandem arrangement is the minimal configuration in which jet--jet interaction in crossflow can be studied systematically. The central questions addressed in this paper are how the two jets mutually alter their trajectories and mean velocity field, and how the interaction depends on the jet spacing and the velocity ratio.
The structure of the single-nozzle JICF is controlled by the nozzle diameter ($d$) and the momentum-flux ratio $J=\rho_j U^2_j/(\rho_0 U^2_0)$; $r=U_j/U_0$ is the velocity ratio, so that $J=r^2$ for equal densities. The bending trajectory of the single jet has been studied extensively (reviewed in~\citep{karagozian_jet_2014, zhang_flow_2017}): different algebraic combinations have been proposed as its length-scaling factor, with $rd$ confirmed experimentally and numerically to give the best collapse~\citep{smith_mixing_1998,muppidi_study_2005}.

The scaling length $rd$ has a clear physical origin: balancing the jet momentum flow rate $\rho_j U_j^2 d^2$ against the crossflow momentum flux $\rho_0 U_0^2$ acting over the characteristic area $\ell_M^2$ gives $\ell_M \sim \sqrt{\rho_j/\rho_0}\, rd$ (equal to $rd$ for equal densities). The subscript $M$ distinguishes it from the mixing length $l_{mix}$ below. It is the height over which the jet's vertical momentum is consumed by the crossflow, within which the trajectory remains quasi-vertical. We adopt $rd$ as the length-scaling factor and write the trajectory of the JICF as a power function
\begin{equation}
\frac{y}{rd}=A\left(\frac{x}{rd}\right)^{B},
\label{trajectory}
\end{equation}
where $A$ ($\approx1.5$--$2.0$ in the classical correlations~\citep{karagozian_transverse_2010}) describes the mixing and penetration efficiency, and $B$ determines the bending, with $B=1/3$ in the point-momentum-source model of \citet{broadwell_structure_1984}.

Along the bending trajectory, the single jet generates its vortex systems (shear-layer vortices, a counter-rotating vortex pair (CVP), horseshoe vortices at the jet foot and wake vortices~\citep{andreopoulos_structure_1985,karagozian_transverse_2010}). For tandem jets in close proximity, the front and rear trajectories merge into a single trajectory shortly after the nozzle exit, which collapses under an $rd$-type scaling with an effective velocity ratio~\citep{new_trajectory_2015}. The $rd$-normalized form of Eq.~\eqref{trajectory} therefore also serves as the baseline for the twin-nozzle trajectories, for which an additional spacing dependence must be introduced.

The trajectory of the twin-nozzle JICF has received much less attention, despite the wide use of tandem jets where rapid mixing is essential, as in gas-turbine afterburners, scramjet engines and the dilution-jet rows of combustor liners. PIV measurements have shown that the downstream jet is shielded by the upstream jet and merges with it into a single trajectory~\citep{zang_particle-image_2014}; the converging, merging and combined regions have been resolved for two tandem planar jets~\citep{zhao_experimental_2018}; the influence of the jet spacing on thermal mixing has been quantified~\citep{naik-nimbalkar_twin_2011}; and time-resolved measurements have concentrated on the low-ratio range ($r\approx1$--$2$) of a three-jet tandem array~\citep{kristo_time-resolved_2021}. The available quantitative single-versus-twin comparisons read the front jet as essentially unchanged: at a single spacing the near-field penetration of the twin front jet was found to be comparable to that of a single jet, with the rear jet shielded~\citep{gutmark_dynamics_2011}, and a two-spacing series described the upstream jet as fairly similar to the single jet~\citep{zang_particle-image_2014}. What is missing is the front jet measured against a single jet of the same velocity ratio across a systematic spacing series, the comparison that isolates the interaction itself. The measurements below supply it.

In the tandem configuration, the rear-jet blockage closes a recirculation bubble in the gap, whose reversed flow couples the two jets through an enhanced turbulent momentum exchange. The exchange itself, and the unsteady shear-layer dynamics it sustains, are measured and analyzed in the companion paper~\citep{companion_part2}, where the front-jet modification is traced to that exchange. Read in those terms, the mechanism carries three signatures that the mean fields examined here can test: the front-jet trajectory should be drawn below the single-jet reference as its momentum is diluted, the shielded rear jet should rise above it, and the velocity distribution should change accordingly, with a deficit on the front-jet centerline and a widened gap-side layer. These signatures are checked against the measurements below.

Here the interaction is examined with 40\,kHz time-resolved PIV measurements of tandem jets in crossflow. The measurements cover the nine combinations of velocity ratio ($r\approx3$, $5$ and $7$) and spacing ($L=2d$, $3d$ and $4d$), together with an 11-record single-jet baseline. Three questions organize the analysis. Is the front jet passive? Are the classical candidate causes (the reverse-flow region and the low-pressure attraction) consistent with the measured mean fields? Can the modification be condensed into a quantitative law, and if so, in which control parameter, and how far does that law extend beyond the measured regime?

The analysis proceeds from the outcome to its origin. Automated trajectory extraction with segmented power-law fits quantifies the front--rear asymmetry. The velocity fields, straightened into their natural coordinates, yield the velocity decay, the cross-track spreading and the reverse-flow structure in the gap. The mean pressure field, reconstructed from the time-resolved fields through a Poisson equation, tests the low-pressure-attraction mechanism. The modification is then condensed into a one-parameter scaling law; the turbulent momentum exchange that carries it, and the unsteady shear-layer dynamics it sustains, are the subject of the companion paper.

This paper makes three contributions, following the steady/dynamic division of the two-part study of a single transverse jet~\citep{zhang_flow_2017,yang_zhang_part2_2017}. First, the front jet is not passive: its trajectory lies below the single-jet reference and its centerline decay departs from the single-jet law in all nine conditions. Second, neither classical candidate is supported by the measurements: the reverse-flow region does not follow the ordering of the modification, and the reconstructed mean pressure is inconsistent with an attraction. Third, the modification obeys a one-parameter dilution law in the velocity ratio alone, the spacing entering through the near-field capacity, while the rear jet is simply shielded. The mechanism behind the dilution (an enclosed-recirculation momentum exchange with the unsteady shear-layer dynamics it sustains) is left, by design, to the companion paper~\citep{companion_part2}.

\section{\label{sec:expt}Experimental measurement}

\begin{figure}
\centerline{\includegraphics[width=\textwidth]{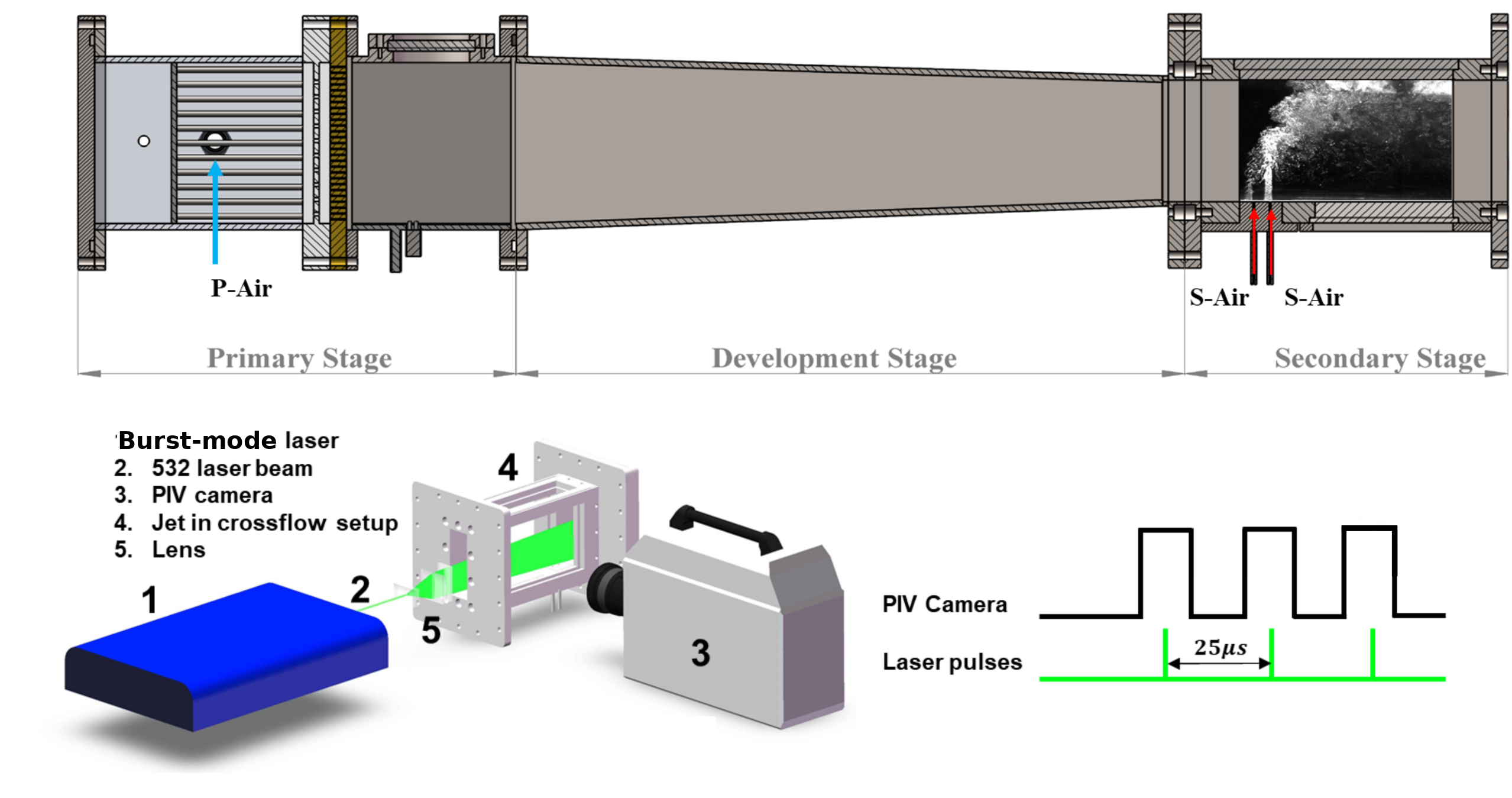}}
\caption{The twin-nozzle JICF test section (structural diagram: primary, development and secondary stages) with the 40\,kHz burst-mode PIV system and its timing sequence (camera exposure window and laser pulse pairs, separated by 25\,$\mu$s).}
\alttext{Schematic of the twin-nozzle test section with its primary, development and secondary stages, the 40 kHz burst-mode PIV laser and camera arrangement, and the 10 ms burst time sequence.}
\label{fig:setup}
\end{figure}

Figure~\ref{fig:setup} shows the twin-nozzle JICF setup, comprising the primary stage (mainstream supply), the development stage (a contraction) and the secondary stage (40\,mm wide $\times$ 115\,mm high $\times$ 295\,mm long, with quartz-glass walls for optical access). The two $d$=4\,mm nozzles are installed in the secondary stage with jet spacing $L$. This is the same nozzle pair as in the reacting twin-jet studies of \citet{wang_experimental_2022,wang_lift-off_2023}; the outlet is open to the atmosphere.

The experimental conditions comprise test cases with nominal velocity ratios of 3, 5 and 7 and jet spacings of $2d$, $3d$ and $4d$. The main-flow velocity varies from about 1.5\,m/s to 3.6\,m/s, corresponding to a jet Reynolds number $Re_d\approx2.9\times10^{3}$ and crossflow Reynolds numbers $U_0d/\nu\approx0.4$--$1.0\times10^{3}$. The jet and the main flow have the same density (both streams are air at 1.293\,kg/m$^3$), so the momentum-flux ratio equals $r^2$.

The 40\,kHz high-repetition-rate PIV system (Fig.~\ref{fig:setup}) uses a flashlamp-pumped Nd:YAG burst-mode laser (Spectral Energies, Quasimodo 1000) operating at 100\,kHz with a 10\,ms burst duration and about 125\,mJ per pulse. The 532\,nm output is expanded by three cylindrical lenses into a laser sheet about 45\,mm high and 0.2\,mm thick, which enters the secondary stage from the rear.

The images are captured by a high-speed CMOS camera (Photron SA-Z) with a normal lens (Nikon 50\,mm f/1.4\,G) for the far-field view and a telephoto lens (Nikon AF-200/4d) for the near field of each jet, which resolves the reverse-flow region and the shear-layer vortices. Al$_{2}$O$_{3}$ tracer particles of 1\,$\mu$m diameter are used; their Stokes number ($\approx0.04$) satisfies the tracer-following requirement. Each 10\,ms burst of 400 image pairs constitutes one record, i.e.\ one of the cases entering the statistics below.
In total, 68 twin-nozzle cases enter the statistics, 5--8 per condition as listed in Table~\ref{table:conditions}, together with the 11-record single-jet baseline (comprising 4, 3 and 4 records at $r\approx3$, 5 and 7), all acquired with the same system.

The long straight pipe upstream of each nozzle produces a fully developed turbulent pipe-flow exit velocity profile, as in the flush-pipe injector of \citet{getsinger_structural_2014}; the centerline velocity therefore exceeds the bulk velocity. The velocity ratio $r=V_{j}/U_{\infty}$ combines the documented bulk jet velocity $V_{j}=10$\,m/s with the free-stream velocity measured in the clean upstream region of each record (supplementary material). The three nominal ratio groups of Table~\ref{table:conditions} therefore have measured velocity ratios of $r\approx2.8$, $4.6$ and $6.5$ (hereafter $r\approx3$, $5$ and $7$ for brevity). The bulk convention is retained throughout because it characterizes the global momentum-flux ratio.

\begin{table*}
\caption{Experimental conditions: the nine combinations of the nominal velocity ratio and jet spacing, numbered 1--9 as referenced in the text, with the number of cases entering the statistics in parentheses (of the 71 acquired records, three are excluded: one without exported velocity fields and two with off-specification nozzle positions; the case accounting is in table ST5 of the supplementary material); the measured velocity ratios are $r\approx2.8$, $4.6$ and $6.5$ in the three ratio groups at the documented bulk jet velocity $V_j=10$\,m/s, with crossflow velocities $\approx3.6$, $2.2$ and $1.5$\,m/s and $Re_d\approx2.9\times10^{3}$ (jet, bulk), $U_0d/\nu\approx0.4$--$1.0\times10^{3}$ (crossflow)}\centering\alttext{Table of the nine experimental conditions arranged by velocity ratio and jet spacing, with the number of cases entering the statistics in parentheses.}
\label{table:conditions}
\begin{tabular}{lccc}
\hline
 & \multicolumn{3}{c}{jet spacing $L$} \\
\cmidrule(lr){2-4}
velocity ratio $r$ & $L=2d$ & $L=3d$ & $L=4d$ \\ \hline
$r=3$ & 1 (8) & 4 (8) & 7 (7) \\ $r=5$ & 2 (8) & 5 (8) & 8 (8) \\ $r=7$ & 3 (8) & 6 (8) & 9 (5) \\ \hline
\end{tabular}
\end{table*}

Unlike the bulk velocity, the centerline exit velocity governs the near-field shear-layer instability: over the 11 single-jet records the measured near-exit centerline velocity is $13.3$--$15.1$\,m/s. Because the near-field shedding frequency is governed by the peak velocity difference across the shear layer, this centerline value ($u_j\approx14$\,m/s) is used for the shear-layer normalizations of the companion paper~\citep{companion_part2}. In the twin configuration the two jets differ: averaged over the 136 jet exits, the rear-jet near-exit velocity is $15.4\pm0.5$\,m/s, close to the single-jet values, whereas the front-jet value is only $12.8\pm0.7$\,m/s. This front-jet deficit is already present at the first resolvable station and is in part the near-exit signature of the interaction rather than of the supply (Section~\ref{sec:limitations}).

\begin{figure}
\centerline{\includegraphics[width=5.0in]{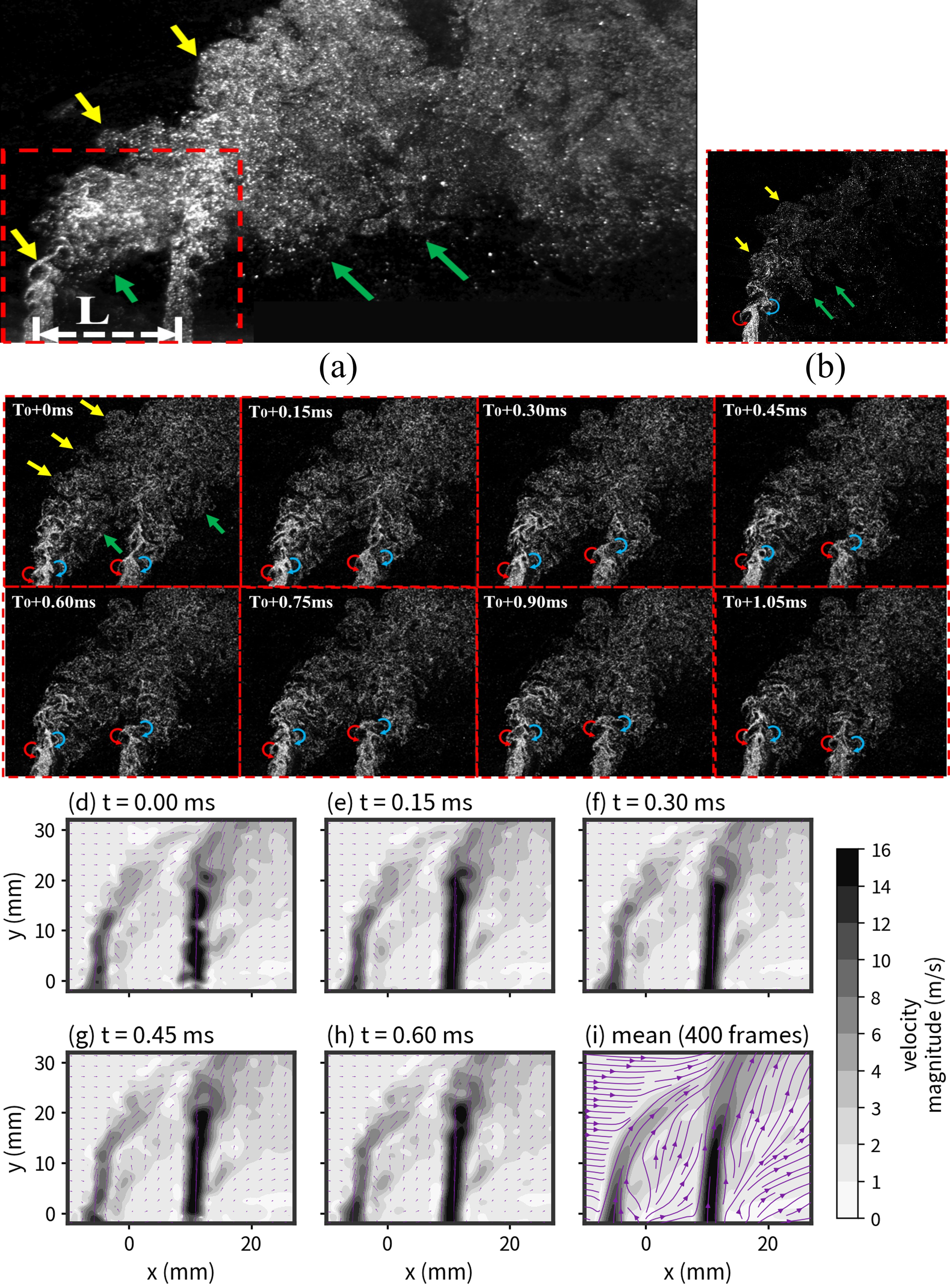}}
\caption{Instantaneous PIV images and velocity fields of the twin-nozzle JICF ($r\approx7$, $L=4d$). (a) Wide field of view with both jets (arrows: crossflow and jet directions; red rectangle: near field; white double arrow: nozzle spacing $L$); (b) near field of the single nozzle (curved arrows: shear-layer rotation); (c) near field of the twin nozzles as an 8-frame sequence ($0.15$\,ms apart, $T_0$--$T_0+1.05$\,ms). (d)--(h) Five instantaneous velocity fields at $t=0.00$--$0.60$\,ms and (i) the time-averaged field with streamlines, averaged over the 400 instantaneous fields of a case.}
\alttext{Instantaneous PIV images of the twin-nozzle jet in crossflow (wide view, near-field views and an eight-frame sequence) and five instantaneous velocity fields with the time-averaged field of the same condition.}
\label{fig:new-s1}\label{fig:new-s2}
\end{figure}

Representative instantaneous PIV images are shown in Fig.~\ref{fig:new-s1}(a)--(c): the wide field of view captures both jets and provides the global trajectories, while the near-field view (the red dashed rectangle in Fig.~\ref{fig:new-s1}a) resolves the two jets, the shear-layer vortices and the wake vortices between the jet and the wall~\citep{fric_vortical_1994}. The time-resolved analyses use the near-field view, where the two jets interact before merging downstream.

The PIV images are processed with commercial software (DaVis 8.4, LaVision) using multi-pass cross-correlation with 24$\times$24-pixel interrogation windows and 75\% overlap, sampling the vector fields on a uniform grid with 0.35\,mm spacing. The first resolvable station therefore lies about 0.5\,mm downstream of the exit. The instantaneous and time-averaged fields of a record are shown for one condition in Fig.~\ref{fig:new-s2}(d)--(i): the shear layers on both sides of each jet are resolved in the instantaneous vectors. The time-averaged field shows the bent trajectories and the recirculation region between the two jets.
The image scale is set by imaging the crossflow alone: with the jets switched off, the uniform crossflow is recorded over the same field of view. Dividing the measured mean particle displacement by the known free-stream velocity and the inter-pulse delay gives the magnification (pixels per millimeter). The resulting scale ($0.0575$\,mm per pixel, i.e.\ $0.345$\,mm per vector with the 75\% overlap of the final pass) is used for all records of a session. The residual of this uniform-flow calibration over the common field of view, measured as the departure of the uniform-flow displacement field from a rigid-body fit, propagates directly into every velocity magnitude and into the velocity ratio $r$. It is a common-mode error and therefore largely cancels in the front-to-rear and twin-to-single comparisons, but it shifts the absolute velocities.

Within the DaVis processing the vector fields are validated by the standard criteria: a correlation-peak-ratio threshold, a median test against the $3\times3$ neighborhood filter, and a maximum-displacement limit set from the expected local velocity. Rejected vectors and vectors outside the illuminated region are flagged invalid and are never interpolated or filled. The validated fraction is reported per record by the DaVis export and is high over the field of view; no gap filling is applied, and the rare rejected vectors are excluded from every statistic. The random uncertainty of a validated vector is set by the correlation statistics of a 24-pixel window, which for this configuration corresponds to a few percent of the local velocity for a single vector. Averaging the 400 fields of a record reduces the random term of the record-mean velocity well below one percent. The mean-field quantities used in the analysis (trajectories, natural-coordinate profiles, pressure sources) are therefore limited by the systematic calibration term rather than by random noise.

\section{\label{sec:results}Results and discussion}

Because the burst-mode sequence resolves the turbulent fluctuations directly, the same records also carry the Reynolds-stress and momentum-mixing statistics that close the mechanism chain in the companion paper; this paper works with the time-averaged fields and the quantities derived from them.
The presentation follows one chain of reasoning, from the measured outcome to its origin. The trajectory extraction and the single-jet baseline that calibrate every later comparison come first (Section~\ref{sec:trajectory}); read against that family, the front-jet trajectory yields the effective velocity ratio -- the tool against which the two classical candidates are then tested (Section~\ref{sec:reff-def}). The velocity decay measured independently in the natural coordinates of the same trajectories confirms that ratio, so that the bending and the decay turn out to be two readouts of one dilution (Sections~\ref{sec:natfield} and~\ref{sec:lam-reff}). 

The chain then eliminates the two classical candidates for the cause in turn: the reverse-flow region first (Section~\ref{sec:velocity}) and the reconstructed mean pressure second (Section~\ref{sec:pressure}) are each shown to be inconsistent with the observed ordering of the modification. With both excluded, the dependence of the modification on the spacing and the velocity ratio is condensed into a one-parameter scaling law (Section~\ref{sec:scaling-fit}), whose spacing dependence is carried by the near-field height (Section~\ref{sec:spacing}); the turbulent momentum exchange that carries the modification and the unsteady shear-layer dynamics behind it are identified in the companion paper. A summary table of the quantitative indicators computed for all cases is given at the end of the section. The mechanism chain is then assembled, together with its generality and three testable predictions (Section~\ref{sec:mechanism}), and its limitations are collected (Section~\ref{sec:limitations}); Section~\ref{sec:concl} draws the conclusions.

\subsection{\label{sec:trajectory}Trajectories and the modification metrics}

\begin{figure}[p]
\centerline{\includegraphics[width=\textwidth]{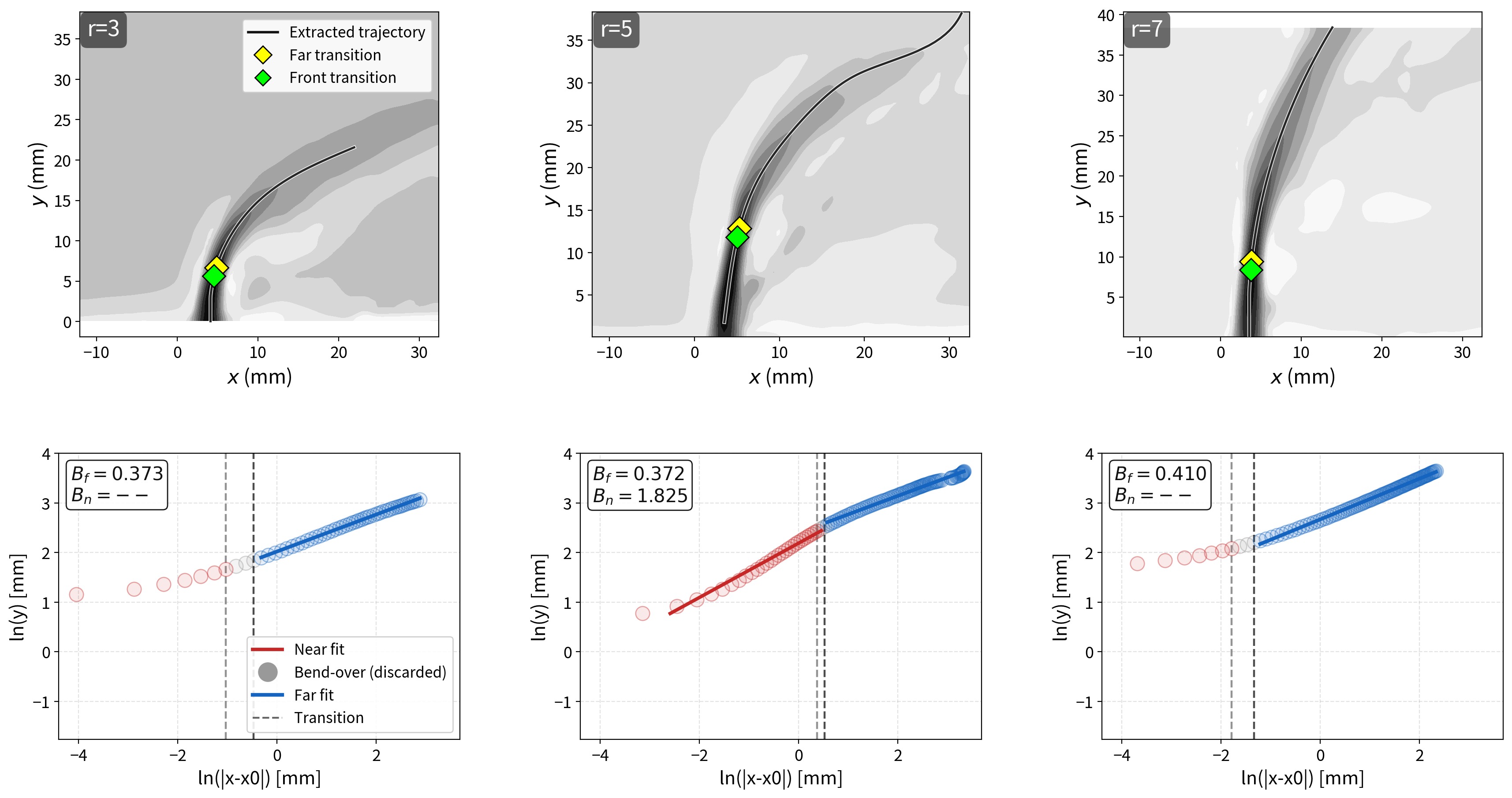}}
\centerline{\includegraphics[width=\textwidth]{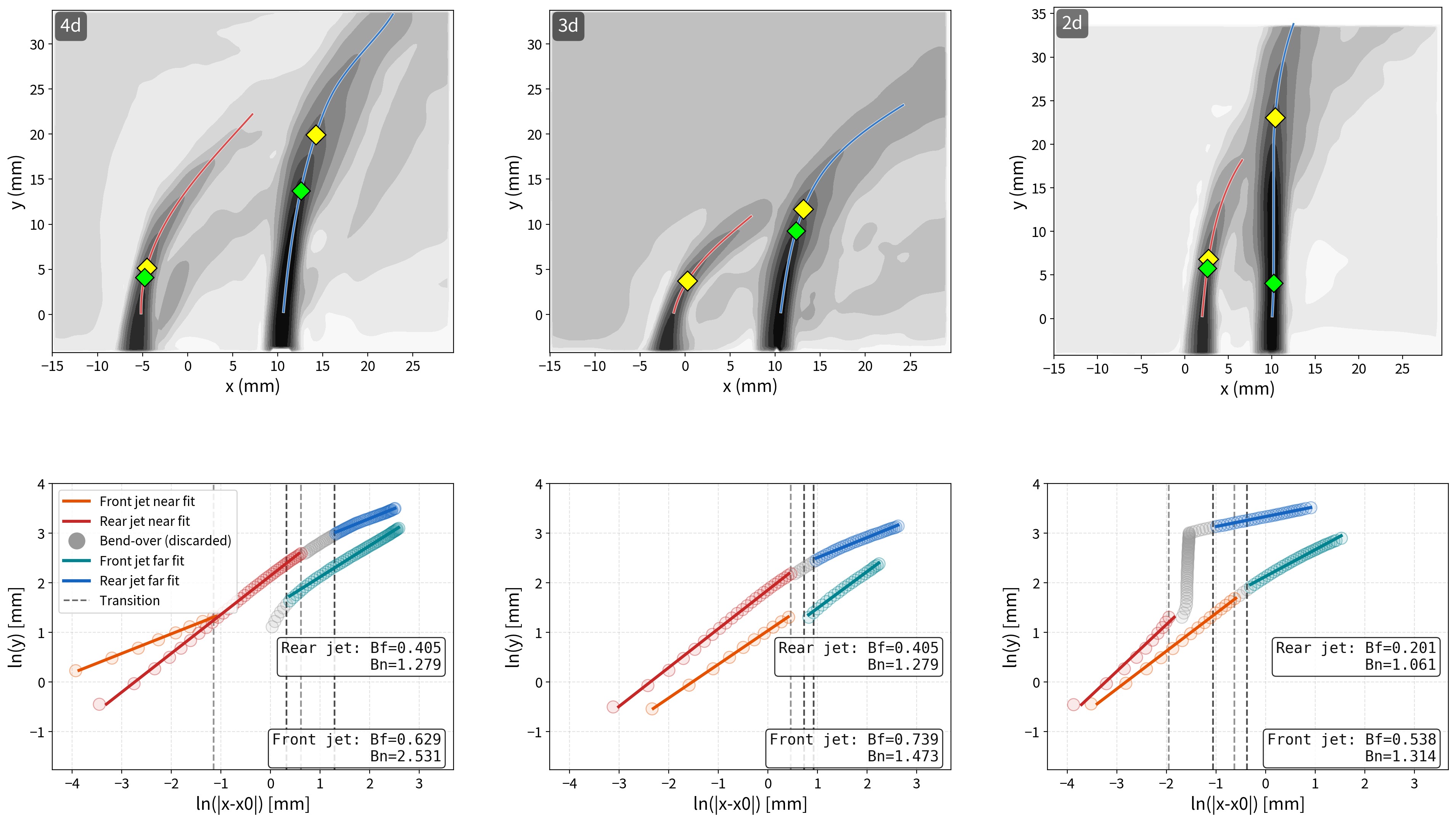}}
\caption{Automated trajectory extraction, near/far transition detection and power-law fitting. Top: single-jet examples at $r\approx3$, 5, 7 (one peak tracked per row). Bottom: twin-jet examples at $L=4d$, $3d$, $2d$ (from the nominal $r=5$, $3$ and $7$ groups; the two strongest peaks of each row assigned to the front and rear jets). Upper panels: mean velocity fields with the extracted trajectories and the near-field (green diamond) and far-field (yellow diamond) transitions; lower panels: the corresponding log--log fits of Eq.~\eqref{trajectory}, with fitted exponents.}
\alttext{Automated trajectory extraction, transition detection and power-law fitting applied to single-jet and twin-jet examples.}
\label{fig:new-s3}
\end{figure}

Each jet trajectory is extracted from the time-averaged velocity-magnitude field by an automated algorithm that tracks the jet row by row, detects the near- and far-field transitions, and fits the two segments separately to the power law of Eq.~\eqref{trajectory} in its log--log form. The far-field segment is fitted in the $y$-on-$x$ orientation of Eq.~\eqref{trajectory}, and the near-field segment in the $x$-on-$y$ orientation $x-x_0=A_n y^{B_n}$. Each trajectory is thereby reduced to a straight near-field rise and a self-similar far-field segment specified by $(B_n,A_n)$ and $(B_f,A_f)$ (Algorithm 1 of the supplementary material). For the twin jets the two strongest peaks of each row are assigned to the front and rear jets, and the same algorithm recovers both trajectories (the far-field points follow the fitted power law with $R^2\ge0.98$). The extraction, transition detection and fitting sequence is illustrated in Fig.~\ref{fig:new-s3}. The single jet is the measuring rod, so its baseline is established first: pooling the ten retained records (eleven acquired; one $r\approx7$ record excluded as an outlier) gives $B_f=0.393\pm0.017$ and $A_f=1.85\pm0.22$, consistent with the classical single-nozzle scaling~\citep{smith_mixing_1998,hasselbrink_transverse_2001}.

The fit thereby compresses each trajectory into a parameter set: the near-field transition height $t_{y,n}$, the near-field pair $(B_n,A_n)$ of the $x$-on-$y$ fit and the far-field pair $(B_f,A_f)$ of Eq.~\eqref{trajectory}. The set also contains the exponential decay rate $\lambda$ of the excess velocity along the arc length (Section~\ref{sec:natfield}), the turning angle $\Delta\theta$ between the initial-segment and the final-segment directions (supplementary material), and the physical penetration depth $H$ read directly from the trajectory. One further parameter is extracted from the trajectory as a whole, not from any single fitted coefficient, and is central to the analysis of this paper: the \emph{effective velocity ratio} $r_{\mathrm{eff}}$ of the front jet, defined and measured below. The change in each parameter is measured against the single jet of the same velocity-ratio group, never against a fixed value. The velocity ratio $r$ is the primary control of the penetration, so the comparison against the single-jet baseline removes the $r$ dependence and leaves the net contribution of the jet--jet interaction. Where the ratio measures the fraction of the single-jet value that survives the interaction, the change is reported as the percentage loss, one minus the ratio: the near-field shortening
\begin{equation}
E_{ty}\equiv1-\frac{t_{y,n}^{\rm front}}{t_s(r)},
\label{eq:ety}
\end{equation}
the penetration loss, the decay-rate excess and the penetration-depth loss
\begin{equation}
A_{\rm loss}\equiv1-\frac{A_f^{\rm front}}{A_s(r)}, \qquad E_{\rm lam}\equiv\frac{\lambda}{\lambda_s}-1, \qquad H_{\rm loss}\equiv1-\frac{H}{H_s(r)},
\label{eq:losses}
\end{equation}
The single-jet reference quantities that enter these definitions are collected in Table~\ref{tab:single-ref}.

\begin{table}
\centering
\caption{Single-jet reference quantities of the three velocity-ratio groups ($r\approx3$, 5 and 7), used throughout as the baseline for the relative measures of Eqs.~\eqref{eq:ety} and~\eqref{eq:losses}. Entries are the group means of the retained single-jet records; the far-field exponent is also given pooled over the three groups, and $A_s(r)$ is the fitted penetration law. The single-jet depth $H_s$ is a lower bound: the reference jet is still rising at the window edge at $r\approx7$.}
\alttext{Table of the single-jet reference quantities of the three velocity-ratio groups used as the baseline of the relative modification measures.}
\label{tab:single-ref}
\footnotesize
\begin{tabular}{lccc}
\toprule
Quantity & $r\approx3$ & $r\approx5$ & $r\approx7$ \\
\midrule
Transition height $t_s$ (mm) & $4.77$ & $7.94$ & $9.17$ \\
Penetration depth $H_s$ (mm) & $21.89$ & $37.79$ & $38.12$ \\
Decay rate $\lambda_s$ (mm$^{-1}$) & $0.0912$ & $0.0726$ & $0.0572$ \\
Penetration law & \multicolumn{3}{c}{$A_s(r)=0.1433\,r+1.2384$} \\
Far-field exponent & \multicolumn{3}{c}{$B_s=0.393\pm0.017$ (pooled, $n=10$)} \\
\bottomrule
\end{tabular}
\end{table}

The transition height $t_s(r)$ and the penetration depth $H_s(r)$ are the group means of the retained single-jet records. The penetration law $A_s(r)=0.1433\,r+1.2384$ is a least-squares fit to the same ten retained records, and the decay rate $\lambda_s$ is interpolated between their group means at the measured ratio of each case (supplementary material). Two further reference quantities are used in the same relative form: the far-field exponent enters as the ratio $B_f/B_s$, where $B_s$ is the pooled single-jet exponent (its group means vary by less than $7\%$, so this ratio is close to the bare front-jet exponent). The turning angle enters as the difference $\Delta\theta^{\rm front}-\Delta\theta^{\rm single}$.

\begin{figure}
\centerline{\includegraphics[width=6.0in]{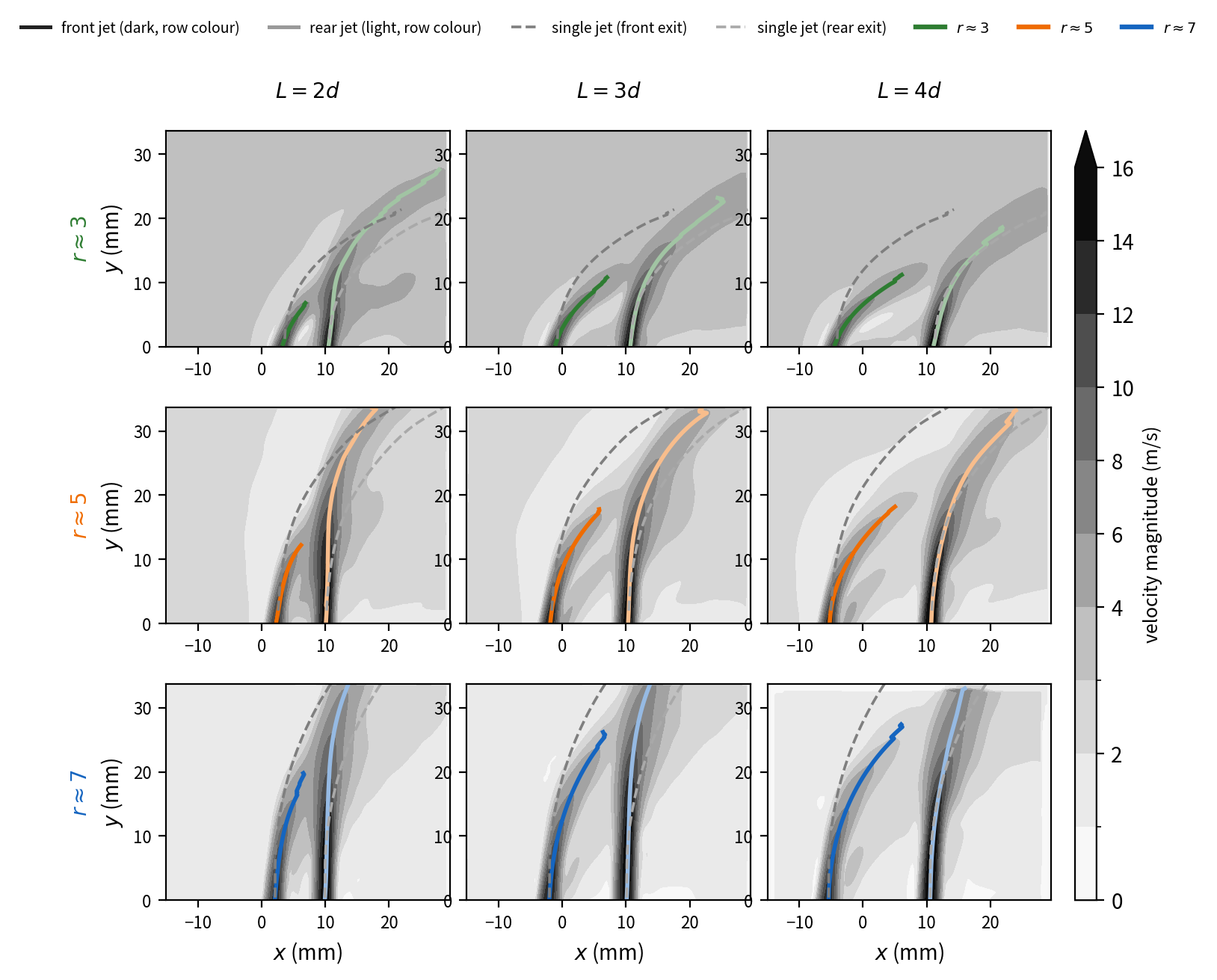}}
\caption{Mean velocity-magnitude fields of the nine conditions (averages over the cases of each condition; filled contour bands of $1$\,m/s below $4$\,m/s and $2$\,m/s above, the same absolute scale as Fig.~\ref{fig:new-s2}), with the mean front (dark solid) and rear (lighter solid) trajectories in the color of the velocity-ratio group of the row. Single-jet mean trajectories at the same nominal ratio: dashed light-gray lines from the front (darker gray) and rear (lighter gray) exits, each from the first resolved point of the corresponding twin trajectory. Rows: $r\approx3$, 5, 7; columns: $L=2d$, $3d$, $4d$ (Table~\ref{table:conditions}).}
\alttext{Grid of nine condition-averaged velocity magnitude fields of the twin jets with the mean front and rear trajectories and the single-jet references, arranged by velocity ratio rows and jet spacing columns.}
\label{fig:avg-single}
\end{figure}

The qualitative picture of Section~\ref{sec:intro} yields a first testable prediction: the front jet bends earlier and the rear jet rises. Figure~\ref{fig:avg-single} gives an overview of the nine conditions in direct physical terms: in every condition the front-jet mean trajectory lies below the single-jet reference, while the rear-jet mean trajectory lies above it in eight of the nine conditions. The interaction therefore lowers the front jet and raises the rear jet.

\begin{figure}
\centerline{\includegraphics[width=\textwidth]{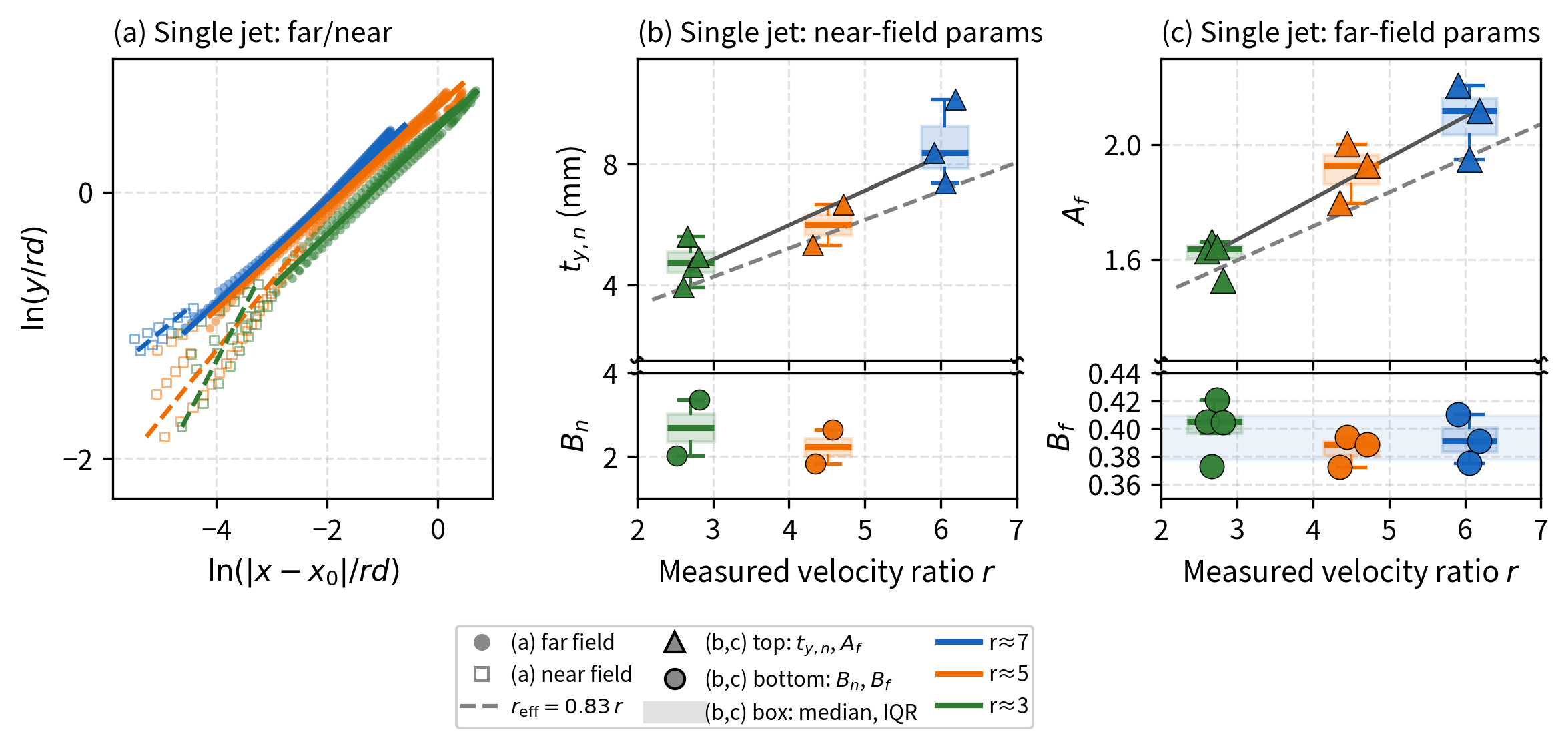}}
\caption{Single-jet trajectory and fitted parameters. (a) Trajectory points in $rd$-normalized log--log coordinates: far field (filled circles, solid fits) and near field (open squares, dashed fits), grouped by measured ratio (green $r\approx3$, orange 5, blue 7). (b) Near-field $t_{y,n}$ (upper) and $B_n$ (lower); (c) far-field $A_f$ (upper) and $B_f$ (lower) versus $r$. Symbols: one record each (offset within group); boxes: median and IQR; bands: pooled mean $\pm1$ SD; gray line: fit through group means. Upper panels: counterfactual baseline $r_{\rm cf}=0.83\,r$ (gray dashed). Twin-jet counterparts: Fig.~\ref{fig:traj-grid}.}
\alttext{Single-jet trajectories in scaled log--log coordinates and the fitted near-field and far-field parameters versus the measured velocity ratio, with medians and interquartile boxes of each ratio group.}
\label{fig:traj-single}
\end{figure}

\begin{figure}
\centerline{\includegraphics[width=5.9in]{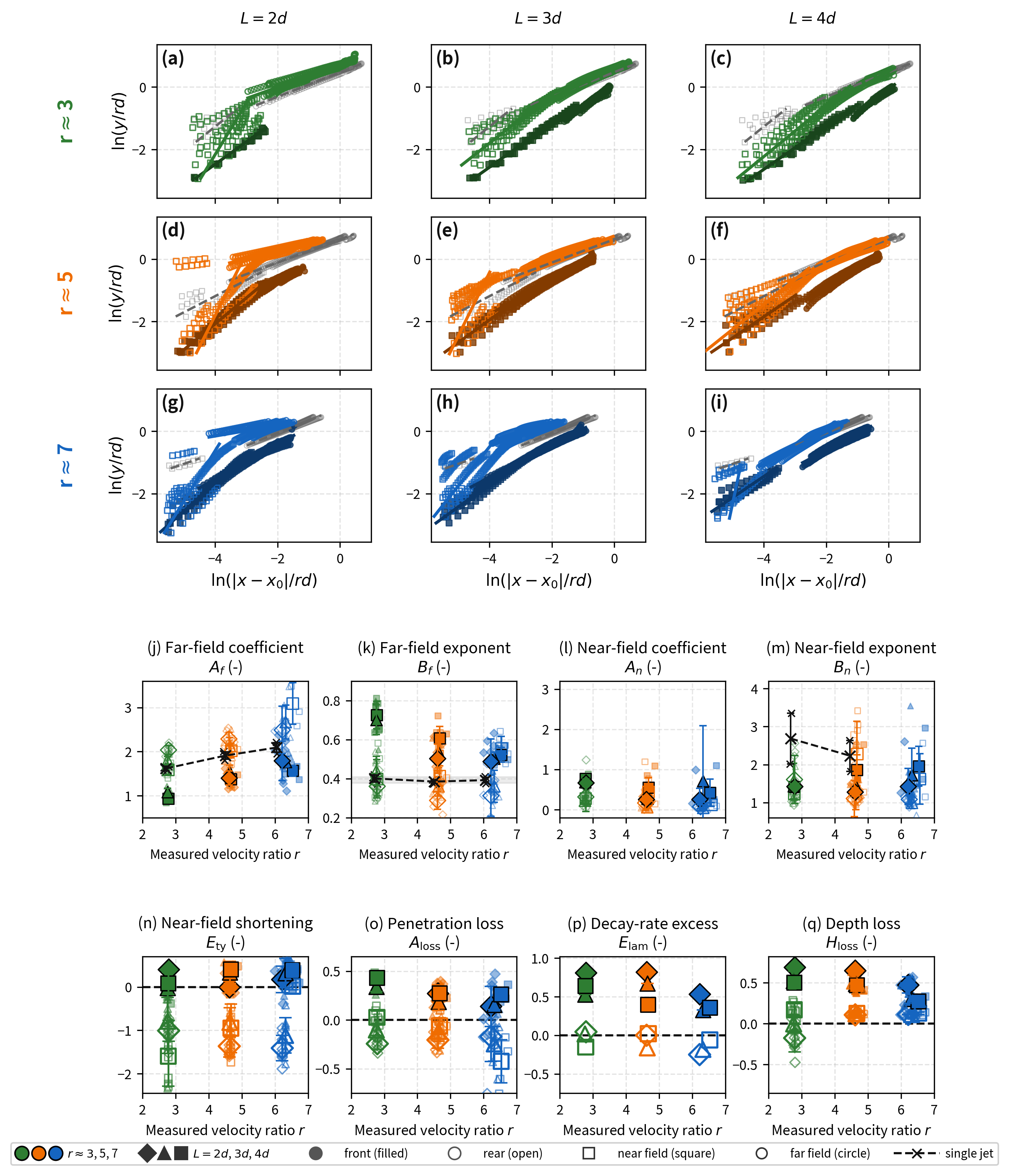}}
\caption{Twin-jet trajectory summary, fitted coefficients and modification metrics. (a)--(i) $rd$-normalized trajectories of the nine conditions (rows $r\approx3$, 5, 7; columns $L=2d$, $3d$, $4d$): near field (squares), far field (circles), front (filled) and rear (open) jets, single-jet references (dashed gray). (j)--(m) Fitted coefficients versus $r$: far-field $A_f$ and $B_f$, near-field $A_n$ and $B_n$ (dashed: single-jet reference; crosses: single-jet records). (n)--(q) Modification metrics of Eqs.~\eqref{eq:ety} and~\eqref{eq:losses}: $E_{ty}$, $A_{\rm loss}$, $E_{\rm lam}$, $H_{\rm loss}$ (filled: front jet; open: the same ratio formed with the rear-jet value; dashed: zero). Per-case values: table ST5; per condition: Table~\ref{tab:traj-params}.}
\alttext{Nine-panel grid of the front and rear twin-jet trajectories with their single-jet references, plus panels of the fitted far-field and near-field coefficients and of the four percentage modification metrics versus the velocity ratio.}
\label{fig:traj-grid}
\end{figure}

The same trajectories, quantified by the fitted power laws of Eq.~\eqref{trajectory}, are collected together with the fitted near- and far-field parameters in Fig.~\ref{fig:traj-single} for the single jet and in Fig.~\ref{fig:traj-grid} for the nine twin conditions.
The pooled far-field exponent of the single jet sits slightly above the point-momentum-source prediction $B=1/3$ of \citet{broadwell_structure_1984} but within the literature range for deflected jets~\citep{pratte_profiles_1967,smith_mixing_1998,hasselbrink_transverse_2001}. Across the nine conditions the fitted lines form a clear hierarchy: rear above front at every $r$, front below the single jet, and rear above single in eight of the nine conditions (the exception is $L=4d$, $r\approx3$, where the rear-jet condition mean falls marginally below the reference); the rear jet is bent more sharply in eight of the nine conditions, while the front jet is likewise not passive.
The spacing acts mainly on the rear-jet $A_f$, and none of these comparisons changes when the measured per-jet exit velocity is used instead of the documented bulk convention (supplementary material).

Panels (j) and (k) of Fig.~\ref{fig:traj-grid} and Table~\ref{tab:traj-params} collect the far-field pair: the penetration coefficient $A_f$ grows with the velocity ratio for both jets, the rear-jet $A_f$ exceeding the front-jet one in every condition where both are defined (factor $1.4$--$2.0$). The exponent $B_f$ falls as $r$ grows for the front jet and shows only a weak, mixed trend for the rear jet (supplementary material). At high velocity ratios the front jet is therefore bent over more, while the rear-jet bending stays near the single-jet level: the shielding in quantitative form.

\begin{table}
\centering
\caption{Far-field trajectory parameters of the nine conditions: the measured velocity ratio and the fitted penetration coefficient $A_f$ and exponent $B_f$ of Eq.~\eqref{trajectory} ($y/(rd)=A\,(x/(rd))^{B}$) for the front and rear jets, mean $\pm$ SD over the valid cases of each condition. A larger $A_f$ means a deeper normalized penetration, and a smaller $B_f$ a trajectory bent over more strongly. No far-field fit exists for the front jet at $L=2d$, $r\approx3$ (fully deflected inside its near-field segment); the last row is the single-jet baseline (ten retained records of the eleven acquired; one $r\approx7$ record excluded as an outlier), and the per-case near-field parameters are in table ST5.}
\alttext{Table of the far-field trajectory parameters of the nine conditions for the front and rear jets and of the single-jet baseline.}
\label{tab:traj-params}
\footnotesize
\begin{tabular}{llcccc}
\toprule
Condition & $r$ & $A_f^{\rm front}$ & $B_f^{\rm front}$ & $A_f^{\rm rear}$ & $B_f^{\rm rear}$ \\
\midrule
$L=2d$, $r\approx3$ & 2.79 & -- & -- & $2.04\pm0.10$ & $0.36\pm0.05$ \\
$L=2d$, $r\approx5$ & 4.62 & $1.38\pm0.17$ & $0.48\pm0.08$ & $2.27\pm0.15$ & $0.28\pm0.05$ \\
$L=2d$, $r\approx7$ & 6.22 & $1.79\pm0.45$ & $0.49\pm0.12$ & $2.49\pm0.54$ & $0.31\pm0.11$ \\
$L=3d$, $r\approx3$ & 2.76 & $1.08\pm0.08$ & $0.70\pm0.08$ & $1.80\pm0.09$ & $0.44\pm0.07$ \\
$L=3d$, $r\approx5$ & 4.64 & $1.53\pm0.16$ & $0.54\pm0.08$ & $2.13\pm0.19$ & $0.36\pm0.05$ \\
$L=3d$, $r\approx7$ & 6.32 & $1.77\pm0.12$ & $0.50\pm0.03$ & $2.67\pm0.38$ & $0.36\pm0.08$ \\
$L=4d$, $r\approx3$ & 2.78 & $0.94\pm0.04$ & $0.73\pm0.07$ & $1.59\pm0.11$ & $0.40\pm0.10$ \\
$L=4d$, $r\approx5$ & 4.67 & $1.38\pm0.09$ & $0.61\pm0.06$ & $2.02\pm0.17$ & $0.41\pm0.05$ \\
$L=4d$, $r\approx7$ & 6.53 & $1.56\pm0.12$ & $0.52\pm0.04$ & $3.10\pm0.47$ & $0.56\pm0.06$ \\
\midrule
single jet (ten retained records) & 2.70--6.06 & \multicolumn{4}{c}{$A_f=1.85\pm0.22$, \quad $B_f=0.393\pm0.017$} \\
\bottomrule
\end{tabular}
\end{table}

Panels (l) and (m) show the near-field pair of the $x$-on-$y$ fit for completeness. The pair is analyzed below but is not used in the scaling analysis; the single jet offers no $A_n$ reference because its near field is quasi-vertical (supplementary material).

The two physical-space counterparts in table ST1 of the supplementary material, the penetration depth $H$ and the far-field deflection $\varphi$, tell the same story without any normalization: the front-jet penetration depth grows from $6.8\pm0.5$\,mm at $L=2d$, $r\approx3$ to $27.7\pm1.6$\,mm at $L=4d$, $r\approx7$. Its far-field deflection $\varphi=\arctan\langle\mathrm{d}x/\mathrm{d}y\rangle$, measured over the last $40\%$ of the tracked trajectory, spans $21.8^\circ$--$54.7^\circ$, falling as the velocity ratio grows and rising with the spacing. By contrast, the rear jet reaches $18$--$34$\,mm and stays within $11$--$49^\circ$: the shielded jet is still rising steeply where the front jet has already turned into the crossflow.

The percentage form of Eqs.~\eqref{eq:ety} and \eqref{eq:losses} condenses the trajectory side of the interaction into four metrics: the trajectory metrics $E_{ty}$, $A_{\rm loss}$ and $H_{\rm loss}$, together with the distribution metric $E_{\rm lam}$ for the along-trajectory decay (Section~\ref{sec:natfield}; Fig.~\ref{fig:traj-grid}(n)--(q)); the transition-height ratio $t_{y,n}^{\rm front}/t_s(r)=1-E_{ty}$ is the near-field proxy of the effective ratio $r_{\rm eff}$ defined below. All four metrics are positive wherever the interaction is detected, so the front jet is modified in the same sense everywhere, yet they rank the nine conditions differently: $E_{ty}$ peaks at $L=4d$, $r\approx5$ and 7 and at $L=2d$, $r\approx3$, $H_{\rm loss}$ grows toward the low-$r$, small-$L$ corner, and $E_{\rm lam}$ is largest at the smallest spacing. Of the four, the following subsections use $E_{ty}$, whose complement $1-E_{ty}$ is the near-field operational proxy; the remaining three serve as consistency checks of the same sign.

Each metric is dominated by its own axis, and the rear-jet counterparts (open markers in Fig.~\ref{fig:traj-grid}(n)--(q)) fall on the opposite side of the single-jet baseline on average, with a longer near-field segment, a deeper penetration and a decay at or below the single-jet rate. The same percentage form therefore captures the whole front--rear asymmetry.
The correlation structure of these metrics with the force-level and pressure quantities is examined in the companion paper~\citep{companion_part2}; here they serve as consistency checks of the same sign.

A stricter question is whether the lower front-jet trajectory merely reflects its lower exit velocity ($12.8$ versus $15.4$\,m/s). Panels (b) and (c) of Fig.~\ref{fig:traj-single} answer this with an explicit counterfactual, a single jet at the reduced ratio $r_{\rm cf}=0.83\,r$: interpolating the single-jet parameter laws at that ratio covers any supply-side deficit between zero and the measured $17\%$.

The measured front-jet values lie far below that prediction: at $r\approx7$ the low-momentum single jet would lose only $7\%$ in $A_f$, whereas the measured front jet loses $A_{\rm loss}=16$--$28\%$. The deficit thus accounts for about one quarter to one half of the measured loss at $r\approx7$, and for $22$--$32\%$ of the loss at $r\approx5$ and $10$--$12\%$ at $r\approx3$. The exit-velocity deficit therefore accounts for at most about one half of the front-jet modification, and the baseline is itself conservative, because the near-exit velocity is sampled where the front jet already decelerates under the interaction.

The near field is analyzed in the same format (panel (b) of Fig.~\ref{fig:traj-single} and Figs.~\ref{fig:traj-grid}(l),(m)): the front-jet near-field segment is up to $40\%$ shorter than that of the single jet in seven of the nine conditions. The rear jet stays vertical longer at small spacing, consistent with the shielding. The turning angle confirms the same asymmetry without relying on the transition detection: the front jet turns less than the single jet of its ratio group in seven of the nine conditions ($\Delta\theta^{\rm front}-\Delta\theta^{\rm single}$ from $-28.5^\circ$ to $+7.1^\circ$; supplementary material). The near-field coefficients are not used further, and the fit-gating criteria, the windward--leeward and spacing dependence of the near-field results and the corresponding statistics are given in the supplementary material.
\subsection{\label{sec:reff-def}The effective velocity ratio}
The first result of this work is that the front jet is diluted, and the dilution is measured without relying on any single fitted coefficient: $r_{\mathrm{eff}}$ is extracted from the trajectory as a whole, by \emph{curve matching} against the single-jet family. The premise is that the diluted front jet traces, over its entire resolvable length, the trajectory that a single jet of a lower velocity ratio would trace. Each front-jet trajectory is therefore re-scaled by the trial ratio $r'$ in the single-jet coordinates $u=(x-x_{\rm exit})/(r'd)$ and $v=y/(r'd)$, and matched against a reference library built from the single-jet records, so that
\begin{equation}
r_{\mathrm{eff}}=\arg\min_{r'}\sum_i\bigl[\,\tilde v_i(r')-F\bigl(\tilde u_i(r');r'\bigr)\bigr]^2,
\qquad
\tilde u_i(r')=\frac{x_i-x_{\rm exit}}{r'd},\quad
\tilde v_i(r')=\frac{y_i}{r'd},
\label{eq:reff-match}
\end{equation}
where the sum runs over the resolved trajectory points inside the matching window $\tilde u_i\in[0.05,2.30]$. Here $F(\cdot\,;r')$ is the single-jet reference curve at ratio $r'$ (the mean of the independent records of the three measured ratio groups, interpolated in $r'$ between them and scaled outside their range by the single-jet penetration law $A_s(r)$). The one-dimensional minimum is located by a golden-section search over $r'\in[0.25r,1.45r]$. The protocol, validation and window robustness are given in the supplementary material. 

\begin{figure}
\centerline{\includegraphics[width=5.7in]{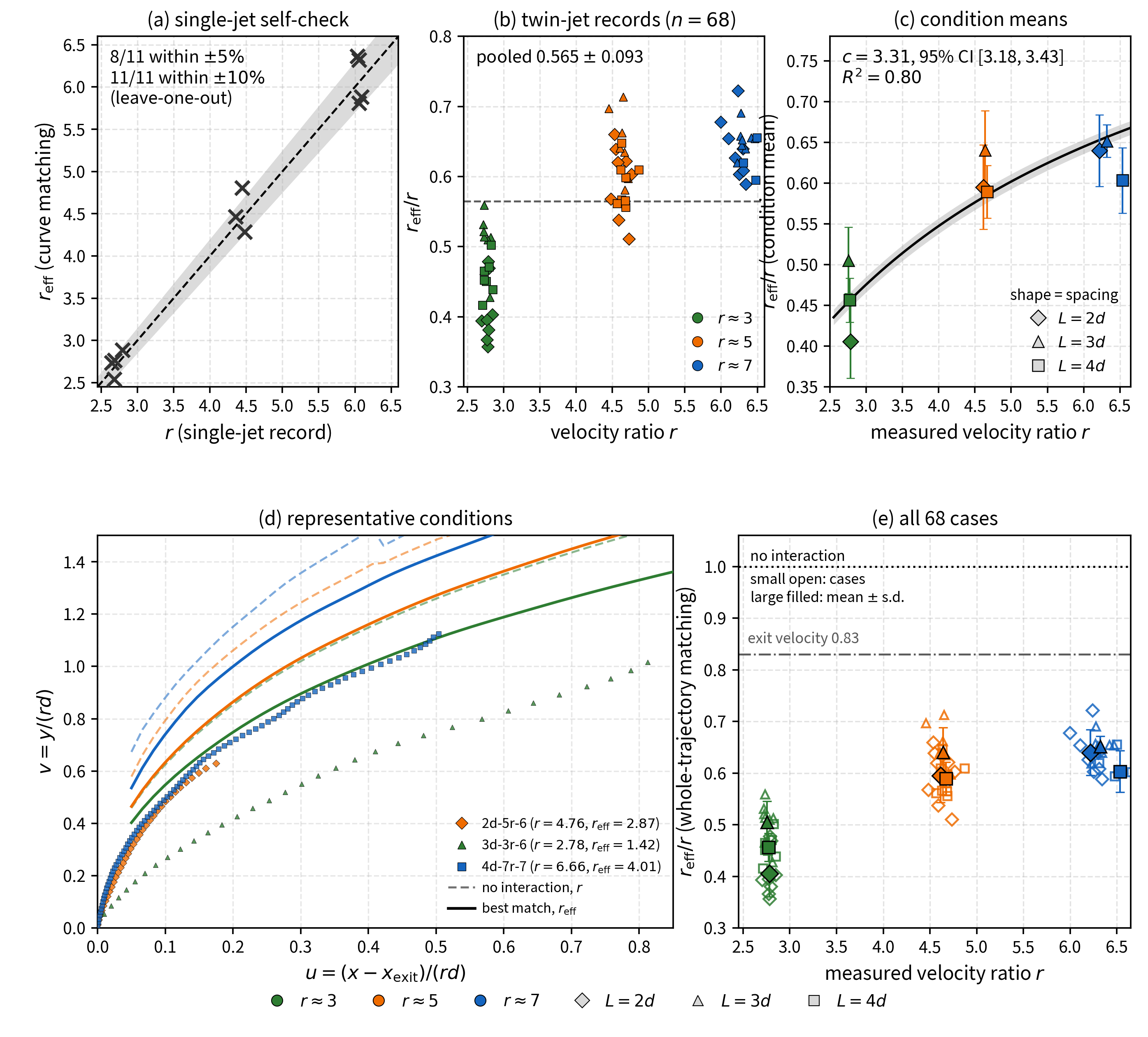}}
\caption{Curve-matching measurement of the effective velocity ratio. (a) Single-jet validation (leave-one-out; dashed: 1:1); (b) per-case twin-jet ratios (pooled $0.565\pm0.093$, all cases $<0.75$); (c) condition means with the hyperbolic fit $r/(r+c)$, $c=3.31$, and its 95\% bootstrap band. (d) Front-jet trajectories of three representative conditions (2d-5r-6, 3d-3r-6, 4d-7r-7; colour: ratio group, shape: spacing) in single-jet scaling coordinates $(u,v)$, against the best-matching reference at $r_{\mathrm{eff}}$ (solid, same colour) and the no-interaction reference at $r$ (dashed, same colour); (e) $r_{\mathrm{eff}}/r$ versus $r$ (color: ratio group; shape: spacing; small open: cases, large filled: condition means $\pm$ SD).}
\alttext{Curve-matching measurement of the effective velocity ratio: the leave-one-out single-jet validation, the per-case and condition-mean ratios, the estimator applied to representative trajectories, and the resulting effective-ratio distribution.}
\label{fig:new-s14}\label{fig:reff-def}
\end{figure}

The estimator uses every resolvable point, so it is insensitive to any single fitted parameter -- including the near-field transition detection, which fails for 12 cases. It recovers the single-jet records themselves with $r_{\mathrm{eff}}\approx r$ (leave-one-out: 8 of 11 within $5\%$, all 11 within $10\%$; Fig.~\ref{fig:new-s14}(a)).
The near-field proxy $1-E_{ty}$ of Eq.~\eqref{eq:ety} tracks the curve-matching ratio at $L=3d$ and $4d$ for $r\approx5$ and 7, but returns near-unity artifacts at $r\approx3$ and in the early-merging condition at $L=2d$ (figure S4(a) of the supplementary material).

Figure~\ref{fig:reff-def} shows the estimator at work and the distribution of $r_{\mathrm{eff}}/r$ over all cases.
The ratio so defined is an effective parameter of the single-jet trajectory family, a shape parameter certified by the independent readouts of Sections~\ref{sec:lam-reff} and~\ref{sec:scaling-fit}, rather than a literal supply ratio; the flux-budget diagnostics of the companion paper quantify this distinction. The dilution is present in all nine conditions; its primary control is the velocity ratio, and the spacing contributes only a secondary modulation (Section~\ref{sec:scaling-fit}). The correspondence is bidirectional: the measured decay rate, inverted through the single-jet decay law, returns an estimate of the same ratio from the decay side alone ($r_{\mathrm{eff},\lambda}$, Section~\ref{sec:lam-reff}), so the trajectory and the decay are two readouts of one dilution parameter rather than two separate effects. Two shortening measures are used below and should not be conflated: the near-field height shortening $E_{ty}=1-t_{y,n}^{\rm front}/t_s(r)$, read from the transition height alone (56 cases), and the dilution-scale bending $\mathcal{B}\equiv1-r_{\mathrm{eff}}/r$, read from the whole trajectory (all cases). They agree in trend but not in value, because they use different references: $E_{ty}$ enters the near-field comparisons and $\mathcal{B}$ enters the decay correspondence of Section~\ref{sec:lam-reff}. The exit-velocity baseline $r_{\rm cf}\approx0.83\,r$ is used to decompose the interaction contribution in Section~\ref{sec:trajectory}.

\subsection{\label{sec:natfield}Velocity decay in natural coordinates}

The examination begins with the time-averaged velocity field of the two jets, straightened into the natural coordinates of their extracted trajectories. This subsection establishes the along-trajectory velocity decay and the cross-track spreading of each jet, and the departure of both from the single-jet reference of the same velocity-ratio group. The decay correspondence that this subsection establishes is confirmed by the independent decay-side estimator in the next subsection, and the reverse flow that the two jets generate between them is examined in Section~\ref{sec:velocity}.

\begin{figure}
\centerline{\includegraphics[width=6.5in]{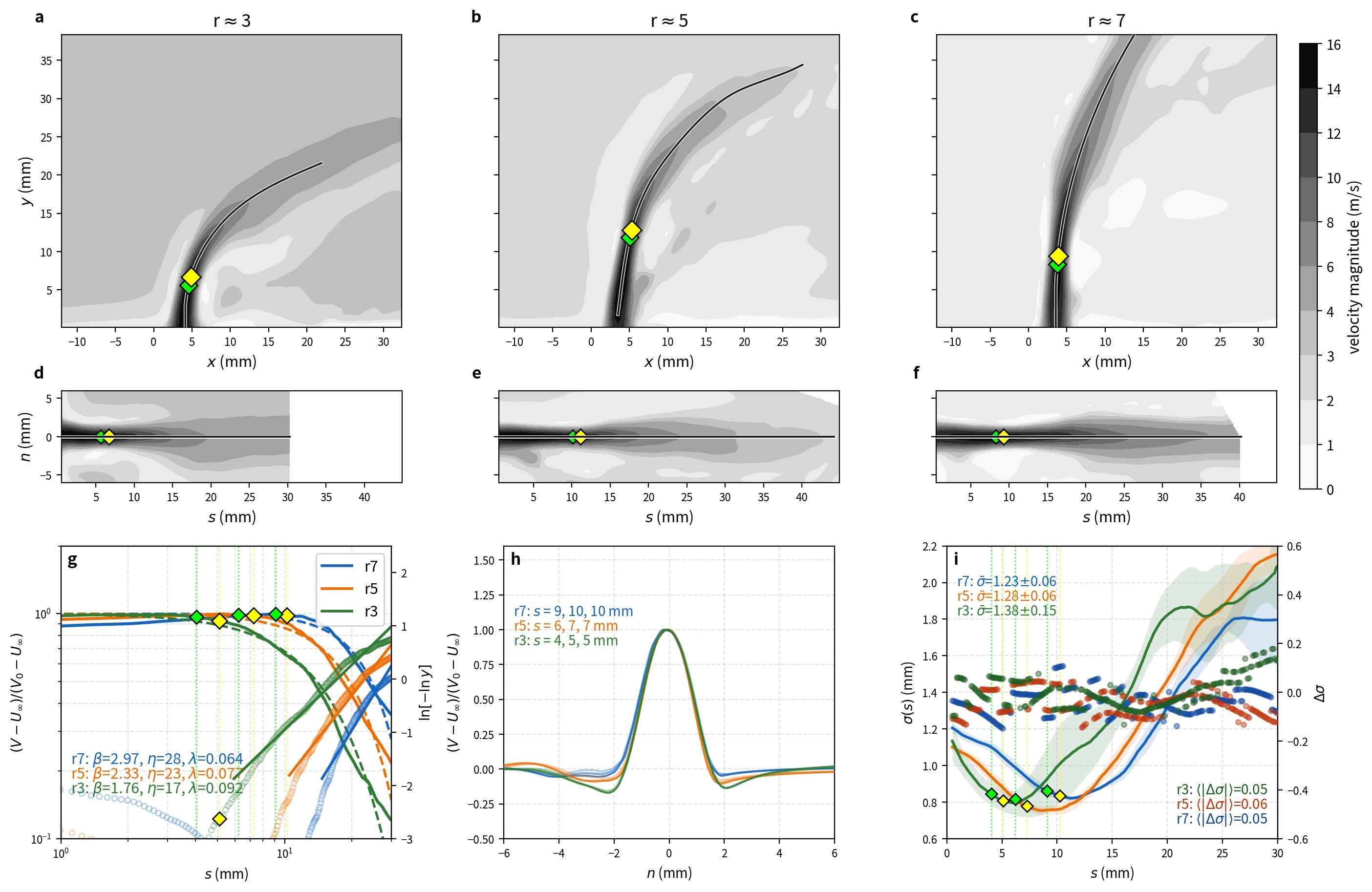}}
\caption{Single-jet examples and the natural-coordinate method. (a)--(c) The three single-jet cases ($r\approx3$, 5, 7) in physical $(x,y)$ coordinates; (d)--(f) the same in straightened $(s,n)$ coordinates (trajectory: $n=0$; diamonds: near/far transitions). (g) Excess-velocity decay with Weibull fits (dashed) and the linearized form, listing $\beta$, $\eta$, $\lambda$; (h) normalized perpendicular profiles; (i) cross-track width $\sigma(s)$ with asymmetry $\Delta\sigma$ (right axis). Ranges: $s\in[0.5,44.9]$\,mm, $n\in[-6,6]$\,mm.}
\alttext{Single-jet natural-coordinate analysis: physical and straightened fields, the excess-velocity decay with Weibull fits, and the cross-track width.}
\label{fig:new-s8}
\end{figure}

\begin{figure}
\centerline{\includegraphics[width=\textwidth]{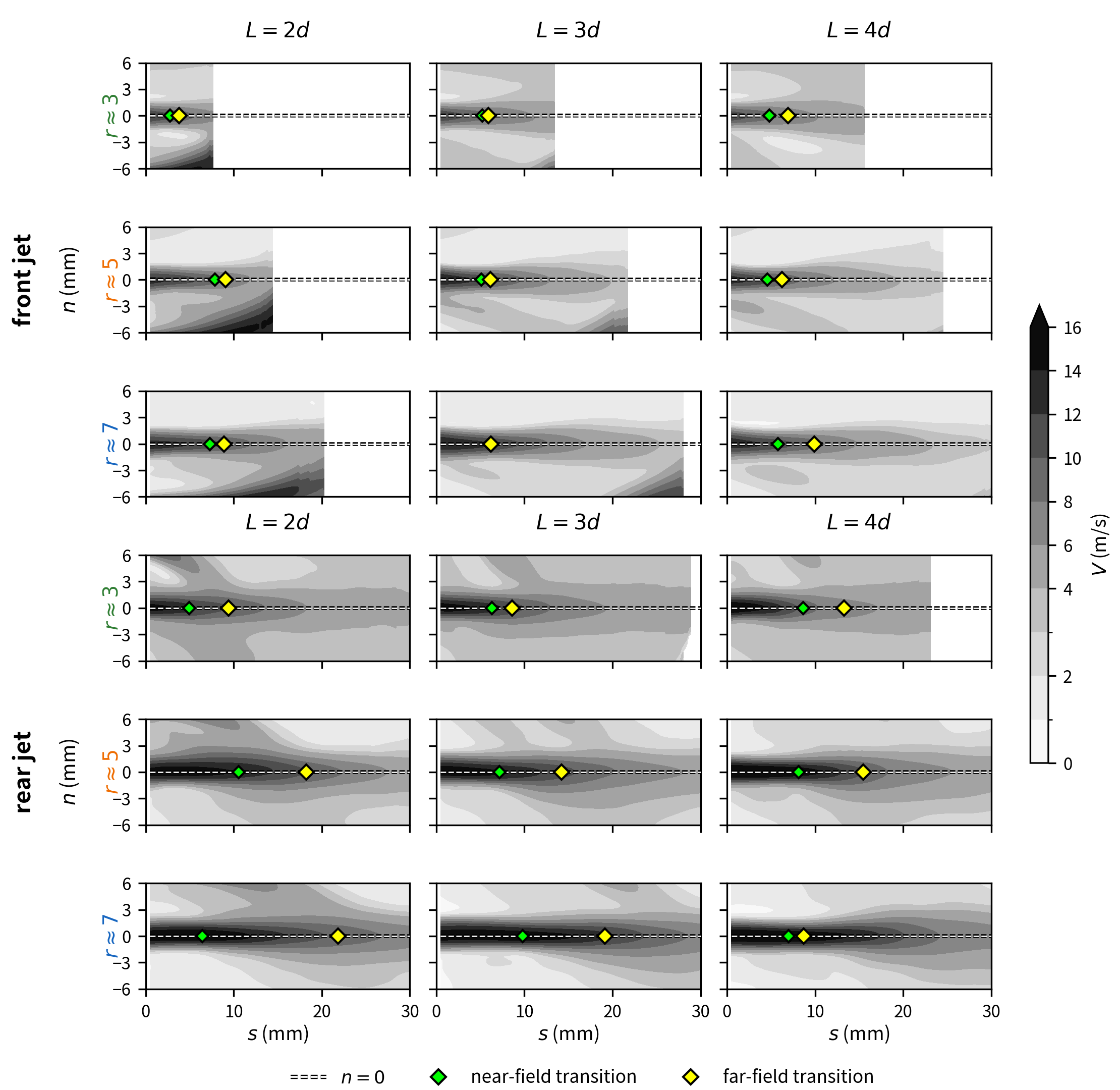}}
\caption{Condition-mean velocity-magnitude fields of the nine conditions in the natural coordinates $(s,n)$ of each trajectory (each case straightened along its own trajectory, averaged cell by cell; physical units, shared color scale, filled contour bands of $1$\,m/s below $4$\,m/s and $2$\,m/s above). Front jets (upper block) and rear jets (lower block); rows $r\approx3$, 5, 7; columns $L=2d$, $3d$, $4d$; $n=0$ dashed, diamonds: near/far transitions. The exit-excess-velocity normalization is not applied here (see Fig.~\ref{fig:nat-decay}); method: Fig.~\ref{fig:new-s8} and Algorithm~2.}
\alttext{Grid of condition-averaged velocity magnitude fields of the nine conditions in the natural coordinates of each trajectory, front jets in the upper block and rear jets in the lower block.}
\label{fig:nat-mean}
\end{figure}

To analyze the velocity field along each jet, the time-averaged field is transformed into the natural coordinates $(s,n)$ of the extracted trajectory (arc length $s$ and signed perpendicular offset $n$, with $n>0$ toward the other jet). Each case is normalized by its jet-exit excess velocity $(V_0-U_\infty)$ (Algorithm~2 of the supplementary material, which also defines the width and decay quantities below). Because every case is straightened along its own trajectory and scaled by its own exit excess velocity, the cases of a condition share a single frame and can be averaged cell by cell, with the independent single-jet records providing the reference on the same grid.
Fig.~\ref{fig:new-s8} shows the single-jet examples of Fig.~\ref{fig:new-s3} in physical and natural coordinates together with the analysis method.
The twin-jet fields are straightened in the same way, with the front and rear jets on their own trajectories (condition-mean fields in Fig.~\ref{fig:nat-mean}).

The peak velocity of the rear jet is systematically larger than that of the front jet, the velocity-field counterpart of the deeper penetration of the shielded rear jet and of the stronger crossflow impact on the front jet. In the gap between the two nozzles the near-wall flow is reversed ($u<0$), while above the recirculation the crossflow accelerates over the rear jet. These are the signatures, respectively, of the front jet bending into the gap recirculation and of the rear jet developing in the shielded region.

Under the effect of the main flow the jet cross-sectional area increases and the initial momentum is redistributed about the bent trajectory: the counter-rotating vortex pair develops in the deflection region and entrains crossflow fluid. The centerline velocity therefore declines along the trajectory~\citep{karagozian_transverse_2010}.

\begin{figure}
\centerline{\includegraphics[width=\textwidth]{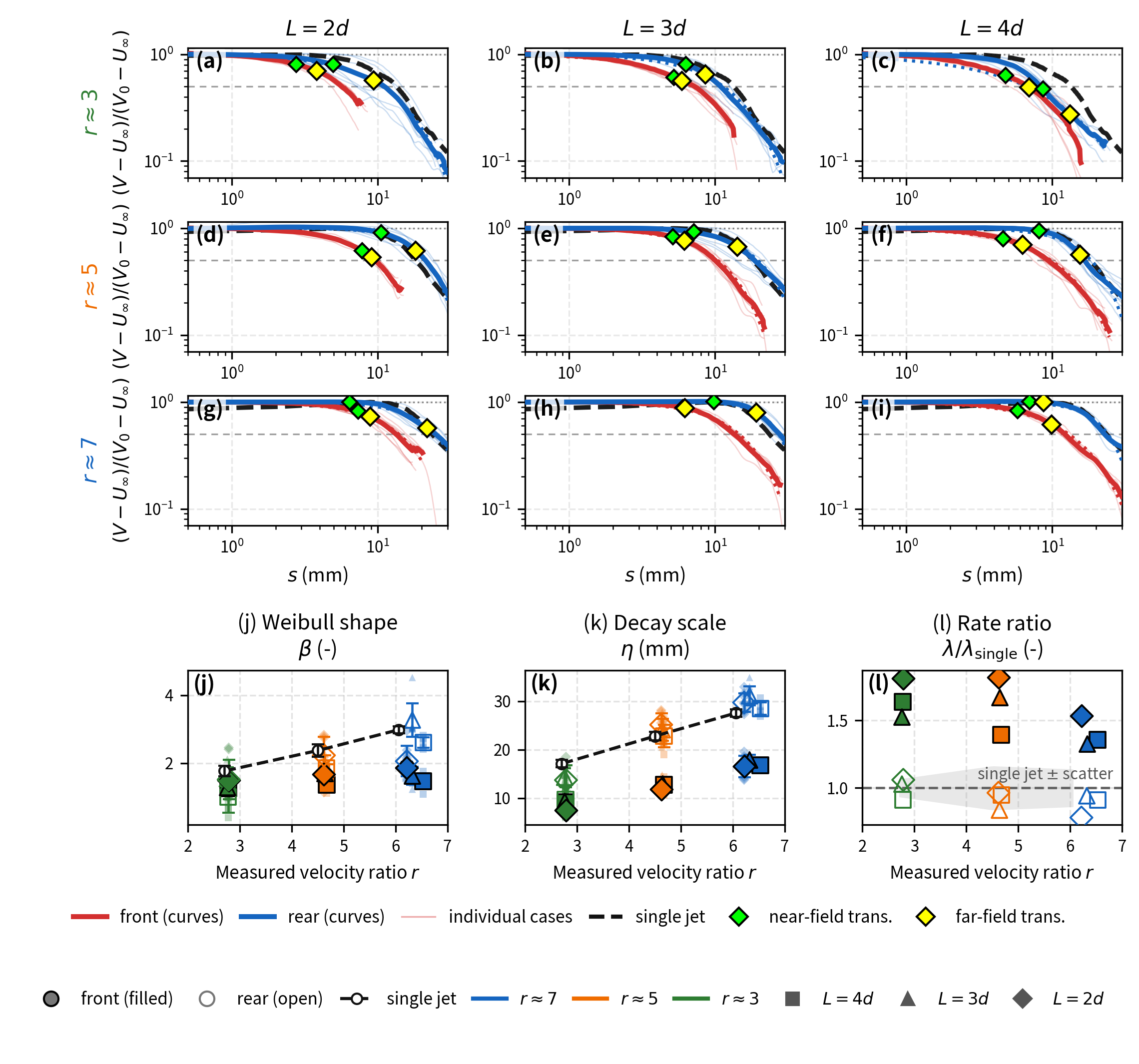}}
\caption{Condition-mean excess-velocity decay in the natural coordinates (rows $r\approx3$, 5, 7; columns $L=2d$, $3d$, $4d$). (a)--(i) $(V-U_\infty)/(V_0-U_\infty)$ versus arc length $s$ (front red, rear blue; light lines: individual cases; heavy dashed: single-jet reference; diamonds: transitions). (j)--(l) Weibull shape $\beta$, decay scale $\eta$ and rate ratio $\lambda/\lambda_{\rm single}$ versus measured ratio (symbol convention of Fig.~\ref{fig:traj-grid}). Width curves: Fig.~\ref{fig:new-s910}.}
\alttext{Nine-panel grid of the condition-mean excess-velocity decay along the natural-coordinate arc length for the front and rear jets with Weibull fits and single-jet references, plus the fitted Weibull parameters versus the velocity ratio.}
\label{fig:nat-decay}
\end{figure}

Figure~\ref{fig:nat-decay} shows the excess-velocity decay of the front and rear jets together with the single-jet reference: the interaction cleanly separates the two jets. The front jet decays faster than the single jet in every condition, while the shielded rear jet follows the single-jet curve within the case-to-case scatter. Quantitatively, the front-jet exponential rate exceeds the single-jet value by the distribution metric $E_{\rm lam}=33$--$82\%$ ($\lambda/\lambda_{\rm single}=1.3$--$1.8$, Eq.~\eqref{eq:losses}). The excess is largest at the smallest spacing and at low-to-mid velocity ratios, whereas the rear jet decays at $0.8$--$1.1$ times the single-jet rate (per-condition values in table ST3). The decay side therefore condenses into a single percentage, $E_{\rm lam}=33$--$82\%$ (panel (p) of Fig.~\ref{fig:traj-grid}); the width side condenses into no percentage at all, because the width parameters follow the single jet within the scatter (below).

The front jet is already the outlier at the nozzle exit: its excess velocity leaves the unity plateau at the first sampled station ($s\approx0.5$\,mm) and falls by up to $19\%$ within the first 2\,mm of trajectory in every condition (supplementary material). It therefore has essentially no potential core, whereas the rear and single jets keep the plateau over the first few millimeters.

An excess-velocity decay faster than that of the single jet is the velocity-field signature of momentum mixing: the gap-side recirculation replaces high-momentum front-jet fluid by low-momentum gap fluid along the leeward layer. The front-jet centerline velocity therefore falls faster than in the single jet, and faster still where the exchange is stronger (small spacing, low velocity ratio). The measured front-jet decay therefore supports the momentum-exchange picture of Section~\ref{sec:intro}, tested directly in the companion paper. The rear jet, developing in the weakened crossflow and in the low-momentum wake of the front jet, retains a single-jet-like decay, the signature of the shielding. The front-jet core is thus destroyed by the gap-side coupling already at the first resolvable station, the earliest manifestation of the recirculation-driven exchange analyzed in the companion paper: the near-field deflection is the trigger and the faster far-field decay the feedback.

\begin{figure}
\centerline{\includegraphics[width=\textwidth]{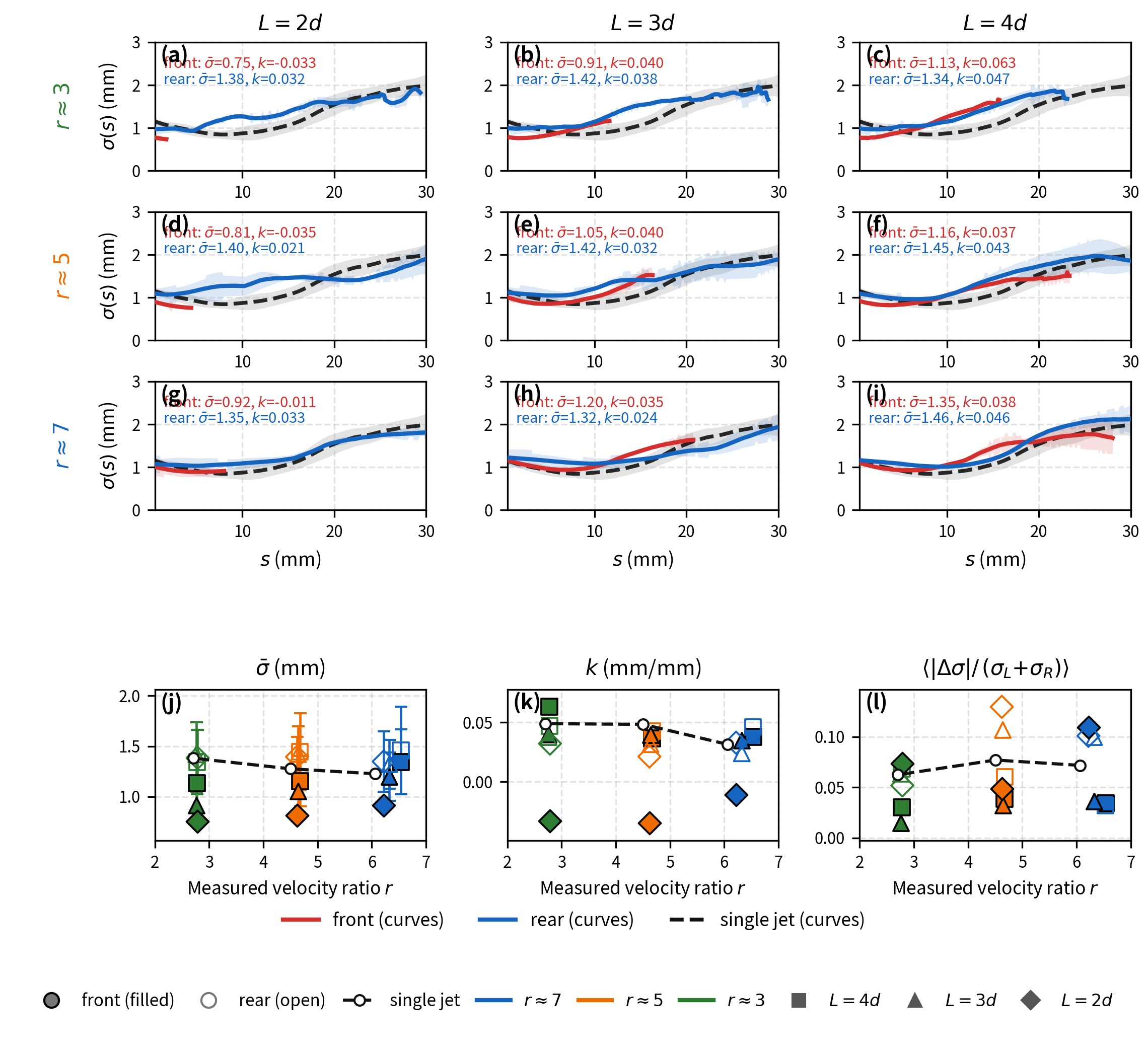}}
\caption{Condition-mean cross-track width in the natural coordinates (front red, rear blue; light bands: per-side widths $\sigma_L$, $\sigma_R$; heavy dashed with gray zone: single-jet mean $\pm$ SD). (a)--(i) $\sigma(s)=(\sigma_L+\sigma_R)/2$ versus $s$ (rows $r\approx3$, 5, 7; columns $L=2d$, $3d$, $4d$), annotated with the overall width $\bar\sigma$ and growth rate $k$; (j)--(l) $\bar\sigma$, $k$ and the relative width asymmetry $\langle|\Delta\sigma|/(\sigma_L+\sigma_R)\rangle$ (evaluated over the jet core, where the profile peak exceeds half its maximum) versus $r$ (symbol convention of Fig.~\ref{fig:traj-grid}).}
\alttext{Condition-mean cross-track width of the nine twin conditions in the natural coordinates with the single-jet reference.}
\label{fig:new-s910}
\end{figure}

The cross-track width shows how the jet column deforms rather than how it slows down: the curves of $\sigma(s)=(\sigma_L+\sigma_R)/2$ and their three fitted parameters are given for all nine conditions in Fig.~\ref{fig:new-s910}. All conditions share a V-shaped evolution, narrowing under crossflow compression to a minimum close to the near-field transition, then spreading slowly downstream. The centerline decay is therefore controlled by mixing, while the cross-section evolves through compression and deformation rather than self-similar spreading. The width parameters vary little across the conditions and the single jet follows the same V shape, so the deformation is intrinsic to the deflected jet; the systematic exception is the $L=2d$ front jet, which keeps contracting as it merges with the rear jet.

The present measurements therefore revise the conclusion of \citet{gutmark_dynamics_2011}, reached at $r=3$ and $L=3.55d$, that the front jet behaves like a single jet. Already in the near field the front-jet trajectory falls below the single-jet reference at that corner of the parameter space (the exit-relative height at $\Delta x=3$\,mm is $0.49$--$0.55$ at $r\approx3$), and in all nine conditions the whole trajectory lies below the single-jet reference while the centerline velocity decays at $1.3$--$1.8$ times the single-jet rate (table ST3 of the supplementary material). The shielded rear jet deviates in the opposite sense in eight of the nine conditions, the exception being the $L=4d$, $r\approx3$ condition, where the rear trajectory falls slightly below the reference (Table~\ref{tab:traj-params}). Both deviations share one origin: the jet--jet interaction closes off a recirculation bubble in the gap, whose reversed flow shortens the front-jet near-field segment and accelerates its decay (pressure signature in Section~\ref{sec:pressure}). The decay-rate deviation also grows as the spacing decreases, a dependence that a single-spacing comparison cannot capture.

\subsection{Decay confirms the dilution}\label{sec:lam-reff}
\begin{figure}
\centerline{\includegraphics[width=\textwidth]{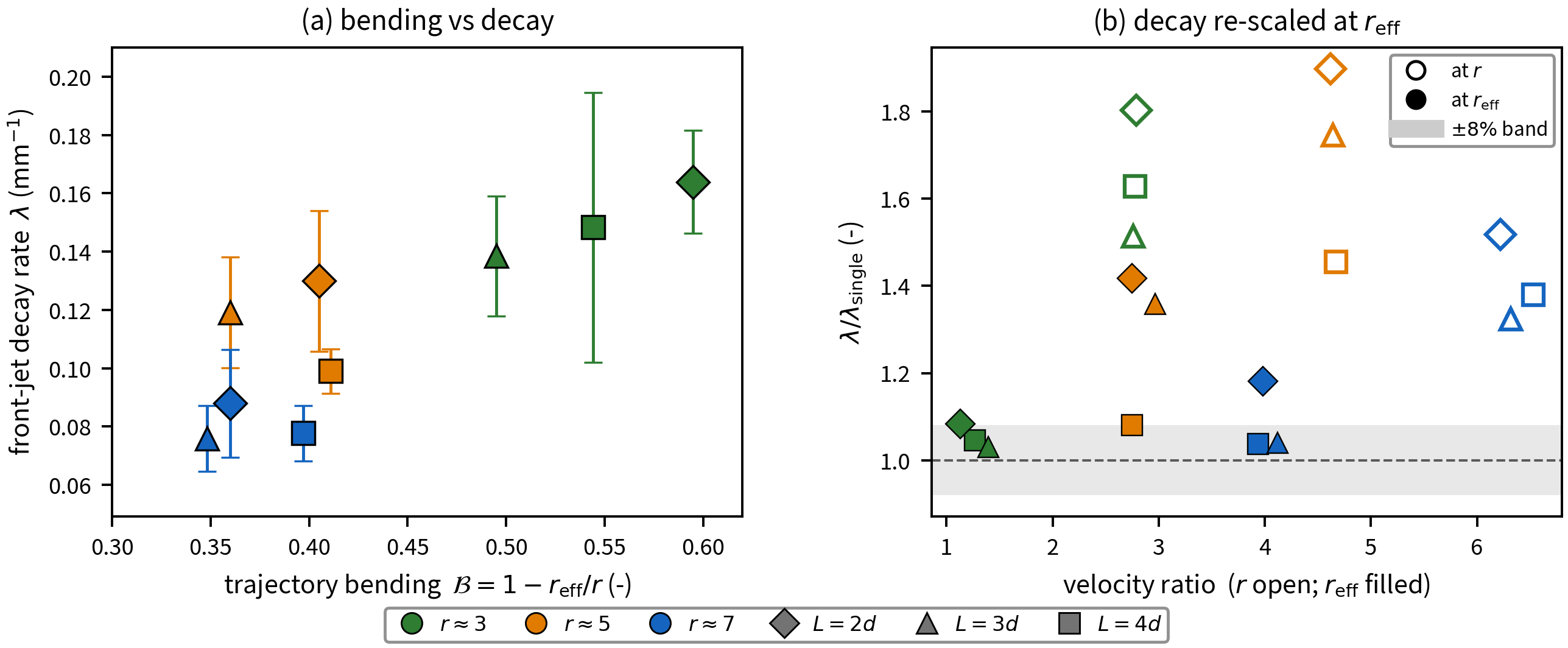}}
\caption{Bending and decay as two readouts of one dilution. (a) Front-jet decay rate $\lambda_{\rm front}$ versus the dilution-scale bending $\mathcal{B}\equiv1-r_{\mathrm{eff}}/r$ (nine condition means $\pm$ SD; condition level $\rho=+0.87$, $p=0.002$). (b) The same rates re-scaled by the single-jet law at the measured ratio (open) and at the effective ratio (filled; shaded: $\pm8\%$ band): median ratio $1.52$ at $r$ versus $1.08$ at $r_{\mathrm{eff}}$, the two largest excesses ($35$--$42\%$) falling on the $r\approx5$ conditions at $L=2d$ and $3d$. Condition-mean version: figure S6 of the supplementary material.}
\alttext{Correspondence between the trajectory bending and the decay rate: the front-jet decay rate against the dilution-scale bending, and the same rates re-scaled by the single-jet law at the measured and at the effective velocity ratio; the two largest residuals belong to the $r\approx5$ conditions at $L=2d$ and $3d$.}
\label{fig:bend-decay}
\end{figure}

The dilution picture defined by the trajectory makes a sharp, falsifiable prediction for the velocity decay measured here: the front jet should decay not at the rate of its true velocity ratio, $\lambda_{\rm single}(r)$, but at the rate of the diluted ratio, $\lambda_{\rm single}(r_{\mathrm{eff}})$. The single-jet rate is a decreasing function of $r$. Its three group means follow $\lambda_{\rm single}\propto r^{-0.56}$; the same power law is extrapolated below the lowest calibrated ratio ($r\approx2.7$), where the effective ratios of the $r\approx3$ conditions fall. Tested at the true ratio, the nine condition means of table ST3 exceed the single-jet rate by $33$--$82\%$ (the $E_{\rm lam}$ of Eq.~\eqref{eq:losses}, median $54\%$). Re-evaluating the same nine rates at $r_{\mathrm{eff}}$ collapses the excess: seven of the nine conditions fall within $2$--$18\%$ of the single-jet rate at their effective ratio (median $8\%$). The two conditions that retain a larger excess ($35$--$42\%$) are the $r\approx5$ conditions at $L=2d$ and $3d$ (the two highest filled symbols in Fig.~\ref{fig:bend-decay}b) -- the first of which is also the condition where the transition-height proxy returns its near-unity artifact (supplementary material). Both the residual pattern and the proxy failure therefore point to the same corner of the parameter space. Figure~\ref{fig:bend-decay}b shows the collapse: the paired rates sit above the single-jet calibration curve at their measured $r$ and move onto it at their $r_{\mathrm{eff}}$, within $\pm8\%$ in the median. The condition-mean version, with the calibration curve at both ratios, is given in figure S6 of the supplementary material. The interaction-induced excess decay is therefore not an additional effect superposed on a dilution. To within $8\%$ in the median, it is the single-jet decay of a jet that arrives at the far field with its ratio already reduced to $r_{\mathrm{eff}}$ by the near-field entrainment. The agreement works in both directions.

Inverted through the single-jet decay law, the measured front-jet rate returns an estimate of the effective ratio from the decay side alone, $r_{\mathrm{eff},\lambda}=(\lambda_{\rm front}/0.162)^{-1/0.563}$. Its condition means track the trajectory-based $r_{\mathrm{eff}}$ at rank correlation $0.88$ over the eight conditions of the three-estimator common subset ($0.67$ per case), with a median downward bias of $8\%$ ($10\%$ per case). The bias is concentrated in the two $r\approx5$ conditions identified above. The far-field trajectory coefficient gives a third route: the ratio form $r_{\mathrm{eff},A}/r=A_f^{\rm front}/A_s(r)$, with the single-jet law $A_s(r)=0.1433\,r+1.2384$ evaluated at the measured $r$ (always inside its calibration domain), reproduces the condition ranking at rank $0.89$ but at a uniform $19$--$31\%$ upward offset; the highest-ratio twin conditions ($r\approx6.2$--$6.5$) lie slightly beyond the single-jet calibration range ($r\le6.06$), where the law is extrapolated (panels (b) and (c) of figure S4 of the supplementary material).

The far-field coefficient retains a memory of the interaction that the whole-trajectory matching removes. All three estimators agree on the ordering of the conditions over their common subset of 52 cases, the far-field coefficient differing only in level, as noted above (figure S5 of the supplementary material).

The two readouts are also linked directly, without any ratio inversion: across the 56 cases with both measurements the trajectory bending and the decay rate rise together, the nine condition means following $\rho=+0.87$ ($p=0.002$; per case $+0.52$; Fig.~\ref{fig:bend-decay}a). The single-jet records (not shown) sit at the unbent, slowest-decaying corner. Re-scaling the same rates by the trajectory-side $r_{\mathrm{eff}}$ moves their ratios onto unity within $\pm8\%$ in the median (Fig.~\ref{fig:bend-decay}b). The jet that bends most also decays fastest, by the same dilution factor. The trajectory and the decay are therefore two readouts of one dilution parameter, and the decay-rate collapse, not any single fitted coefficient, is what certifies them. This subsection and Section~\ref{sec:trajectory} measure the interaction through two different observables (one trajectory-integrated, one local to the decaying excess velocity), and their agreement is what elevates $r_{\mathrm{eff}}$ from a trajectory descriptor to a parameter of the flow. The momentum budget of the companion paper closes the chain by locating the removed momentum in the gap.

\subsection{\label{sec:velocity}Reverse-flow region}

\begin{figure}
\centerline{\includegraphics[width=6.1in]{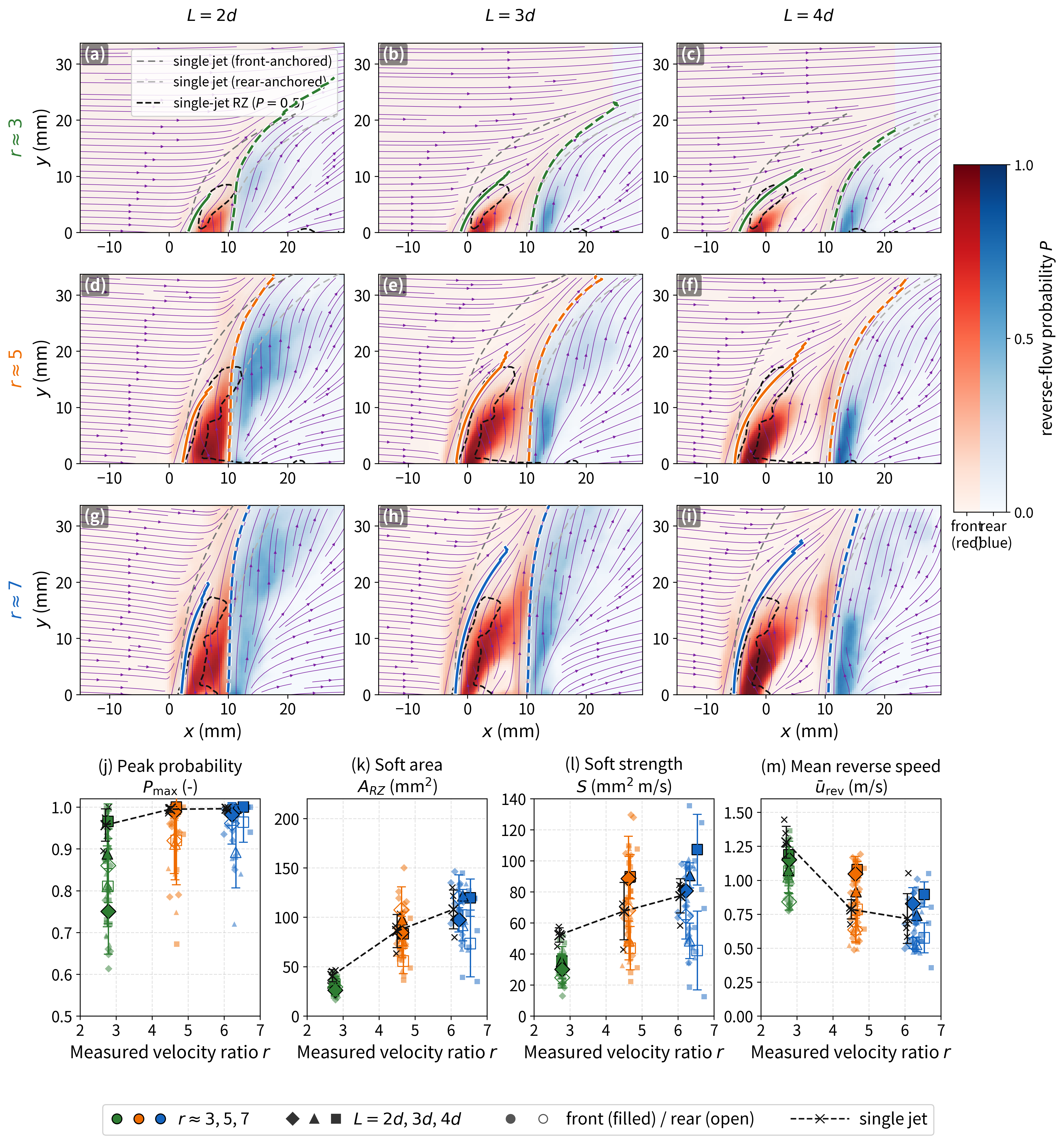}}
\caption{Reverse-flow fields and statistics of the nine conditions (rows $r\approx3$, 5, 7; columns $L=2d$, $3d$, $4d$), with the single-jet reference. (a)--(i) Condition-mean fields with streamlines and the reverse-flow probability $P$ (red: front side of the rear-jet trajectory; blue: rear side; two-sided scale, $0$--$1$), mean trajectories (front solid, rear dashed; row color) and single-jet references (gray dashed, drawn at both exits; black dashed: persistent $P=0.5$ boundary). (j)--(m) Statistics versus measured ratio with $u_{\rm thr}=0.05$\,m/s: $P_{\max}$, soft area $A_{RZ}$, soft strength $S$, mean reverse speed $\bar u_{\rm rev}$ -- per-case values (small symbols), condition means $\pm$ SD (large symbols) and single-jet records (crosses), in the symbol convention of Fig.~\ref{fig:traj-grid}.}
\alttext{Grid of condition-averaged velocity magnitude fields with streamlines and the reverse-flow probability of the nine conditions, plus panels of the peak probability, soft area, soft strength and mean reverse speed versus the velocity ratio.}
\label{fig:rz-combined}\label{fig:rz-soft}
\end{figure}

The inter-jet flow structure is examined next, in terms of whether the rear-jet blockage creates an enclosed gap recirculation, how large and strong the reverse flow is on each side of the gap, and how it compares with the wake of the single jet. The rear jet blocks part of the main flow approaching the gap between the two jets, and the streamwise velocity locally changes sign in the recirculating region that forms in the gap. The two-dimensional center-plane cross-section represents this three-dimensional region (Fig.~\ref{fig:rz-combined}a--i).

Quantitatively, the reverse-flow zone is identified on the time-averaged velocity field as the connected region in which the streamwise velocity satisfies $u<-u_{\rm thr}$, with $u_{\rm thr}=0.05$\,m/s, and its strength is defined as
\begin{equation}
RZ=\iint_{\mathcal{A}}u\,\mathrm{d}\sigma,
\label{eq:rz}
\end{equation}
where $\mathcal{A}$ is the domain of the time-averaged reverse-flow zone and $u$ is the streamwise velocity component. Because $u<0$ throughout $\mathcal{A}$, the magnitude of the integral is reported as the strength.

The threshold is chosen above the residual fluctuation level of the time-averaged fields (the robustness of this choice is documented in table ST4).
The domain is divided by the rear-jet trajectory: the reverse flow upstream of it, including the inter-jet recirculation, is attributed to the front jet, and only the reverse flow downstream of it is attributed to the rear jet.

\begin{table}
\centering
\caption{Quantitative indicators of the twin-nozzle JICF (mean $\pm$ SD over the valid cases of each condition unless noted). The far-field decay rate is the per-case exponential rate along the natural-coordinate arc length, fitted past the far-field transition; table ST3 of the supplementary material lists the rates fitted from the near-field transition, which are systematically larger; the soft reverse-flow area and strength are computed from the frame-resolved reverse-flow probability and attributed by the rear-jet trajectory; the intersection pressure is evaluated at the closest approach of the two trajectories. The per-case rates are listed in table ST3; the unsteady-dynamics indicators are collected in the companion paper~\citep{companion_part2}.}
\alttext{Table of the quantitative indicators of the nine conditions: far-field decay rate, soft reverse-flow area and strength, and the intersection pressure.}
\label{tab:metrics}
\footnotesize
\begin{tabular}{llccc}
\toprule
\textbf{Indicator} & \textbf{Jet} & \textbf{$L=4d$} & \textbf{$L=3d$} & \textbf{$L=2d$} \\
\midrule
Far-field decay rate & front & $0.076\pm0.018$ & $0.081\pm0.017$ & $0.101\pm0.034$ \\
$\lambda$ (per-case, mm$^{-1}$) & rear  & $0.046\pm0.014$ & $0.051\pm0.008$ & $0.050\pm0.014$ \\
Soft RZ area        & front & $73.7\pm37.3$  & $84.1\pm38.1$  & $69.4\pm33.4$ \\
(mm$^2$)            & rear  & $51.9\pm25.1$  & $66.6\pm28.9$  & $85.5\pm44.6$ \\
Soft RZ strength $S$ & front & $75.1\pm33.3$  & $71.8\pm27.5$  & $66.5\pm32.6$ \\
(mm$^2$\,m/s)        & rear  & $40.0\pm17.2$  & $43.2\pm12.8$  & $52.4\pm24.6$ \\
Intersection pressure & ---   & $14.8\pm4.6$   & $15.5\pm5.7$   & $34.0\pm14.8$ \\
(Pa)                 &       &                &                &               \\
\bottomrule
\end{tabular}
\end{table}

Because the reverse-flow zone fluctuates in position and strength, the reported metric is built from the frame-resolved reverse-flow probability. Thresholding the $N=400$ instantaneous fields of a record at the same level gives the probability field
\begin{equation}
P(x,y)=\frac{1}{N}\sum_{n=1}^{N}\mathbf{1}\!\left[u_n(x,y)<-u_{\rm thr}\right],
\label{eq:rz-prob}
\end{equation}
averaged over the cases of each condition (Table~\ref{tab:metrics}). The soft reverse-flow area is the integral of this probability field over the gap-side zone,
\begin{equation}
A_{RZ}=\iint P(x,y)\,\mathrm{d}\sigma,
\label{eq:rz-soft-area}
\end{equation}
and the soft reverse-flow strength is the probability-weighted reverse volume flux per unit span, with $\langle\cdot\rangle$ the frame average over the $N=400$ fields,
\begin{equation}
S=\iint\langle|u|\,\mathbf{1}_{u<-u_{\rm thr}}\rangle\,\mathrm{d}\sigma,
\label{eq:rz-soft-strength}
\end{equation}
whose ratio,
\begin{equation}
\bar{u}_{\rm rev}=S/A_{RZ},
\label{eq:rz-urev}
\end{equation}
separates the intensity of the reverse flow from its extent; the frame-resolved strength of Eq.~\eqref{eq:rz-soft-strength} is the counterpart of the time-averaged strength of Eq.~\eqref{eq:rz}.

Fig.~\ref{fig:rz-combined}(a--i) shows the condition-mean velocity magnitude and streamlines, with the reverse-flow probability shaded red on the front side and blue on the rear side. The main flow impinges on the front jet and moves around it, forming the front reverse-flow region between the jets; it then passes around the rear jet and forms the rear reverse-flow region on the leeward side. The inter-jet part is the recirculating zone identified in Section~\ref{sec:natfield}.

Fig.~\ref{fig:rz-soft}(j--m) shows the four statistics versus the measured velocity ratio: the per-case values (small symbols), the condition means $\pm$ SD (large symbols) and the single-jet records (crosses).
Combining area and intensity through the soft strength $S$, the front jet dominates at every spacing, the rear-side mean reverse speed being only about two-thirds of the front-side value at $r\approx5$ and 7. The peak reverse-flow probability $P_{\max}$ separates the two sides in the same sense, the front-side peak approaching unity as the spacing and the velocity ratio grow. The rear zone is therefore the more intermittent of the two even where its soft area is the larger.
The rear jet thus sustains the stronger front reverse-flow region; the single-nozzle JICF lies between the front and rear values at $r\approx5$ and 7 but exceeds both at $r\approx3$ (Fig.~\ref{fig:rz-soft}).

\begin{figure}
\centerline{\includegraphics[width=\textwidth]{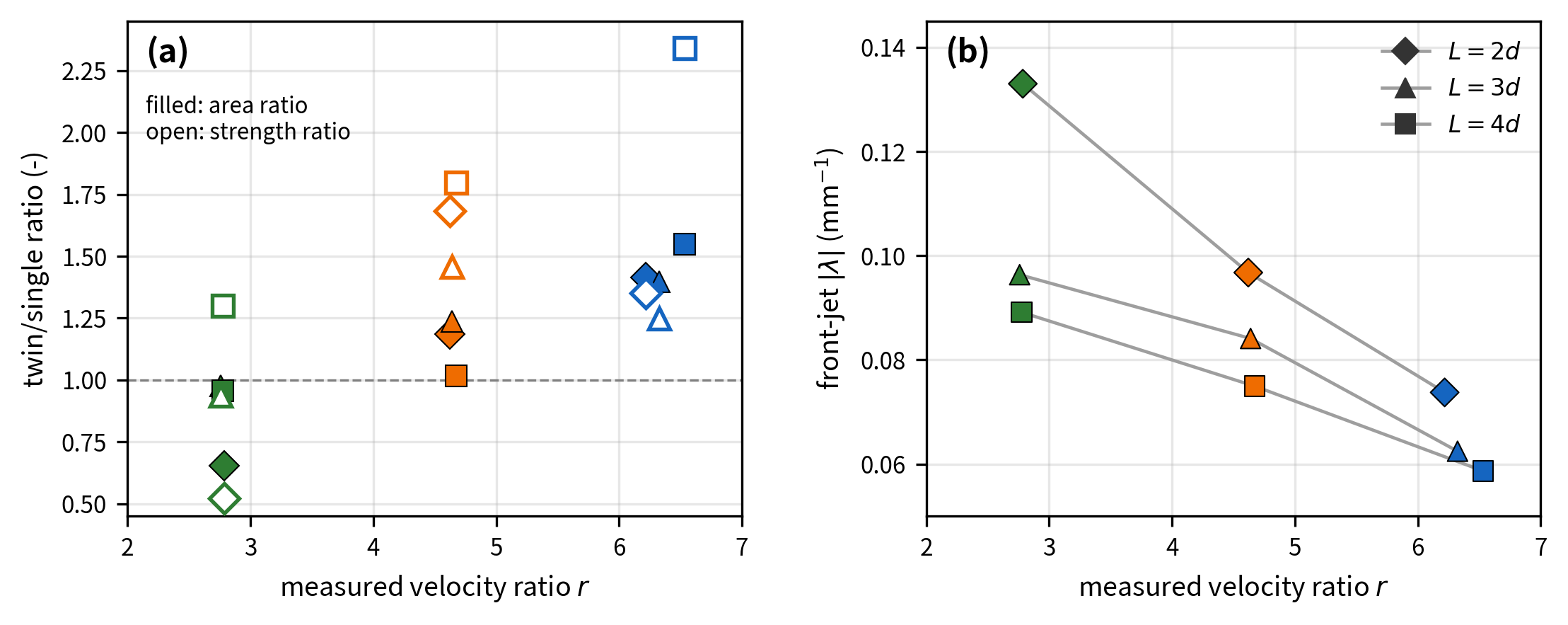}}
\caption{Recirculation impact on the front jet. (a) Twin/single area (filled) and strength (open) ratios of the front-side recirculation versus measured ratio (rear jet acting on the recirculation; color: velocity-ratio group; shape: spacing). (b) Front-jet far-field decay rate versus measured ratio, one series per spacing (recirculation acting on the front jet). The hard-threshold absolute statistics behind these ratios, compared with the single-jet reference, are given in figure S7 of the supplementary material.}
\alttext{Twin-to-single-jet ratios of the reverse-flow area and strength versus measured ratio, and the front-jet far-field decay rate versus measured ratio for the three spacings.}
\label{fig:new-s10}\label{fig:new-s11}
\end{figure}

The larger front reverse-flow region does not imply a deeper front jet: it is the separation bubble on the front jet's leeward side, into which the front jet bends, so its size reflects the blockage rather than the penetration itself.
As the spacing decreases, the front and rear zones merge at $L=2d$. The reverse-flow statistics relative to the single-jet baseline of the same ratio are given in figure S7 of the supplementary material, and their twin/single ratios in Fig.~\ref{fig:new-s10}(a). The reverse-flow metrics are not normalized by the momentum flux there, so the test is repeated with normalized quantities: dividing the front-side area by the momentum-flux ratio $J$ (and the strength by $JU_0d^2$) leaves no significant correlation with any of the modification metrics (Spearman $\rho=0.08$--$0.50$, all $p\ge0.17$ over the nine conditions), and the raw and normalized areas are themselves almost uncorrelated ($r=-0.05$). The exclusion of the recirculation as a direct driver therefore does not depend on the normalization.

Quantitatively, the time-averaged blockage reshapes the recirculation rather than merely enlarging it (same $u<-0.05$\,m/s threshold, largest connected component; the condition means are robust to this choice, table ST4). Compared with the single-jet wake at the same velocity ratio, only at $r\approx7$, where the momentum-flux ratio is largest, does the blockage add a clear excess ($1.40$--$1.55$ times the single-jet area and $1.25$--$2.34$ times its strength; Fig.~\ref{fig:new-s11}(a)).
In both configurations the recirculation grows with $r$, its upstream edge hugging the front-jet leeward side, and the bubble stays enclosed between the two jets.

The signature of the blockage is therefore not primarily the size of the recirculation but its position and strength relative to the front jet. This is exactly the geometry that focuses the reversed gap flow onto the front-jet leeward layer and turns the two gap-side layers counter-current. The quantitative characterization of this counter-current shear-layer configuration ($R_1$, defined in the companion paper~\citep{companion_part2}) and its consequences for the momentum exchange are developed there. In short, the rear jet replaces the self-induced open wake of a single jet with a stronger, enclosed gap bubble pinned to the front-jet leeward side. Stated precisely, the ordering excludes the recirculation as a mean-flow entity (its aggregate area and strength) from acting as the direct driver of the modification. What remains in play is the turbulent exchange across its gap-side layer, measured in the companion paper.

The blockage thus amplifies the recirculation where the momentum-flux ratio is high and the gap is wide, while the recirculation sharpens the bending and accelerates the decay of the front jet most where the two jets are closest and the ratio is low (Fig.~\ref{fig:new-s11}(b)); the far-field exponent follows the same ordering (Fig.~\ref{fig:traj-grid}(k)). Quantitatively, the twin/single area and strength ratios of the front-side recirculation rise well above unity at $r\gtrsim5$, while at $r\approx3$ the gap recirculation is not larger than the single-jet wake; the strongest excess occurs at $L=4d$, $r\approx7$. Whether this coupling is driven by the mean pressure force or by the momentum and entrainment exchange of the recirculating shear layers cannot be decided from the mean velocity field alone. These questions are addressed next through the reconstructed pressure field (Section~\ref{sec:pressure}) and then, in the companion paper, through the turbulent momentum exchange of the gap flow.

\subsection{Low-pressure attraction\label{sec:pressure}}

The reconstructed pressure field now tests the classical candidate for the coupling: the low-pressure attraction known from twin parallel jets. At every spacing a recirculation zone occupies the gap, into which the front jet bends immediately downstream of the exit, while the rear jet rises in the shielded wake and penetrates deeper.

The pressure distribution is reconstructed from the time-resolved PIV velocity fields by solving the Poisson equation with the mean-velocity and Reynolds-stress sources, following the standard PIV-based pressure reconstruction approach~\citep{vanoudheusden_piv-based_2013,dekat_instantaneous_2012,charonko_assessment_2010},
\begin{equation}
\nabla^{2}P = -\rho\,\nabla\cdot[(\mathbf{U}\cdot\nabla)\mathbf{U}]
-\rho\,\nabla\cdot\nabla\cdot\langle\mathbf{u}'\mathbf{u}'\rangle,
\label{eq:poisson}
\end{equation}
where $\mathbf{U}=(U,V)$ is the mean velocity, $\mathbf{u}'=(u',v')$ the fluctuating velocity, $\rho$ the air density and $\langle\cdot\rangle$ the time average. The pressure is set to $P=0$ on the domain boundary, so only pressure differences within the field are meaningful.

\begin{figure}
\centerline{\includegraphics[width=6.35in]{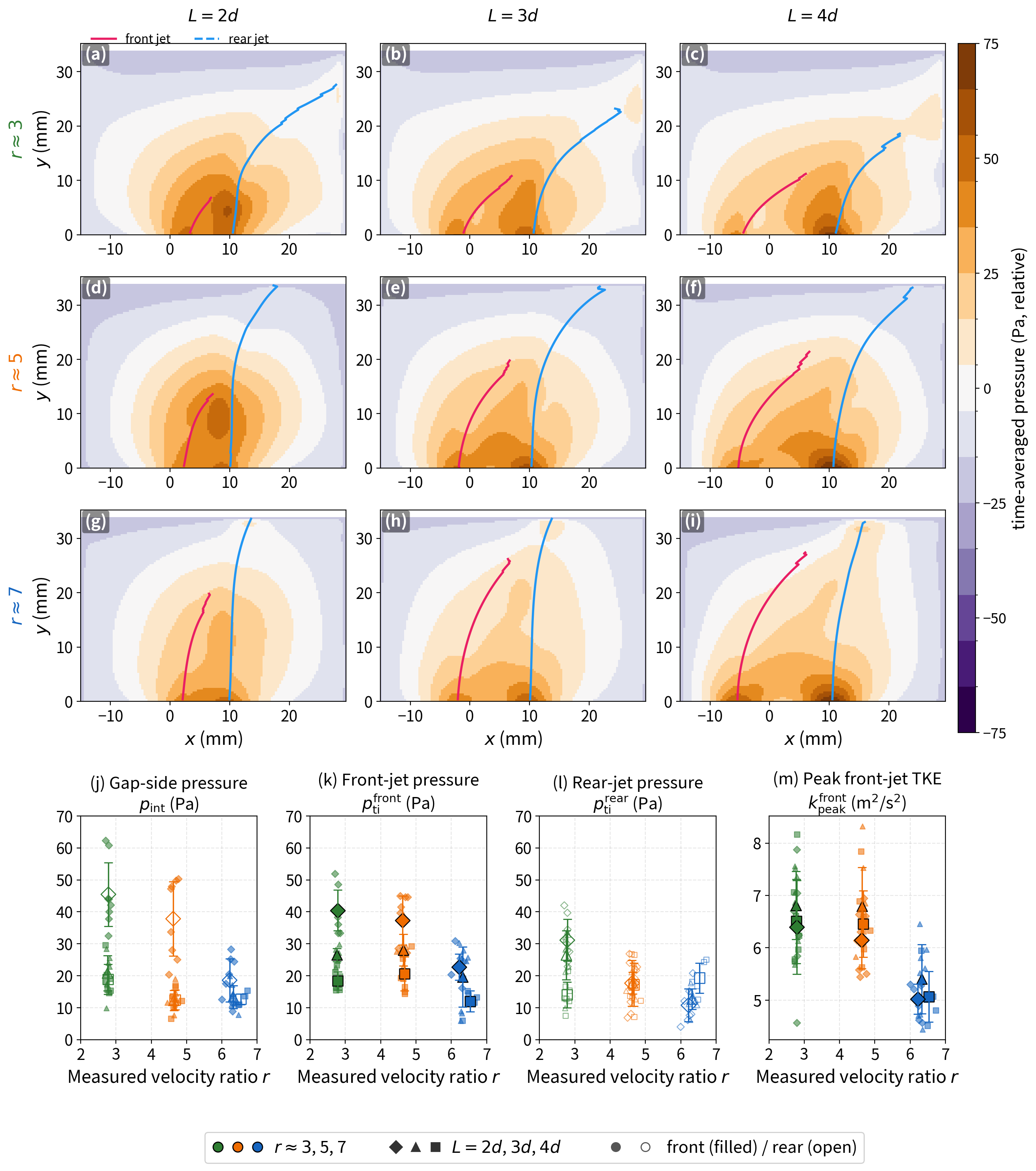}}
\caption{Condition-mean pressure fields (Pa, relative) of the nine conditions (rows $r\approx3$, 5, 7; columns $L=2d$, $3d$, $4d$), with the mean front (red) and rear (blue) trajectories. (j)--(m) Metrics versus measured ratio, per case (small symbols) and as condition means $\pm$ SD (large symbols): (j) closest-approach pressure $p_{\rm int}$; (k), (l) pressure on the front- and rear-jet trajectories at the near/far transition; (m) peak front-jet turbulent kinetic energy (own scale); (j)--(l) share the ordinate $0$--$70$\,Pa.}
\alttext{Grid of the condition-averaged reconstructed pressure fields of the nine conditions with the mean trajectories, plus panels of the gap-side pressure, the trajectory pressures, and the peak front-jet turbulent kinetic energy versus the velocity ratio.}
\label{fig:pressure-ave}
\end{figure}

In twin parallel jets without crossflow, the classical explanation of the mutual attraction is a sub-atmospheric pressure region created between the jets by their entrainment, which draws the two jets together~\citep{miller_force-momentum_1960,tanaka_interference_1970,nasr_two_1997,anderson_experimental_2001}; the reconstructed inter-jet pressure field tests that mechanism directly.
Figure~\ref{fig:pressure-ave} shows the mean pressure fields of the nine conditions (numbered 1--9 in Table~\ref{table:conditions}): the pressure is positive on the windward side of each nozzle and in the confined region between the two jets, and negative in the wake and above the nozzles. This distribution is consistent with the pressure difference between the upstream and downstream sides of the jet column associated with CVP formation~\citep{karagozian_jet_2014}.

The per-case field range $p_{\max}-p_{\min}$ is $190\pm36$\,Pa over all cases, about three times the jet-exit dynamic pressure evaluated with the bulk jet velocity ($\frac{1}{2}\rho V_j^2\approx65$\,Pa). The physically meaningful differences used throughout are the trajectory-relative values $p_{\rm int}$ ($22\pm13$\,Pa) and the windward--leeward contrasts, several times smaller than the field range.

When split at the near/far transition, the pressure sampled along the trajectories is elevated for both jets in the near field ($20.6$--$46.7$\,Pa) and relaxes toward the freestream reference in the far field (Fig.~\ref{fig:pressure-ave}k,l).
The positive inter-jet pressure is the blockage signature of the rear jet: acting as a bluff body, the rear jet decelerates the crossflow confined between the two nozzles and raises the pressure in the gap (Table~\ref{tab:metrics}). The measured pressure between the jets is elevated rather than depressed, strongest where the blockage is strongest. The elevation is not an artifact of the zero-pressure boundary condition: recomputing the gap pressure against a local reference leaves the gap-side excess positive in all cases (supplementary material).

The front-jet lowering is therefore unlikely to result from the Bernoulli low-pressure attraction that merges twin parallel jets in a quiescent ambient~\citep{miller_force-momentum_1960,tanaka_interference_1970,nasr_two_1997,anderson_experimental_2001}: the sub-atmospheric inter-jet pressure required by that mechanism is not observed in the reconstructed mean fields at any spacing. Instead, the gap pressure is elevated by the crossflow blockage. The reconstruction uncertainty (Section~\ref{sec:limitations}) prevents a strict exclusion, but the force-level comparison of the companion paper~\citep{companion_part2} shows that the lateral pressure force across the front jet is not the carrier of the extra bending. The coupling is associated with the momentum/entrainment effect of the recirculating gap flow.

The pressure along the front-jet trajectory near the nozzle exit is slightly elevated relative to the single jet, while the lateral pressure difference across the front jet at low heights is reduced, and in most conditions reversed in sign. Over $y=2$--$10$\,mm the windward-minus-leeward difference (sampled $\pm2$\,mm from the trajectory) averages $-2.9\pm4.8$\,Pa, not significantly different from zero, negative in seven of the nine conditions -- whereas the single jet keeps a positive difference at every ratio.
The cross-jet pressure difference that would drive the extra bending is thus absent in the reconstructed fields, which are consistent with the mean pressure playing no driving role, as shown in the companion paper.

As the spacing decreases, the inter-jet stagnation strengthens and the front jet bends more sharply. The two trends do not, however, track each other across the parameter set: the inter-jet pressure rises from $14.8$\,Pa at $L=4d$ to $34.0$\,Pa at $L=2d$ (Table~\ref{tab:metrics}). The whole-trajectory dilution, by contrast, is not monotone in $L$ and the near-field height ratio is deepest at $L=4d$ (see the corresponding table in Section~\ref{sec:scaling-fit}), opposite to the pressure ordering. The elevated inter-jet pressure is therefore a signature of the blockage that sustains the recirculation, not the driver of the deflection.

Reading the three pressures of Fig.~\ref{fig:pressure-ave}j--l on their common scale settles the question directly. The gap pressure reaches its largest value where the blockage is strongest -- $34.0$\,Pa at $L=2d$ against $14.8$\,Pa at $L=4d$ -- and all three pressures remain positive at every spacing, so the front- and rear-jet trajectory pressures never fall to the sub-atmospheric level that an attraction would require. The region between the jets is therefore pressurized, never sub-atmospheric, which is the opposite of what a low-pressure attraction between the two jets requires. The sign of the variation is also the reverse of what that mechanism requires: the gap pressure peaks where the front jet is bent the most, whereas the deepest near-field shortening occurs at the widest spacing. The pressure ordering is therefore anti-correlated with the shortening rather than driving it. Because the three pressures are drawn on one ordinate, these statements are read off a single panel instead of being assembled from separate figures. 

The ordering is that of a response rather than of a forcing: the mean pressure of an incompressible flow is a diagnostic field, fixed by the velocity and stress distributions through Eq.~\eqref{eq:poisson}. The blockage and the turning therefore set the pressure distribution rather than the reverse; as a diagnostic response field the reconstruction cannot strictly exclude a mechanism that itself generates the pressure, and the force-level comparison of the companion paper supplies that independent test. The gap-pressure level is anti-correlated with the shortening across the conditions in the companion paper.

\begin{table}
\centering
\caption{Resolved force-level comparison of the near-field gap-side exchange (condition means; $s=2$--$12$\,mm zone means), reproduced from the companion paper~\citep{companion_part2}. $X_{\rm turb}$ is the near-field turbulent transport, $X_{\rm pres}$ the reconstructed pressure force and $E_{ty}$ the near-field shortening of Eq.~\eqref{eq:ety}; the turbulent term exceeds the pressure term in eight of the nine conditions.}
\alttext{Table of the resolved force-level comparison of the near-field gap-side exchange in the nine conditions: the near-field turbulent transport, the reconstructed pressure force and the near-field shortening of the front jet.}
\label{tab:force-compare}
\footnotesize
\begin{tabular}{llccc}
\toprule
Condition & $r$ & $X_{\rm turb}$ (m/s$^2$) & $X_{\rm pres}$ (m/s$^2$) & $E_{ty}$ \\
\midrule
$L=2d$, $r\approx3$ & 2.79 & 3.82 & 2.72 & 0.40 \\
$L=2d$, $r\approx5$ & 4.62 & 3.19 & 1.84 & $-0.01$ \\
$L=2d$, $r\approx7$ & 6.22 & 2.77 & 1.69 & 0.18 \\
$L=3d$, $r\approx3$ & 2.76 & 2.29 & 1.90 & $-0.02$ \\
$L=3d$, $r\approx5$ & 4.64 & 3.67 & 2.36 & 0.34 \\
$L=3d$, $r\approx7$ & 6.32 & 3.33 & 2.13 & 0.32 \\
$L=4d$, $r\approx3$ & 2.78 & 2.51 & 1.68 & 0.08 \\
$L=4d$, $r\approx5$ & 4.67 & 3.32 & 2.55 & 0.41 \\
$L=4d$, $r\approx7$ & 6.53 & 2.86 & 2.91 & 0.38 \\
\bottomrule
\end{tabular}
\end{table}

The reconstructed field therefore disfavors a low-pressure attraction but does not by itself specify the exchange mechanism. The force-level comparison of the companion paper settles the question with the directly measured terms (Table~\ref{tab:force-compare}): the near-field turbulent transport on the front-jet gap side ($2.3$--$3.8$\,m/s$^2$) exceeds the reconstructed pressure force ($1.7$--$2.9$\,m/s$^2$) in eight of the nine conditions. The lateral pressure force is therefore not the carrier of the extra bending. With the two classical candidates disfavored, the remaining candidates for the modification are the turbulent momentum exchange of the gap-side shear layers and a self-excited global mode of the recirculation (the exit-velocity deficit, the gap pressure and the early merging contribute as well at $r\approx3$, Section~\ref{sec:limitations}). The mean fields cannot discriminate between the two main candidates, and that discrimination is the task of the time-resolved analysis of the companion paper, where the enclosed-recirculation momentum exchange of the gap flow is quantified.

\subsection{The scaling law\label{sec:scaling-fit}}
The effective velocity ratio was defined and measured in Section~\ref{sec:reff-def} (Fig.~\ref{fig:reff-def}), and the decay collapse at $r_{\mathrm{eff}}$ (Section~\ref{sec:lam-reff}) has confirmed it as a parameter of the flow rather than a trajectory descriptor. This section does not redefine the estimator: it fits the law and discriminates the competing forms; the channel through which the spacing enters is isolated in Section~\ref{sec:spacing}. Under the equal-density condition the effective ratio carries the momentum-flux dilution $J_{\mathrm{eff}}/J=(r_{\mathrm{eff}}/r)^2$, and the law is stated below in both forms.

The condition means of the curve-matching ratio, $r_{\mathrm{eff}}/r$ -- already displayed per case in Fig.~\ref{fig:reff-def}(e) -- are collected numerically in Table~\ref{tab:reff-stats}.
\begin{table}
\centering
\caption{Per-condition effective velocity-ratio reduction $r_{\mathrm{eff}}/r$ from the whole-trajectory curve-matching estimator (mean $\pm$ SD over the cases), the near-field height ratio $y_t/y_s$ at $\Delta x=3$\,mm, and the one-sided $t$-test against the exit-velocity baseline $0.83$. All twin cases yield a curve-matching ratio; the fifth column ($n=56$) is the transition-height estimator, whose near-unity artifacts at $r\approx3$ and at $L=2d$, $r\approx5$ mark the 12 cases without a detectable transition; the $\dagger$ on the $L=2d$, $r\approx3$ value marks the condition formerly excluded from the fits.}
\alttext{Table of the per-condition effective velocity-ratio reduction from the whole-trajectory curve matching and of the near-field height ratio at the fixed downstream offset.}
\label{tab:reff-stats}
\begin{tabular}{lcccccc}
\toprule
Condition & $r$ & $n$ & $r_{\mathrm{eff}}/r$ & $t_{y,n}^{\rm front}/t_s$ & $y_t/y_s$ ($\Delta x=3$\,mm) & $p$ (vs 0.83) \\
\midrule
$L=2d$, $r\approx3$ & 2.79 & 8 & $0.405\pm0.045$ & $0.60^{\dagger}$ & $0.554\pm0.042$ & $<0.001$ \\
$L=2d$, $r\approx5$ & 4.62 & 8 & $0.595\pm0.052$ & $1.00$ & $0.637\pm0.045$ & $<0.001$ \\
$L=2d$, $r\approx7$ & 6.22 & 8 & $0.640\pm0.044$ & $0.83$ & $0.703\pm0.047$ & $<0.001$ \\
$L=3d$, $r\approx3$ & 2.76 & 8 & $0.505\pm0.041$ & $1.02$ & $0.535\pm0.043$ & $<0.001$ \\
$L=3d$, $r\approx5$ & 4.64 & 8 & $0.640\pm0.048$ & $0.66$ & $0.631\pm0.042$ & $<0.001$ \\
$L=3d$, $r\approx7$ & 6.32 & 8 & $0.652\pm0.020$ & $0.68$ & $0.695\pm0.042$ & $<0.001$ \\
$L=4d$, $r\approx3$ & 2.78 & 7 & $0.456\pm0.027$ & $0.92$ & $0.490\pm0.044$ & $<0.001$ \\
$L=4d$, $r\approx5$ & 4.67 & 8 & $0.589\pm0.032$ & $0.60$ & $0.565\pm0.033$ & $<0.001$ \\
$L=4d$, $r\approx7$ & 6.53 & 5 & $0.603\pm0.040$ & $0.65$ & $0.657\pm0.041$ & $<0.001$ \\
\bottomrule
\end{tabular}
\end{table}
Three regularities reverse the ordering suggested by the transition-height estimator. First, every case has $r_{\mathrm{eff}}/r<0.75$ (pooled mean $0.565\pm0.093$): the dilution is a universal feature of the twin configuration, not a threshold phenomenon. Second, the primary control is the velocity ratio. A two-way decomposition of the per-case variance across the nine conditions attributes 73\% to the $r$ groups and 7\% to the $L$ groups, with the remaining 20\% in their interaction; that the two main-effect fractions sum to the fit $R^2=0.80$ of the next paragraph is a numerical coincidence, not a derivation. The front jet is \emph{least} diluted at the highest velocity ratio (the opposite of the ordering suggested by the transition-height estimator), because the jet momentum flux grows as $r^2$ while the integrated exchange is nearly flat across the ratio groups in the companion paper: the same exchange removes a diminishing fraction of a growing momentum supply. Third, the spacing contributes only a secondary modulation at fixed $r$, non-monotone across the ratio groups.

The transition-height ratio, by contrast, fails exactly where the interaction is strongest: it tracks the curve-matching ratio at $L=3d$ and $4d$ for $r\approx5$ and 7, but clusters near unity wherever the near-field segment that the detection latches onto is already distorted. The distortion is caused by the strong bending at $r\approx3$ or the early merging at $L=2d$. It therefore underestimates the dilution there. The scaling correlation built on this ratio -- a power law in the combined scale $\Pi=rL/d$ -- inherits the failure. Its apparent success ($R^2=0.84$) is an artifact of the estimator: retested against the curve-matching ratios it collapses, and against the decay-side ratios it fails outright, with the free exponent of the combined scale fitting as $+0.25$ rather than $-1/2$. The failure is structural, and the physical reading -- a momentum redistribution over an interaction area $\ell_M L$ -- is withdrawn in favor of the velocity-dilution form below (the full diagnosis and the window-robustness tests are given in the supplementary material).

The hyperbolic form is motivated by the mass-conserving dilution of a momentum-conserving jet -- an analogy that fixes the functional form, not a literal near-field mechanism, as the flux budget of the companion paper~\citep{companion_part2} shows. In the analogy, the gap-side exchange strips high-momentum fluid from the front jet and injects low-momentum gap fluid. The jet fluid that crosses the gap-side layer carries its momentum essentially unchanged, so the excess momentum of the front jet is conserved even as its mass flux grows with the entrained gap fluid. The direct flux budget of the companion paper shows this conservation to hold only where the bending is weak, because the crossflow removes along-trajectory momentum where the bending is strongest. 

Let $V_j=rU_\infty$ be the issuing jet velocity and $Q_j=V_j d$ the issuing volume flux per unit span, and let $Q_e$ be the entrained gap-fluid flux per unit span. 
Conserving the excess momentum flow rate per unit span, $\rho V_j Q_j=\rho V_j^2 d$, while the carrying mass flux grows from $\rho Q_j$ to $\rho(Q_j+Q_e)$, gives a diluted velocity $V_{\mathrm{eff}}=V_j\,Q_j/(Q_j+Q_e)$, so the effective velocity ratio of the trajectory is reduced to
\begin{equation}
\frac{r_{\mathrm{eff}}}{r}=\frac{V_{\mathrm{eff}}}{V_j}=\frac{Q_j}{Q_j+Q_e}=\frac{r}{r+c},
\qquad c \equiv \frac{Q_e}{U_\infty d},
\qquad \frac{J_{\mathrm{eff}}}{J}=\left(\frac{r}{r+c}\right)^2,
\label{eq:reff-hyperbola}
\end{equation}
where the single empirical constant $c$ is the effective capacity of the gap, measured in units of the crossflow flux through a nozzle-size area, $\rho U_\infty d$ per unit span. The same momentum is then carried by a larger mass flux, so the effective velocity ratio falls as $r/(r+c)$: a hyperbolic saturation toward unity at large $r$, not a power-law decay. The same law read in momentum units states that the effective momentum-flux ratio -- the quantity the trajectory and the decay jointly measure -- is diluted to $J_{\mathrm{eff}}/J=[r/(r+c)]^2$, the square of the velocity-dilution factor under the equal-density condition $J\propto r^2$. The conserved quantity of the derivation is the momentum flow rate $\rho V_j^2 d$, not the ratio $J=r^2$, so the conservation and the diluted $J$ are consistent. This is the finite-supply version of the classical turbulent-jet dilution argument -- the centerline dilution of \citet{hawthorne_mixing_1949} and the finite-reservoir saturation of \citet{thring_newby_1953}, whose theory gives exactly this hyperbolic form -- with $c$ playing the role of the entrainment coefficient of \citet{becker_concentration_1978}. The single capacity $c$ can be decomposed into a near-field part, measured independently by the exit-relative height ratio, and a downstream increment: writing $c=c_h(L)+\Delta c$, the height ratio gives $y_t/y_s=r/[r+c_h(L)]$ with $c_h=2.50$, $2.63$ and $3.30$ at $L=2d$, $3d$ and $4d$, and one constant increment $\Delta c=0.54$ then reproduces the nine condition means of $r_{\mathrm{eff}}/r$ with $R^2=0.78$, against $R^2=0.80$ for the constant capacity $c=3.31$. Replacing the constant by $c_h(L)$ therefore does not improve the fit: the spacing effect is concentrated in the first 3\,mm of the trajectory and is partly absorbed by the downstream mixing, so the law is stated with a constant capacity calibrated over the whole trajectory, and $c_h(L)$ is reported as its independently measured near-field component. This is also why the whole-trajectory ratio does not order monotonically with the spacing even though its near-field components do: the $2d$ and $3d$ capacities differ by less than their propagated scatter.

\begin{figure}
\centerline{\includegraphics[width=\textwidth]{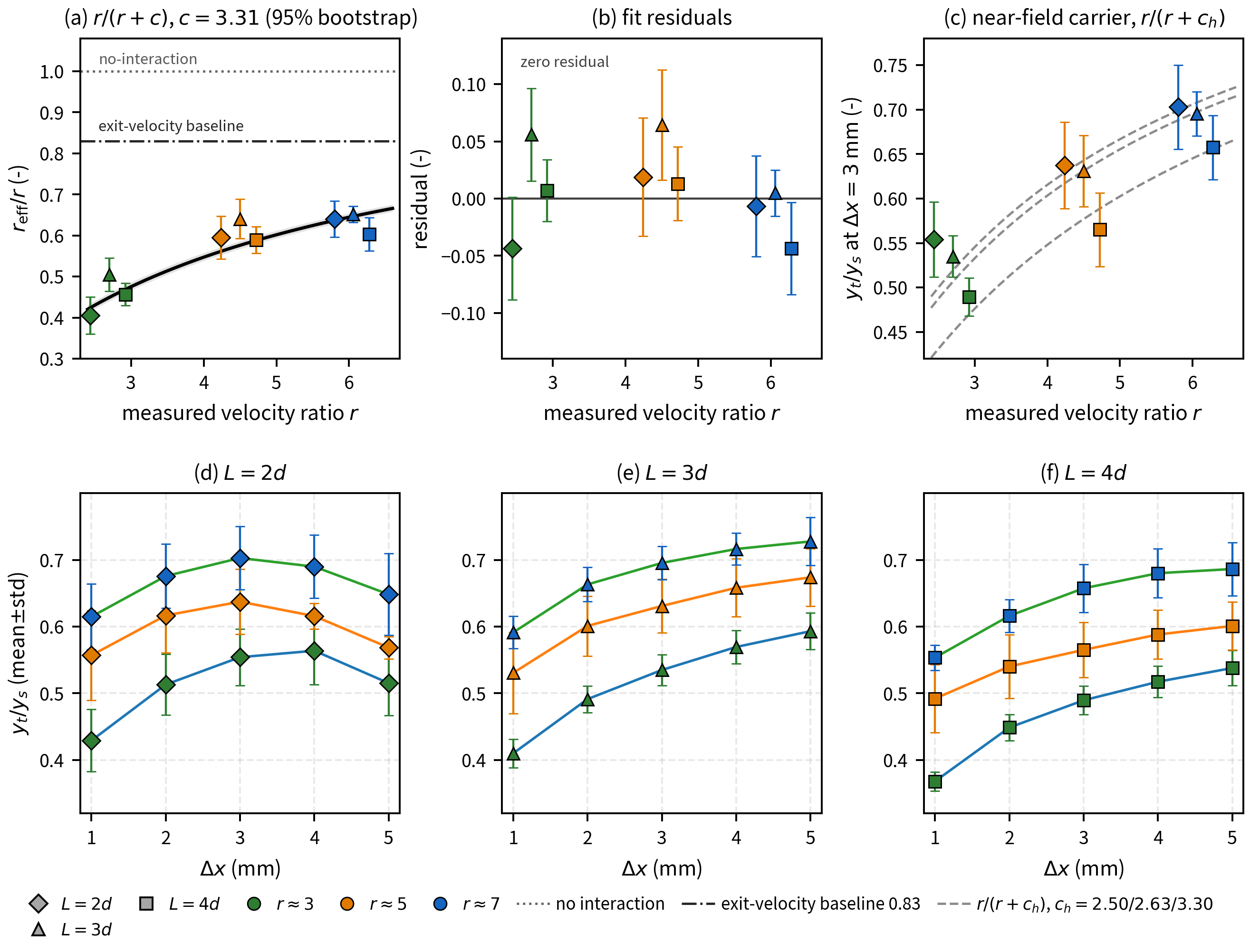}}
\caption{Velocity-dilution scaling and the near-field height carrier. (a) $r_{\mathrm{eff}}/r$ from whole-trajectory curve matching versus measured $r$ over the nine conditions (symbol convention of Fig.~\ref{fig:traj-grid}; dotted: no-interaction reference $1$; dash-dotted: exit-velocity baseline $0.83$), with the hyperbolic fit $r/(r+c)$ ($c=3.31$, 95\% bootstrap band); (b) fit residuals; (c) near-field height carrier $y_t/y_s$ at $\Delta x=3$\,mm versus measured $r$, per spacing (dashed: $r/(r+c_h)$ with $c_h=2.50/2.63/3.30$ for $2d/3d/4d$, evaluated at the group-mean measured ratio of the ratio groups; the markers sit at their own measured ratios). (d)--(f) $y_t/y_s$ versus the downstream offset $\Delta x=1$--$5$\,mm from the exit, one panel per spacing; color: velocity-ratio group; mean $\pm$ SD. Within the first 3\,mm downstream of the exit the ratio increases with $r$ and decreases with $L$ in every condition.}
\alttext{Velocity-dilution scaling of the effective ratio: the hyperbolic fit, the fit residuals and the near-field height carrier versus measured ratio, and the height ratio versus the downstream offset for the three spacings.}
\label{fig:reff-correlation}\label{fig:new-s16}
\end{figure}

Over the nine condition means, the one-parameter form of Eq.~\eqref{eq:reff-hyperbola} explains 80\% of the variance with $c=3.31$. The case-resampling bootstrap 95\% confidence interval is $[3.18,\ 3.43]$, and the total 95\% confidence interval is $[3.03,\ 4.81]$ once the reference library and the matching window are resampled as well (Fig.~\ref{fig:reff-correlation}a). The single constant is not merely fitted to the same nine conditions that test it: refitting $c$ to eight conditions and predicting the ninth (leave-one-condition-out) reproduces the held-out ratio to within $15\%$ in $r_{\mathrm{eff}}/r$ (RMSE $0.042$, predictive $R^2=0.75$).

The form is fitted with a single constant to the nine conditions, which span both controls, $2\le L/d\le4$ and $2.7\le r\le6.7$. The law is a statement at the condition level: the per-case ratios scatter by $\pm0.05$--$0.12$ about their condition means, wider than the fit residuals of $\pm0.03$ (supplementary material), so ``holding for all analyzed cases'' is read at the condition level. The interval is the case-resampling uncertainty at the fixed primary window; the estimator's calibration freedom is examined next (figure S4(c) of the supplementary material).

The competing two-parameter affine alternative explains less of the leave-one-out prediction error, and the fit residuals carry no systematic structure in either control (Fig.~\ref{fig:reff-correlation}b): the form absorbs both controls with no systematic leftover. The parameter drift diagnostic reinforces the reading: fitting $c$ separately to each velocity-ratio group returns a 22\% spread, whereas the momentum-dilution variant -- the square-root form in which the conserved quantity is the momentum flux itself -- drifts 29\% and fits worse. The velocity-dilution form is both the better fit and the more stable one. The flux-budget diagnostics of the companion paper settle the status of the motivating analogy: within the resolvable pre-merging window the dilution appears as a centerline velocity deficit and a loss of along-trajectory momentum flux to the crossflow rather than as a net growth of excess mass flux. Hence $c$ is the window-calibrated effective capacity constant of the law, and the mass-addition bookkeeping above holds only as the integrated description of the post-merging evolution. 

The decay side supports the same form through the trajectory-side ratio itself rather than through an independent estimator: re-scaling the measured decay rates by $r_{\mathrm{eff}}$ collapses them onto the single-jet law (Fig.~\ref{fig:bend-decay}b and Section~\ref{sec:lam-reff}). The bending and the decay rate rise together across the conditions (Fig.~\ref{fig:bend-decay}). The trajectory and the decay are therefore two readouts of the one dilution parameter.

The matching window is the single calibration choice of the estimator, and its sensitivity is quantified directly. The primary window $\tilde u\in[0.05,\,2.30]$ covers the near field and the early bending region of all ratio groups. Shrinking it to the narrow near-field window $\tilde u\in[0.05,\,0.50]$, which excludes the bending region altogether, extracts deeper ratios at $r\approx3$ ($0.27$--$0.30$ against $0.41$--$0.51$) while leaving the $r\approx5$ and 7 groups nearly unchanged. The ordering of the ratio groups ($r\approx3<5<7$) is therefore preserved within every spacing, and the fit constant moves from $c=3.31$ ($R^2=0.80$) to $c=4.45$ ($R^2=0.75$). The form of the law is therefore a property of the flow, while the constant carries a window systematic of order $+34\%$, an order of magnitude larger than the case-resampling confidence interval of the primary fit. Resampling the reference library and the matching window together with the cases (400 draws per layer) widens it to the total 95\% interval $[3.03,\ 4.81]$ reported above. 

Accordingly $c$ is a window-calibrated effective capacity constant, and its uncertainty is stated as the total interval. The affine two-parameter alternative, which fits marginally better in the narrow window ($R^2=0.83$ against $0.75$) but worse in the primary one, is disfavored on parsimony (the full comparison of the fitted forms is given in figure S4(c) of the supplementary material).

The constant $c\approx3.3$ admits a reading as the entrainment capacity of the gap: the dilution bookkeeping of Eq.~\eqref{eq:reff-hyperbola} requires an entrained flux $Q_e\approx c\,U_\infty d$ per unit span. Evaluated at the measured free-stream velocities, this flux is nearly $r$-invariant, consistent with the $r$-flat integrated exchange $S_{uv}$ measured in the companion paper. The direct flux budget of the companion paper shows, however, that this flux is not drawn across the gap-side edge within the resolvable pre-merging window (the measured growth there is more than an order of magnitude smaller). The bookkeeping therefore describes the integrated post-merging evolution rather than a literal near-field entrainment. The finite entrainment capacity together with the $r^2$ growth of the jet momentum is what makes the dilution saturate. The same recirculating gap fluid, re-entrained many times, can only dilute the jet by a bounded factor, and the bound is approached from below as $r$ grows.

\subsection{The near-field capacity\label{sec:spacing}}
The velocity ratio carries the primary trend of the dilution, but the spacing $L$ is not without effect: the condition means of Table~\ref{tab:reff-stats} at fixed $r$ spread by up to $\pm0.06$ about the hyperbolic fit. The whole-trajectory ratios order the three spacings non-monotonically across the ratio groups. The $L$-dependence is instead carried cleanly by a near-field measure: the exit-relative height ratio $y_t/y_s$, the height of the front-jet trajectory at a fixed small downstream offset $\Delta x$ from the exit ($1\le\Delta x\le5$\,mm), scaled by the single-jet height at the same offset. At every offset up to $\Delta x=3$\,mm this ratio is strictly ordered in both controls, falling as $L$ grows and rising as $r$ grows (Fig.~\ref{fig:new-s16}(c)--(f)), even though the whole-trajectory ratios are not.
Reading each spacing through the hyperbolic form Eq.~\eqref{eq:reff-hyperbola}, $y_t/y_s=r/(r+c_h)$, returns a spacing-dependent capacity $c_h=2.50$, $2.63$ and $3.30$ at $L=2d$, $3d$ and $4d$ (the $\Delta x=1$\,mm station returns larger values; supplementary material). The wider gap has the larger near-field entrainment capacity -- the widest gap hosts the strongest reversed core, as reported in the companion paper. The near-exit deflection is therefore deepest at $L=4d$ even though the whole-trajectory dilution is not monotone in $L$. The spacing effect is real but localized: it orders the first $3$\,mm ($0.1$--$0.3d$) of the trajectory and is absorbed into the scatter of the whole-trajectory ratio beyond it.

In the near field the measured effect is a shortening of the front jet's straight segment and an earlier deflection of its trajectory. At the exit-relative offsets $\Delta x\le3$\,mm the front-jet height is everywhere below the single-jet reference of the same ratio group, the depth of the deficit growing with the spacing (Fig.~\ref{fig:reff-correlation}c). The first tracked point lies at or slightly downstream of the nozzle exit, so the whole effect is a deflection rather than an upstream displacement of the base.

The same reversed gap flow underlies each front--rear asymmetry: the front-jet trajectory lies below the single-jet reference and its centerline velocity decays faster than the single jet, whereas the rear-jet trajectory lies above it and its decay rate remains close to the single-jet value.
The front-jet decay rate is largest at the smallest spacing for $r\approx5$ and 7, and non-monotone in the spacing at $r\approx3$ (Section~\ref{sec:natfield}), while the near-field turbulent transport varies across the nine-condition grid without a monotone spacing dependence, as reported in the companion paper. The velocity-dilution scaling absorbs this non-monotone spacing modulation only in part -- the whole-trajectory ratio of Table~\ref{tab:reff-stats} does not order monotonically with $L$, while its near-field component $c_h$ does grow with the spacing; the spacing does not order the decay rate and the whole-trajectory ratio consistently (supplementary material). The coupling of the exchange to the trajectory is therefore a function of the velocity ratio alone, to first order.

\subsection{Mechanism, generality and testable predictions}
\label{sec:mechanism}\label{sec:generality}\label{sec:predictions}
The mean fields and their scaling assemble into one chain. The rear jet blocks the crossflow and closes a recirculation in the gap; the front jet, whose leeward layer is swept by the reversed gap flow, bends earlier, decays faster and stays lower, while the shielded rear jet rises higher. The reverse-flow region (Section~\ref{sec:velocity}) and the reconstructed mean pressure (Section~\ref{sec:pressure}) are each inconsistent with the observed ordering of the modification -- the gap pressure is elevated, not depressed, and the lateral pressure difference across the front jet is not enhanced within the reconstruction uncertainty -- so the classical Bernoulli low-pressure attraction of twin parallel jets is disfavored. The quantitative content of this paper is nevertheless the velocity-dilution scaling of Eq.~\eqref{eq:reff-hyperbola} and its verified consequences, which hold irrespective of the causal interpretation: the turbulent momentum exchange that carries the modification is shown in the companion paper~\citep{companion_part2} to scale with the shortening at the condition level, with its resolved turbulent term exceeding the resolved pressure term in eight of the nine conditions, and to be identified there together with the unsteady shear-layer dynamics it sustains. The evidence behind the chain is observational, and the exclusion of the two classical alternatives narrows the remaining candidates to the turbulent momentum exchange of the gap-side layers or a self-excited global mode; the exchange is large in absolute terms exactly where the shortening is largest, and in an experiment without controlled intervention this is the strongest causal statement available -- the reverse pathway, the early bending strengthening the exchange, cannot be excluded from the present records.

The coupling picture yields three expectations for the unsteady dynamics, which the companion paper examines. (i) The gap-side shear layers, being convectively unstable, should roll up into shear-layer vortices that convect downstream, with no evidence of a self-excited global mode. (ii) The front and rear jets, whose layers share the same near-field shear-layer instability modified only weakly by the exchange, should shed at comparable frequencies. (iii) The windward and leeward vortex sheets of the two jets should merge downstream, mirroring the convergence of the trajectories, as already documented for tandem jets~\citep{zang_particle-image_2014}.

The mean trajectory and the unsteady dynamics are coupled in both directions: the Reynolds stresses enter the mean-momentum balance and the pressure reconstruction of Eq.~\eqref{eq:poisson}, so the deflection region is where the turbulent exchange is produced. In the opposite direction, the modified trajectories set the geometry and the strength of the shear layers. The unsteady vortex dynamics, analyzed in the companion paper, provide the dynamic counterpart of the front--rear asymmetry established here from the time-averaged fields.

The dilution law is geometric in origin: a tandem arrangement in which the rear jet closes a confined recirculation against the front jet is expected to exhibit the same front-low/rear-high asymmetry within the measured regime, as long as the spacing is small enough for the recirculation to close, while its invariance to nozzle geometry and Reynolds number remains to be tested. At $r\approx3$ the front-jet trajectory nonetheless lies below the single-jet reference at every spacing, a residual lowering that the present records cannot decompose: the exit-velocity deficit (baseline $0.83$), the elevated gap pressure (largest at $r\approx3$, $L=2d$) and the early merging of the two jets all contribute, and the two upstream supply lines were not instrumented separately. The same corner carries a calibration caveat: the companion paper locates the boundary of the closed-recirculation regime at $r\approx3$, where the exchange enhancement falls to a marginal value, yet the dilution law is deepest there, so the fit is a configuration-level statement rather than a pure exchange measurement.

This regime view reconciles the earlier observations (Table~\ref{tab:lit-compare}). The study of \citet{gutmark_dynamics_2011} ($r=3$, $L=3.55d$) asks the same question with the same peak-velocity trajectory definition, and the present measurements reproduce their findings on the shielded, deeper-penetrating rear jet and on the faster decay of the front jet relative to the rear jet. The whole-trajectory matching then adds the missing front-jet comparison: in all nine conditions the front jet falls below the single-jet reference, the deviation being largest at the low ratios where their measurement was made, and the near-field similarity they reported is not recovered (Section~\ref{sec:spacing}). \citet{zang_particle-image_2014} reported the upstream jet as fairly similar to the single jet from a two-spacing series. \citet{new_trajectory_2015} collapsed the merged tandem trajectory with an effective velocity ratio, whereas the present front-jet effective ratio is a whole-trajectory match of the front jet to a single-jet family; \citet{zhao_experimental_2018} documented the rear-jet shielding at a single spacing, to which the present results add the front-jet modification it implies. The earlier studies therefore differ in scope as well as in method.

\begin{table}
\centering
\caption{Comparison with the earlier twin/tandem jet studies discussed here. Entries marked ``---'' are not varied or not stated in the cited study.}
\alttext{Table comparing the present measurements with the earlier twin and tandem jet studies discussed here: configuration, velocity ratio, jet spacing, diagnostic and the reported front-jet finding.}
\label{tab:lit-compare}
\footnotesize
\resizebox{\textwidth}{!}{%
\begin{tabular}{llllll}
\toprule
Study & Configuration & $r$ & $L/d$ & Diagnostic & Front-jet finding \\
\midrule
\citet{gutmark_dynamics_2011} & two circular JICF & $3$ & $3.55$ & PIV & near-field penetration comparable to single \\
\citet{zang_particle-image_2014} & two tandem circular & $1$--$8$ & $2$, $3$ & PIV, PLIF & fairly similar to single \\
\citet{new_trajectory_2015} & two tandem circular, close & --- & close & PIV & merged-trajectory collapse \\
\citet{zhao_experimental_2018} & two tandem planar & --- & single & PIV, POD & rear-jet shielding \\
\citet{kristo_time-resolved_2021} & three-jet tandem array & $0.9$--$1.7$ & array & TR-PIV, POD & downstream jets similar \\
present & two tandem circular & $2.8$--$6.5$ & $2$, $3$, $4$ & 40\,kHz TR-PIV & diluted in all nine conditions \\
\bottomrule
\end{tabular}}
\end{table}

Three predictions follow from the law and can be tested without repeating the present experiment. First, the effective capacity $c$ is a geometric property of the gap and of the rear-jet blockage, so it should vary monotonically with them while remaining invariant to the velocity ratio. Second, the estimator is portable: the same whole-trajectory matching measures $r_{\mathrm{eff}}$ for any tandem configuration, unequal nozzles and non-circular exits included, with the single-jet family as the only reference library required. Third, the part of the dilution carried by the gap-side exchange must weaken or disappear where the rear jet can no longer close the gap recirculation -- at spacings beyond the closed-recirculation threshold, or when the rear jet is inclined away from the front jet; the exchange measurements of the companion paper locate that threshold, while the total dilution, deepest at the low-ratio end of the present range, carries additional contributions there.

\subsection{Limitations}\label{sec:limitations}
Several limitations qualify the conclusions. The measurement is two-dimensional, so the spanwise fluctuations $w'$, the spanwise stress gradients and the full three-dimensional structure of the recirculation are not resolved (their consequences for the momentum balance are discussed in the companion paper~\citep{companion_part2}), and the pressure is reconstructed relative to a zero boundary condition, so only pressure differences within the field are meaningful, with a case-to-case scatter comparable to some of the reported differences (Section~\ref{sec:pressure}). The front and rear jets also exit at different measured near-exit velocities ($12.8$ versus $15.4$\,m/s at $\approx0.5$\,mm), an asymmetry that enters every front--rear comparison; part of it is the near-exit signature of the interaction itself (the front jet lacks a potential core), and the low-momentum counterfactual (Fig.~\ref{fig:traj-single}) shows that even attributing the entire difference to the supply explains only one quarter to one half of the front-jet penetration loss (Section~\ref{sec:trajectory}), although a residual supply-side contribution cannot be excluded because the two upstream lines were not instrumented separately.

The fixed normalization $u_j=14$\,m/s carries about $\pm10\%$ uncertainty into the shear-layer normalizations of the companion paper, and the effective ratio $r_{\mathrm{eff}}$ is calibrated by a single-jet library spanning only the three measured ratio groups: at $r\approx3$ the diluted ratios fall below the lowest calibrated group, where the reference is the $A_f(r)$ extrapolation of the single-jet penetration law, while the high-ratio end ($r\approx6.2$--$6.5$) lies slightly beyond the single-jet calibration range; the near-field proxy additionally assumes that the single-jet group-mean transition height varies linearly with $r$, and its failure mode is the transition detection itself (12 cases). The two estimators agree at $r\approx5$ and 7, and the proxy's failure corner is exactly where the curve-matching values are the authority. Finally, two conditions show no detected shortening of the front-jet near-field segment ($L=2d$, $r\approx5$ and $L=3d$, $r\approx3$) and in one condition the near-field transition detection is scattered ($L=3d$, $r\approx7$); they are kept in the statistics rather than excluded, which makes the reported correlations conservative.

\section{Conclusions}\label{sec:concl}
Time-resolved PIV at 40\,kHz examined tandem jets in crossflow over nine spacing--velocity-ratio conditions ($L=2d$--$4d$, $r\approx3$--$7$) with a single-jet baseline. This paper has characterized the mean-flow phenomenology and condensed the modification into a one-parameter scaling law.

\textbf{The front jet is not passive.}
Contrary to the finding of \citet{gutmark_dynamics_2011} at $r=3$ and $L=3.55d$, its trajectory lies below the single-jet reference in every condition, and the rear-jet trajectory lies above it in eight of the nine conditions, with the two trajectories converging as the spacing decreases. The near-field segment shortens by up to 40\% in seven of the nine conditions and the centerline velocity decays at $1.3$--$1.8$ times the single-jet rate. By contrast, the shielded rear jet penetrates $1.4$--$2.0$ times as deep as the front jet (far-field coefficient $A_f$).

\textbf{The classical candidates are inconsistent with the mean fields.}
The reverse-flow region scales with the geometry but not with the ordering of the modification (Section~\ref{sec:velocity}) -- what is excluded is the recirculation as a mean-flow entity acting as the direct driver, not the turbulent exchange across its gap-side layer measured in the companion paper~\citep{companion_part2}. The reconstructed mean pressure is inconsistent with the classical low-pressure-attraction mechanism: the inter-jet pressure is elevated, not depressed, at every spacing, and the lateral pressure difference across the front jet is weakened, not enhanced (Section~\ref{sec:pressure}). With the two classical candidates disfavored, the remaining candidates for the modification are the turbulent momentum exchange of the gap-side layers and a self-excited global mode; the time-resolved analysis of the companion paper distinguishes the two in favor of the exchange. At $r\approx3$ the exit-velocity deficit and the early merging contribute as well (Section~\ref{sec:limitations}).

\textbf{The dilution follows a one-parameter law.}
Measured by whole-trajectory matching against the single-jet family, the effective ratio follows $r_{\mathrm{eff}}/r=r/(r+c)$ with $c=3.31$ (95\% CI $[3.03,\ 4.81]$; $R^2=0.80$), equivalently a momentum-flux dilution $J_{\mathrm{eff}}/J=[r/(r+c)]^2$, holding at the condition level over the nine conditions, with none of the analyzed cases excluded from the estimator.
The trajectory bending and the decay rate are two readouts of the one dilution parameter, rising together across the conditions ($\rho=+0.87$, $p=0.002$) and collapsing onto the single-jet law once the decay is re-scaled at $r_{\mathrm{eff}}$; a combined-scale power law built on the near-field transition height is superseded ($R^2=-4.8$ against the whole-trajectory ratios).

The dilution is deepest at the lowest ratio and saturates toward unity as $r$ grows. The spacing enters through the near-field capacity $c_h(L)$ ($2.50$, $2.63$ and $3.30$ at $L=2d$, $3d$ and $4d$); as a replacement for the constant capacity it does not improve the fit ($R^2=0.78$ against $0.80$), so the law is stated with the whole-trajectory constant and $c_h(L)$ as its measured near-field component. The near-field capacity separates the $L=4d$ case from the narrower spacings but not $2d$ from $3d$ (supplementary material).

The mechanism behind the law -- the enclosed-recirculation momentum exchange and the unsteady shear-layer dynamics it sustains -- is the subject of the companion paper. The law itself contains a single fitted constant and its only external input is the single-jet reference family; a leave-one-condition-out test predicts each held-out condition to within $15\%$ in $r_{\mathrm{eff}}/r$ (predictive $R^2=0.75$). It is therefore a transferable one-parameter description of the front (diluted) jet in any tandem geometry in which the gap recirculation closes, while its invariance to nozzle geometry and Reynolds number remains to be tested (Section~\ref{sec:generality}).

\subsection*{Supplementary Material}

The supplementary material documents the measurement and processing protocols, the uncertainty and robustness checks, and the supporting statistics, organized by the section of the main text to which each item belongs. The supplementary material of the companion paper~\citep{companion_part2} covers the momentum exchange and the unsteady dynamics.

\subsection*{Acknowledgments}

The author thanks the National Natural Science Foundation of China (NSFC) (52076137); the Taishan Scholar Program of Shandong Province (No. tsqn202306209); and the Natural Science Foundation of Shanghai (21ZR1431300).

\subsection*{Conflict of Interest}

The author has no conflicts of interest to disclose.

\subsection*{Author Contributions}

Xunchen Liu: Conceptualization (lead); Methodology (lead); Investigation (lead); Formal analysis (lead); Software (lead); Validation (lead); Visualization (lead); Writing -- original draft (lead); Writing -- review and editing (lead); Funding acquisition (lead); Supervision (lead).

\subsection*{Data Availability Statement}

The data that support the findings of this study are openly available in the Science Data Bank at https://doi.org/\allowbreak 10.57760/\allowbreak sciencedb.014ij. The release contains the per-case time-averaged velocity fields, the extracted trajectories and fitted parameters, and the Reynolds-stress, mixing-length and pressure statistics of the twin-nozzle cases and the single-jet records.

\subsection*{References}

\bibliographystyle{aipnum4-2-titles}
\bibliography{twin-jet-piv,liutex}

@article{karagozian_transverse_2010,
	title = {Transverse jets and their control},
	doi = {10.1016/j.pecs.2010.01.001},	volume = {36},
	issn = {0360-1285},
	language = {en},
	number = {5},
	urldate = {2020-02-18},
	journal = {Progress in Energy and Combustion Science},
	author = {Karagozian, Ann R.},
	month = oct,
	year = {2010},
	pages = {531--553},
}

@article{karagozian_jet_2014,
	title = {The jet in crossflow},
	doi = {10.1063/1.4895900},	volume = {26},
	issn = {1070-6631},
	number = {10},
	urldate = {2020-02-20},
	journal = {Physics of Fluids},
	author = {Karagozian, Ann R.},
	month = sep,
	year = {2014},
	pages = {101303},
}

@article{broadwell_structure_1984,
	title = {Structure and mixing of a transverse jet in incompressible flow},
	doi = {10.1017/S0022112084002408},	volume = {148},
	issn = {1469-7645, 0022-1120},
	language = {en},
	urldate = {2020-03-11},
	journal = {Journal of Fluid Mechanics},
	author = {Broadwell, J. E. and Breidenthal, R. E.},
	month = nov,
	year = {1984},
	pages = {405--412},
}

@article{smith_mixing_1998,
	title = {Mixing, structure and scaling of the jet in crossflow},
	doi = {10.1017/S0022112097007891},	volume = {357},
	issn = {1469-7645, 0022-1120},
	language = {en},
	urldate = {2020-03-11},
	journal = {Journal of Fluid Mechanics},
	author = {Smith, S. H. and Mungal, M. G.},
	month = feb,
	year = {1998},
	pages = {83--122},
}

@article{hasselbrink_transverse_2001,
	title = {Transverse jets and jet flames. {Part} 1. {Scaling} laws for strong transverse jets},
	doi = {10.1017/S0022112001005146},	volume = {443},
	issn = {1469-7645, 0022-1120},
	language = {en},
	urldate = {2021-06-16},
	journal = {Journal of Fluid Mechanics},
	author = {Hasselbrink, Ernest F. and Mungal, M. G.},
	month = sep,
	year = {2001},
	pages = {1--25},
}

@article{fric_vortical_1994,
	title = {Vortical structure in the wake of a transverse jet},
	doi = {10.1017/S0022112094003800},	volume = {279},
	issn = {1469-7645, 0022-1120},
	language = {en},
	urldate = {2021-07-03},
	journal = {Journal of Fluid Mechanics},
	author = {Fric, T. F. and Roshko, A.},
	month = nov,
	year = {1994},
	pages = {1--47},
}

@article{gutmark_dynamics_2011,
	title = {Dynamics of single and twin circular jets in cross flow},
	doi = {10.1007/s00348-010-0965-2},	volume = {50},
	issn = {0723-4864, 1432-1114},
	language = {en},
	number = {3},
	urldate = {2021-07-16},
	journal = {Experiments in Fluids},
	author = {Gutmark, Ephraim J. and Ibrahim, Irene M. and Murugappan, Shanmugam},
	month = mar,
	year = {2011},
	pages = {653--663},
}

@article{muppidi_study_2005,
	title = {Study of trajectories of jets in crossflow using direct numerical simulations},
	doi = {10.1017/S0022112005003514},	volume = {530},
	issn = {0022-1120, 1469-7645},
	language = {en},
	urldate = {2021-07-29},
	journal = {Journal of Fluid Mechanics},
	author = {Muppidi, Suman and Mahesh, Krishnan},
	month = may,
	year = {2005},
	pages = {81--100},
}

@article{getsinger_structural_2014,
	title = {Structural and stability characteristics of jets in crossflow},
	doi = {10.1017/jfm.2014.605},	volume = {760},
	issn = {0022-1120, 1469-7645},
	language = {en},
	urldate = {2021-10-29},
	journal = {Journal of Fluid Mechanics},
	author = {Getsinger, D. R. and Gevorkyan, L. and Smith, O. I. and Karagozian, A. R.},
	month = dec,
	year = {2014},
	pages = {342--367},
}

@article{andreopoulos_structure_1985,
	title = {On the structure of jets in a crossflow},
	doi = {10.1017/S0022112085002348},	volume = {157},
	issn = {0022-1120, 1469-7645},
	language = {en},
	urldate = {2021-11-03},
	journal = {Journal of Fluid Mechanics},
	author = {Andreopoulos, J.},
	month = aug,
	year = {1985},
	pages = {163--197},
}

@article{pratte_profiles_1967,
	title = {Profiles of the {Round} {Turbulent} {Jet} in {A} {Cross} {Flow}},
	volume = {93},
	copyright = {© 1967 American Society of Civil Engineers},
	url = {https://ascelibrary.org/doi/abs/10.1061/JYCEAJ.0001735},
	doi = {10.1061/JYCEAJ.0001735},
	language = {EN},
	number = {6},
	urldate = {2021-05-11},
	journal = {Journal of the Hydraulics Division},
	author = {Pratte, Bruce D. and Baines, W. Douglas},
	month = nov,
	year = {1967},
	pages = {53--64},
}

@article{zhang_flow_2017,
	title = {Flow {Dynamics} and {Mixing} of a {Transverse} {Jet} in {Crossflow}—{Part} {I}: {Steady} {Crossflow}},
	volume = {139},
	issn = {0742-4795},
	shorttitle = {Flow {Dynamics} and {Mixing} of a {Transverse} {Jet} in {Crossflow}—{Part} {I}},
	url = {https://asmedigitalcollection.asme.org/gasturbinespower/article/139/8/082601/385975/Flow-Dynamics-and-Mixing-of-a-Transverse-Jet-in},
	doi = {10.1115/1.4035808},
	language = {en},
	number = {8},
	journal = {Journal of Engineering for Gas Turbines and Power},
	author = {Zhang, Liwei and Yang, Vigor},
	month = aug,
	year = {2017},
	pages = {082601},
}

@article{wang_experimental_2022,
	title = {Experimental {Study} of a {Novel} {Twin}-{Nozzle} {Configuration} for {Reacting} {Jet} in {Hot} {Crossflow}},
	volume = {144},
	issn = {0195-0738},
	url = {https://doi.org/10.1115/1.4053646},
	doi = {10.1115/1.4053646},
	number = {9},
	journal = {Journal of Energy Resources Technology},
	author = {Wang, Zhen and Wang, Yayao and Liu, Xunchen},
	month = feb,
	year = {2022},
	pages = {092105},
}

@article{wang_lift-off_2023,
	title = {Lift-off region temperature field and planar flow field of a twin-nozzle reacting jet in hot crossflow},
	volume = {39},
	issn = {1540-7489},
	url = {https://www.sciencedirect.com/science/article/pii/S1540748922004850},
	doi = {10.1016/j.proci.2022.08.123},
	number = {1},
	journal = {Proceedings of the Combustion Institute},
	author = {Wang, Zhen and Wang, Yayao and Liu, Xunchen},
	month = jun,
	year = {2023},
	pages = {1269--1278},
}

@article{naik-nimbalkar_twin_2011,
	title = {Twin jets in cross-flow},
	volume = {66},
	issn = {00092509},
	url = {https://linkinghub.elsevier.com/retrieve/pii/S0009250911001849},
	doi = {10.1016/j.ces.2011.03.018},
	language = {en},
	number = {12},
	journal = {Chemical Engineering Science},
	author = {Naik-Nimbalkar, V. S. and Suryawanshi, A. D. and Patwardhan, A. W. and Banerjee, I. and Padmakumar, G. and Vaidyanathan, G.},
	month = jun,
	year = {2011},
	pages = {2616--2626},
}

@inproceedings{zang_particle-image_2014,
	title = {A particle-image velocimetry study on the near-field behaviour of tandem twin jets in cross-flow},
	language = {en},
	author = {Zang, B. and New, T. H.},
	year = {2014},
	booktitle = {Proceedings of the 17th International Symposium on Applications of Laser Techniques to Fluid Mechanics},
	address = {Lisbon, Portugal},
	pages = {1--8},
}

@article{new_trajectory_2015,
	title = {On the trajectory scaling of tandem twin jets in cross-flow in close proximity},
	volume = {56},
	issn = {0723-4864, 1432-1114},
	url = {http://link.springer.com/10.1007/s00348-015-2070-z},
	doi = {10.1007/s00348-015-2070-z},
	language = {en},
	number = {11},
	journal = {Experiments in Fluids},
	author = {New, T. H. and Zang, B.},
	month = nov,
	year = {2015},
	pages = {200},
}

@article{zhao_experimental_2018,
	title = {An {Experimental} {Characterization} of the {Interaction} {Between} {Two} {Tandem} {Planar} {Jets} in a {Crossflow}},
	volume = {140},
	doi = {10.1115/1.4040224},
	journal = {Journal of Fluids Engineering},
	author = {Zhao, Kun and Okolo, Patrick and Wang, Yong and Kennedy, John and Bennett, Gareth},
	month = may,
	year = {2018},
	number = {11},
	pages = {111106},
}

@article{kristo_time-resolved_2021,
	title = {Time-resolved particle image velocimetry measurements of a tandem jet array in a crossflow at low velocity ratios},
	volume = {62},
	issn = {0723-4864, 1432-1114},
	url = {https://link.springer.com/10.1007/s00348-021-03159-x},
	doi = {10.1007/s00348-021-03159-x},
	language = {en},
	number = {4},
	journal = {Experiments in Fluids},
	author = {Kristo, Paul J. and Kimber, Mark L.},
	month = apr,
	year = {2021},
	pages = {67},
}

@article{miller_force-momentum_1960,
	title = {Force-momentum fields in a dual-jet flow},
	volume = {7},
	issn = {0022-1120, 1469-7645},
	url = {https://www.cambridge.org/core/journals/journal-of-fluid-mechanics/article/abs/forcemomentum-fields-in-a-dualjet-flow/C9E62FF844CF5111A5AA35DED47DF4E6},
	doi = {10.1017/S0022112060001468},
	language = {en},
	number = {2},
	urldate = {2026-08-15},
	journal = {Journal of Fluid Mechanics},
	author = {Miller, David R. and Comings, Edward W.},
	month = feb,
	year = {1960},
	pages = {237--256},
}

@article{tanaka_interference_1970,
	title = {The interference of two-dimensional parallel jets: 1st report, experiments on dual jet},
	volume = {13},
	issn = {0021-3764},
	url = {https://doi.org/10.1299/jsme1958.13.272},
	doi = {10.1299/jsme1958.13.272},
	language = {en},
	number = {56},
	urldate = {2026-08-15},
	journal = {Bulletin of JSME},
	author = {Tanaka, Eiichi},
	month = feb,
	year = {1970},
	pages = {272--280},
}

@article{nasr_two_1997,
	title = {Two parallel plane jets: mean flow and effects of acoustic excitation},
	volume = {22},
	issn = {0723-4864, 1432-1114},
	url = {https://link.springer.com/article/10.1007/s003480050044},
	doi = {10.1007/s003480050044},
	language = {en},
	number = {3},
	urldate = {2026-08-15},
	journal = {Experiments in Fluids},
	author = {Nasr, A. and Lai, J. C. S.},
	month = jan,
	year = {1997},
	pages = {251--260},
}

@article{anderson_experimental_2001,
	title = {Experimental and numerical investigation of two-dimensional parallel jets},
	volume = {123},
	issn = {0098-2202, 1528-901X},
	url = {https://asmedigitalcollection.asme.org/fluidsengineering/article/123/2/401/457031},
	doi = {10.1115/1.1363701},
	language = {en},
	number = {2},
	urldate = {2026-08-15},
	journal = {Journal of Fluids Engineering},
	author = {Anderson, Elgin A. and Spall, Robert E.},
	month = jun,
	year = {2001},
	pages = {401--406},
}

@inproceedings{hawthorne_mixing_1949,
	title = {Mixing and combustion in turbulent gas jets},
	doi = {10.1016/S1062-2896(49)80035-3},
	booktitle = {Third Symposium on Combustion and Flame and Explosion Phenomena},
	author = {Hawthorne, William R. and Weddell, Dwight S. and Hottel, Hoyt C.},
	year = {1949},
	pages = {266--288},
	note = {Published 1949 by Williams \& Wilkins, Baltimore},
}

@inproceedings{thring_newby_1953,
	title = {Combustion length of enclosed turbulent jet flames},
	booktitle = {Fourth Symposium (International) on Combustion},
	author = {Thring, Meredith W. and Newby, Michael P.},
	publisher = {Williams \& Wilkins},
	address = {Baltimore},
	year = {1953},
	pages = {789--796},
	doi = {10.1016/S0082-0784(53)80103-7},
}

@article{becker_concentration_1978,
	title = {Entrainment, momentum flux and temperature in vertical free turbulent diffusion flames},
	doi = {10.1016/0010-2180(78)90055-X},
	author = {Becker, Hans A. and Yamazaki, Shintaro},
	year = {1978},
	journal = {Combustion and Flame},
	volume = {33},
	pages = {123--149},
}

@article{vanoudheusden_piv-based_2013,
	title = {{PIV}-based pressure measurement},
	volume = {24},
	doi = {10.1088/0957-0233/24/3/032001},
	number = {3},
	journal = {Measurement Science and Technology},
	author = {van Oudheusden, B. W.},
	year = {2013},
	pages = {032001},
}

@article{dekat_instantaneous_2012,
	title = {Instantaneous planar pressure determination from {PIV} in turbulent flow},
	volume = {52},
	doi = {10.1007/s00348-011-1237-5},
	number = {5},
	journal = {Experiments in Fluids},
	author = {de Kat, R. and van Oudheusden, B. W.},
	year = {2012},
	pages = {1089--1106},
}

@article{charonko_assessment_2010,
	title = {Assessment of pressure field calculations from particle image velocimetry measurements},
	volume = {21},
	doi = {10.1088/0957-0233/21/10/105401},
	number = {10},
	journal = {Measurement Science and Technology},
	author = {Charonko, John J. and King, Cameron V. and Smith, Barton L. and Vlachos, Pavlos P.},
	year = {2010},
	pages = {105401},
}

@article{yang_zhang_part2_2017,
	title = {Flow Dynamics and Mixing of a Transverse Jet in Crossflow---{Part II}: Oscillating Crossflow},
	volume = {139},
	issn = {0742-4795},
	doi = {10.1115/1.4035809},
	number = {8},
	journal = {Journal of Engineering for Gas Turbines and Power},
	author = {Zhang, Liwei and Yang, Vigor},
	year = {2017},
	pages = {082602},
}

@unpublished{companion_part2,
	title = {Tandem jets in crossflow: the enclosed-recirculation momentum exchange and the unsteady shear-layer dynamics},
	author = {Liu, Xunchen},
	year = {2026},
	note = {companion paper, submitted},
}
\end{document}


\sloppy
\maketitle

The figures and tables below accompany the manuscript ``Tandem jets in crossflow: a one-parameter dilution law for the trajectory and the velocity decay''. They are organized section by section in the same order as the main text: each section of this supplement expands the corresponding section of the main text, whose number is given in the section heading, and the figures and tables follow that order. Cross-references to the main text are given by the printed numbers of that document (for example ``Eq.~(1) of the main text''), since the two documents are compiled separately. The pseudocode summaries are given with the results they produce: Algorithm~1 (trajectory extraction) and Algorithm~2 (natural-coordinate transformation) precede their figures in the corresponding sections, and Algorithms~3--6 (single-jet library, condition-mean fields, curve-matching estimator, reverse-flow and pressure reconstruction) are collected with their reproduction notes in the final section. The momentum exchange and the unsteady shear-layer dynamics are treated in the companion paper~\citep{companion_part2}, whose supplementary material carries the corresponding figures and tables.

\section{Experimental measurement (Sec.~II of the main text)}
The measurement and processing details are documented in Sec.~II of the main text. The velocity-ratio convention and the batch notes (the measured free-stream velocity of each batch and the within-batch comparisons) are given under (i) of the trajectory-method section below, together with the velocity-normalization robustness of the fitted trajectory parameters. No supplementary figure belongs to this section.
\clearpage

\section{Trajectories and the modification metrics (Sec.~III A of the main text)}

The trajectory extraction, the transition detection and the power-law fitting of Sec.~III A of the main text are summarized in Algorithm~\ref{alg:track}; the procedure produces the per-case trajectories and fits shown in figures S1--S3, and its five stages are commented step by step in the method section below. The step numbers refer to the logical stages marked in the comments rather than to individual program lines.
\begin{algorithm}[H]
\algtitle{Automated trajectory extraction, near/far transition detection and power-law fitting}{alg:track}
\SetKwInput{KwIn}{Input}
\SetKwInput{KwOut}{Output}
\KwIn{time-averaged velocity field $V(x,y)$; nozzle positions $x_f$, $x_r$; origin $x_0$ (measured jet-exit point); nozzle diameter $d$; grid spacing $\Delta$}
\KwOut{near-field $(A_n,B_n)$; far-field $(A_f,B_f)$; far-field transition index $k_t$; near-field transition index $k_n$}
$F\leftarrow[\,]$, $R\leftarrow[\,]$\; \tcp{Step 1. initialize the two tracks ($F$: front-jet points, $R$: rear-jet points)}
\For{each jet $J\in\{F,R\}$ (tracked independently)}{ \tcp{Steps 2--7. track the jets independently, row by row}
  \For{each row $y_i$}{
    find the peaks of $V(y_i,:)$ within a window of half-width $d$ around the previous point (around the nozzle for the first rows)\;
    refine each peak by a parabolic sub-pixel fit\;
    keep the refined peak nearest to the previous point\;
    validate: monotone $x$, speed jump $\le30\%$\;
    append the point to the jet's track\;
  }
}
refine each point perpendicular to the trajectory ($\le1$\,mm) and smooth by an arc-length B-spline\; \tcp{Steps 8--9. perpendicular refinement and smoothing}
truncate $R$ at 1.2\,mm from the right boundary (low-ratio rear jets at the bend-over end)\; \tcp{Step 10. truncation}
\For{each candidate split $k$}{ \tcp{Steps 11--12. split search}
  fit the near-field segment ($x$-on-$y$) and the far segment ($y$-on-$x$)\;
  score $\gets$ straightness $R^2$ $\times$ exponent stability\;
}
$k_t\gets$ earliest split with far-field $R^2\ge0.9$, stability $\ge0.85$, $\Delta B\le5\%$ (12\% for case 3d-7r-4) and $\mathrm{d}x/\mathrm{d}y\ge0.05$ (rear)\; \tcp{Step 13. far-field transition}
\For{each candidate $k_n$ inside the near field ($<k_t$)}{ \tcp{Steps 14--17. near-field transition search}
  fit the near-field segment $0\ldots k_n$ in the $x$-on-$y$ orientation\;
  score $\gets$ $R^2$ $\times$ length factor $\times$ $y$-span factor\;
}
$k_n\gets$ best near-field transition; discard the bend-over segment $k_n\ldots k_t$\; \tcp{Step 18. near-field transition}
apply near-field gating: $R^2\ge0.5$ (front) or $0.8$ (rear) on segment $0\ldots k_n$\; \tcp{Step 19. near-field gating}
$(A_n,B_n)\gets$ log--log fit on the near-field segment $0\ldots k_n$; $(A_f,B_f)\gets$ log--log fit on the far segment\; \tcp{Steps 20--22. parameter fits}
\Return $(A_n,B_n,A_f,B_f,k_t,k_n)$
\end{algorithm}
\begin{figure}
\centerline{\includegraphics[width=6.2in]{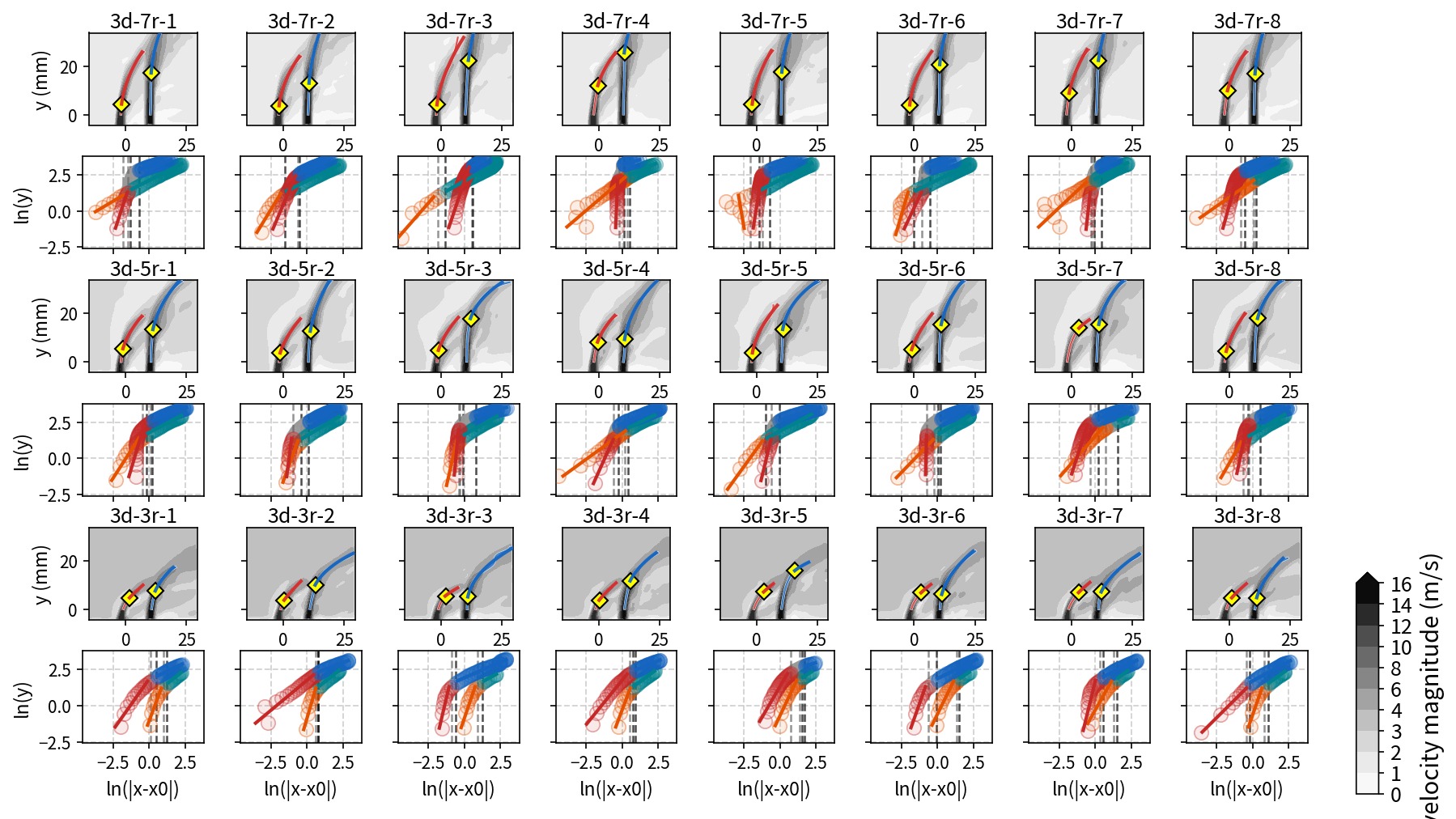}}
\caption{\textbf{Figure S1.} All $3d$ cases: velocity cloud maps and log--log power-law fits, organized by measured velocity ratio (3 rows of 8 cases). Front jet: red; rear jet: blue; yellow diamond: transition point.}\alttext{Per-case velocity cloud maps and power-law fits of all L=3d cases, organized by measured velocity ratio.}
\label{fig:app-3d}
\end{figure}

\begin{figure}
\centerline{\includegraphics[width=6.2in]{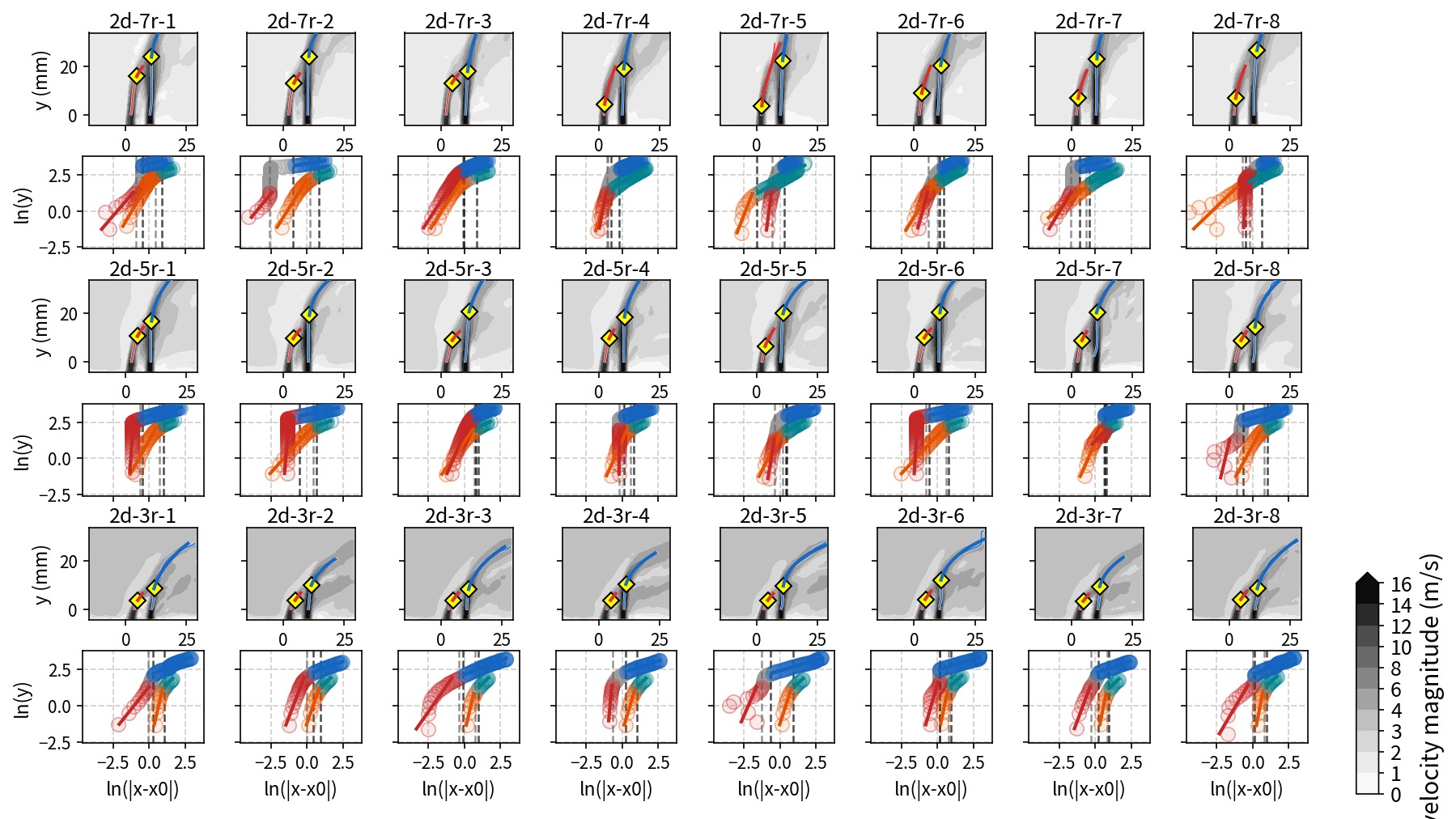}}
\caption{\textbf{Figure S2.} All $2d$ cases: velocity cloud maps and log--log power-law fits, organized by measured velocity ratio (3 rows of 8 cases). Front jet: red; rear jet: blue; yellow diamond: transition point.}\alttext{Per-case velocity cloud maps and power-law fits of all L=2d cases, organized by measured velocity ratio.}
\label{fig:app-2d}
\end{figure}

\clearpage

\begin{figure}
\centerline{\includegraphics[width=6.2in]{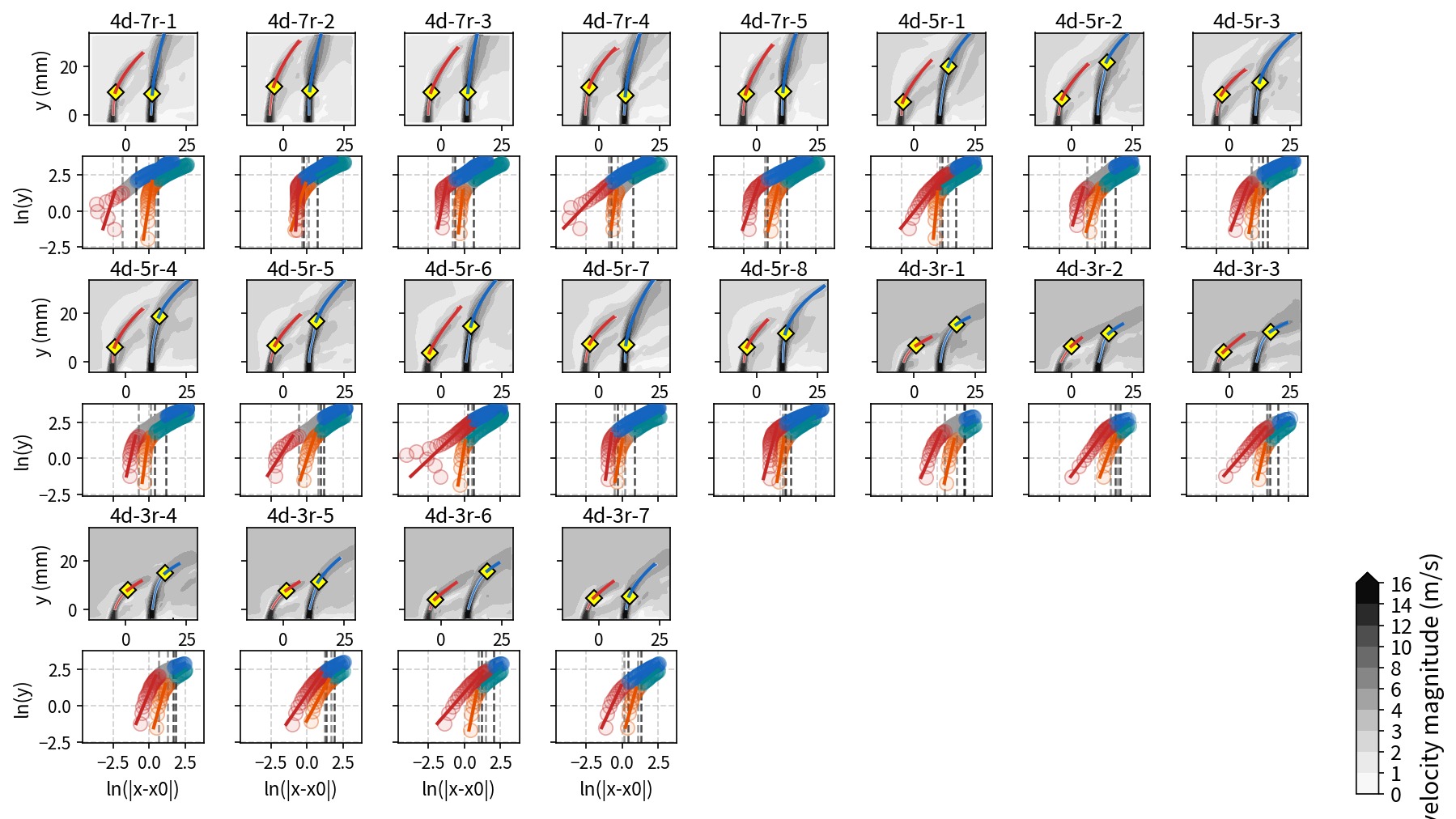}}
\caption{\textbf{Figure S3.} All $4d$ cases: velocity cloud maps and log--log power-law fits, organized by measured velocity ratio (3 rows: 7, 8 and 7 cases; the case accounting, including the records excluded from the statistics, is given in table ST5). Front jet: red; rear jet: blue; yellow diamond: transition point.}\alttext{Per-case velocity cloud maps and power-law fits of all L=4d cases, organized by measured velocity ratio.}
\label{fig:app-4d}
\end{figure}
\begin{table}
\centering
\caption{\textbf{Table ST1.} Physical-space penetration and deflection of the extracted trajectories (Sec.~III A of the main text; the corresponding main-text panels (l) and (m) of Fig.~6 show instead the near-field coefficients $A_n$ and $B_n$): the penetration depth $H$ (the maximum height reached above the nozzle-exit plane) and the far-field deflection $\varphi=\arctan\langle\mathrm{d}x/\mathrm{d}y\rangle$ over the last $40\%$ of the tracked trajectory ($0^\circ$ vertical, $90^\circ$ turned into the crossflow), for the front and the rear jet of the nine conditions (mean $\pm$ SD over the valid cases) and for the eleven single-jet records grouped by their measured velocity ratio. The last column of each jet gives the number of records whose trajectory is still rising at the last resolved station: for those $H$ is a lower bound set by the measurement window, which is why the rear-jet $H$ clusters near the window height at $r\approx5$ and 7 while its deflection stays small -- the shielded jet is still climbing where the window ends.}
\alttext{Table of penetration depth, maximum height and far-field deflection of the extracted trajectories for the front and rear jets and the single-jet records.}
\label{tab:st9-penetration}
\footnotesize
\begin{tabular}{llcccccc}
\toprule
& & \multicolumn{3}{c}{front jet} & \multicolumn{3}{c}{rear jet} \\
\cmidrule(lr){3-5}\cmidrule(lr){6-8}
Condition & $r$ & $H$ (mm) & $\varphi$ (deg) & still rising & $H$ (mm) & $\varphi$ (deg) & still rising \\
\midrule
$L=2d$, $r\approx3$ & 2.79 & $6.79\pm0.48$ & $39.0\pm1.4$ & $4/8$ & $25.73\pm3.69$ & $49.4\pm2.6$ & $1/8$ \\
$L=2d$, $r\approx5$ & 4.62 & $13.19\pm0.67$ & $28.7\pm2.8$ & $5/7$ & $33.72\pm0.14$ & $26.2\pm4.0$ & $7/7$ \\
$L=2d$, $r\approx7$ & 6.22 & $20.03\pm3.73$ & $21.8\pm3.3$ & $8/8$ & $33.82\pm0.09$ & $12.0\pm2.6$ & $8/8$ \\
$L=3d$, $r\approx3$ & 2.76 & $10.45\pm0.78$ & $49.6\pm3.3$ & $1/8$ & $21.97\pm2.45$ & $48.0\pm6.0$ & $0/8$ \\
$L=3d$, $r\approx5$ & 4.64 & $19.32\pm1.74$ & $37.6\pm2.6$ & $5/8$ & $33.24\pm0.23$ & $35.7\pm5.5$ & $2/8$ \\
$L=3d$, $r\approx7$ & 6.32 & $26.54\pm2.13$ & $29.1\pm4.6$ & $7/8$ & $33.89\pm0.09$ & $11.3\pm3.1$ & $8/8$ \\
$L=4d$, $r\approx3$ & 2.78 & $10.93\pm0.58$ & $54.7\pm3.0$ & $0/7$ & $18.15\pm1.63$ & $48.7\pm2.3$ & $0/7$ \\
$L=4d$, $r\approx5$ & 4.67 & $20.03\pm1.85$ & $44.1\pm4.7$ & $2/8$ & $33.07\pm0.72$ & $38.3\pm4.8$ & $5/8$ \\
$L=4d$, $r\approx7$ & 6.53 & $27.70\pm1.58$ & $38.4\pm2.6$ & $1/5$ & $33.30\pm0.49$ & $14.3\pm2.4$ & $5/5$ \\
\midrule
single, $r\approx7$ & 6.06 & $38.12\pm0.20$ & $29.6\pm4.6$ & $3/4$ & -- & -- & -- \\
single, $r\approx5$ & 4.50 & $37.79\pm0.28$ & $48.7\pm0.2$ & $1/3$ & -- & -- & -- \\
single, $r\approx3$ & 2.70 & $21.89\pm0.86$ & $58.3\pm3.3$ & $0/4$ & -- & -- & -- \\
\bottomrule
\end{tabular}
\end{table}
\begin{table*}
\centering
\caption{\textbf{Table ST2.} Per-case fitted far-field parameters ($B_f$, $A_f$) for the 10 retained single-jet records of the 11 acquired (each record is an independent 400-frame average; the $r\approx7$ batch 2 is omitted as an outlier, so the $r\approx7$ group lists three of its four records; the group mean $r=6.05$ is therefore over three records, whereas the main-text reference uses the four-record value $6.06$), with the three velocity-ratio groups listed side by side and the group means in the last row. Fit standard errors are given in parentheses for the per-case entries, and SDs in the group-mean rows; $r$ is the measured velocity ratio. Near-field parameters are not reported because the near-field segment is quasi-vertical (see text).}\alttext{Table of the per-case fitted far-field parameters of the ten retained single-jet records, grouped by velocity ratio.}
\label{tab:single-BA}
\scriptsize
\begin{tabular}{
  l S[table-format=1.2] S[table-format=1.3, table-space-text-post={(18)}] S[table-format=1.3, table-space-text-post={(132)}] |
  l S[table-format=1.2] S[table-format=1.3, table-space-text-post={(18)}] S[table-format=1.3, table-space-text-post={(132)}] |
  l S[table-format=1.2] S[table-format=1.3, table-space-text-post={(18)}] S[table-format=1.3, table-space-text-post={(132)}]
}
\toprule
Case & {$r$} & {$B_f$} & {$A_f$} & Case & {$r$} & {$B_f$} & {$A_f$} & Case & {$r$} & {$B_f$} & {$A_f$} \\
\midrule
7r-1 & 6.04 & 0.410(2) & 2.208(11) & 5r-1 & 4.48 & 0.372(3) & 1.799(7) & 3r-1 & 2.79 & 0.373(2) & 1.663(4) \\
7r-3 & 6.06 & 0.375(1) & 1.949(4) & 5r-2 & 4.45 & 0.394(2) & 2.002(8) & 3r-2 & 2.65 & 0.405(3) & 1.631(6) \\
7r-4 & 6.06 & 0.391(3) & 2.120(12) & 5r-3 & 4.58 & 0.389(1) & 1.928(2) & 3r-3 & 2.69 & 0.421(3) & 1.645(6) \\
--- & \multicolumn{1}{c}{---} & \multicolumn{1}{c}{---} & \multicolumn{1}{c}{---} & --- & \multicolumn{1}{c}{---} & \multicolumn{1}{c}{---} & \multicolumn{1}{c}{---} & 3r-4 & 2.68 & 0.405(2) & 1.528(3) \\
\midrule
7r mean & 6.05 & 0.392(18) & 2.092(132) & 5r mean & 4.50 & 0.385(12) & 1.910(103) & 3r mean & 2.70 & 0.401(20) & 1.617(61) \\
\bottomrule
\end{tabular}
\end{table*}
\begin{sidewaystable}[p]
\centering
\caption{\textbf{Table ST5.} Per-case far-field parameters ($B_f$, $A_f$) and near-field parameters ($B_n$, $A_n$) for all 68 twin-nozzle cases in a landscape layout with the three spacings side by side; standard errors are given in parentheses aligned to the last decimal place. The group-mean and overall-mean rows (mean, SD in parentheses) summarize the fitted parameters. Cases are labeled as $L$-$r$-index within each nominal velocity-ratio group; of the 71 records acquired, three are excluded from the statistics -- one record of the $L=4d$, $r\approx7$ group without exported velocity fields and two of the same group whose nozzles sat at $x\approx9.4$--$9.6$\,mm instead of the standard 10\,mm nozzle position -- and are not listed here; $r$ is the measured velocity ratio, and ``---'' marks parameters that are not reported: the far-field fits of the $L=2d$, $r\approx3$ front jets (whose trajectory is swallowed by the crossflow before a self-similar far field develops) and seven near-field fits of rear jets (whose quasi-vertical near-field segment gives an ill-conditioned fit), so that near-field parameters are reported for 129 of the 136 jet trajectories.}\alttext{Landscape table of the per-case fitted far-field and near-field parameters of all 68 twin cases, grouped by spacing.}
\label{tab:BA_all}
\resizebox{\textwidth}{!}{%
\begin{tabular}{r|l|rrrr|rrrr@{\hspace{1.5em}}||r|l|rrrr|rrrr@{\hspace{1.5em}}||r|l|rrrr|rrrr}
\toprule
\multicolumn{10}{c}{\textbf{$L=4d$}} & \multicolumn{10}{c}{\textbf{$L=3d$}} & \multicolumn{10}{c}{\textbf{$L=2d$}} \\
\cmidrule(lr){1-10}\cmidrule(lr){11-20}\cmidrule(lr){21-30}
\textbf{Case} & \textbf{$r$} & \multicolumn{4}{c}{\textbf{Front jet}} & \multicolumn{4}{c}{\textbf{Rear jet}} & \textbf{Case} & \textbf{$r$} & \multicolumn{4}{c}{\textbf{Front jet}} & \multicolumn{4}{c}{\textbf{Rear jet}} & \textbf{Case} & \textbf{$r$} & \multicolumn{4}{c}{\textbf{Front jet}} & \multicolumn{4}{c}{\textbf{Rear jet}} \\
\cmidrule(lr){3-6}\cmidrule(lr){7-10}\cmidrule(lr){13-16}\cmidrule(lr){17-20}\cmidrule(lr){23-26}\cmidrule(lr){27-30}
 & & $B_f$ & $A_f$ & $B_n$ & $A_n$ & $B_f$ & $A_f$ & $B_n$ & $A_n$ &  & & $B_f$ & $A_f$ & $B_n$ & $A_n$ & $B_f$ & $A_f$ & $B_n$ & $A_n$ &  & & $B_f$ & $A_f$ & $B_n$ & $A_n$ & $B_f$ & $A_f$ & $B_n$ & $A_n$ \\
\midrule
4d-7r-1 & 6.73 & 0.498(5) & 1.366(12) & 2.962(74) & 1.099(143) & 0.528(6) & 2.911(45) & 1.149(43) & 0.095(11) & 3d-7r-1 & 6.35 & 0.514(3) & 1.766(12) & 1.650(17) & 0.221(11) & 0.305(3) & 2.319(23) & 1.100(51) & 0.044(6) & 2d-7r-1 & 6.24 & 0.332(7) & 1.352(18) & 0.984(19) & 0.101(3) & 0.198(2) & 2.008(13) & 1.146(45) & 0.177(21) \\
4d-7r-2 & 6.50 & 0.515(5) & 1.711(15) & 1.489(7) & 0.186(3) & 0.643(2) & 3.486(15) & --- & --- & 3d-7r-2 & 6.23 & 0.503(3) & 1.654(14) & 1.798(5) & 0.246(3) & 0.429(4) & 2.741(27) & 0.986(16) & 0.033(1) & 2d-7r-2 & 6.25 & 0.377(7) & 1.272(20) & 1.191(15) & 0.136(4) & 0.163(2) & 1.837(11) & 0.878(90) & 0.019(4) \\
4d-7r-3 & 6.47 & 0.556(5) & 1.582(14) & 1.935(1) & 0.347(1) & 0.606(9) & 3.801(89) & --- & --- & 3d-7r-3 & 6.42 & 0.562(5) & 1.968(22) & 1.358(6) & 0.142(3) & 0.510(8) & 3.470(65) & 0.982(12) & 0.051(1) & 2d-7r-3 & 6.34 & 0.329(5) & 1.107(11) & 1.172(16) & 0.119(4) & 0.392(5) & 2.602(33) & 0.991(13) & 0.045(1) \\
4d-7r-4 & 6.32 & 0.469(6) & 1.492(15) & 1.760(3) & 0.229(2) & 0.526(5) & 2.691(35) & 1.153(24) & 0.089(5) & 3d-7r-4 & 6.27 & 0.440(7) & 1.607(21) & 1.223(12) & 0.172(4) & 0.262(12) & 2.383(80) & --- & --- & 2d-7r-4 & 6.31 & 0.633(6) & 2.232(37) & 1.617(15) & 0.221(9) & 0.369(5) & 2.700(35) & 1.152(47) & 0.050(6) \\
4d-7r-5 & 6.66 & 0.572(5) & 1.669(16) & 1.621(6) & 0.252(4) & 0.476(1) & 2.584(10) & 2.567(29) & 0.188(9) & 3d-7r-5 & 6.47 & 0.525(3) & 1.871(14) & 1.454(31) & 0.217(19) & 0.305(3) & 2.320(19) & 0.670(16) & 0.008(0) & 2d-7r-5 & 6.11 & 0.595(8) & 2.380(55) & 2.611(61) & 0.989(147) & 0.476(7) & 3.708(67) & 1.157(39) & 0.064(7) \\
 &  &  &  &  &  &  &  &  &  & 3d-7r-6 & 6.29 & 0.490(1) & 1.692(2) & 3.549(176) & 4.358(1696) & 0.286(4) & 2.463(31) & 1.109(49) & 0.084(11) & 2d-7r-6 & 6.30 & 0.570(8) & 2.044(38) & 1.238(12) & 0.152(4) & 0.387(6) & 2.432(29) & 1.181(45) & 0.105(12) \\
 &  &  &  &  &  &  &  &  &  & 3d-7r-7 & 6.28 & 0.504(6) & 1.876(24) & 1.383(13) & 0.184(5) & 0.435(7) & 3.057(48) & 1.148(43) & 0.068(8) & 2d-7r-7 & 6.20 & 0.538(8) & 1.905(38) & 1.314(20) & 0.163(8) & 0.201(0) & 2.157(2) & 1.061(54) & 0.048(7) \\
 &  &  &  &  &  &  &  &  &  & 3d-7r-8 & 6.30 & 0.486(6) & 1.725(19) & 1.339(10) & 0.175(3) & 0.353(4) & 2.609(29) & 0.945(17) & 0.018(1) & 2d-7r-8 & 6.00 & 0.534(7) & 2.025(35) & 1.257(21) & 0.157(8) & 0.314(5) & 2.464(28) & --- & --- \\
\midrule
4d mean & 6.53 & 0.522(42) & 1.564(139) & 1.953(588) & 0.423(382) & 0.556(67) & 3.095(527) & 1.623(818) & 0.124(56) \\
3d mean & 6.32 & 0.503(35) & 1.770(125) & 1.719(762) & 0.714(1473) & 0.361(88) & 2.670(409) & 0.991(161) & 0.044(27) \\
2d mean & 6.22 & 0.488(123) & 1.790(478) & 1.423(511) & 0.255(299) & 0.312(113) & 2.489(575) & 1.081(111) & 0.073(53) \\
\midrule
4d-5r-1 & 4.87 & 0.629(5) & 1.476(12) & 2.531(47) & 1.091(108) & 0.405(5) & 2.057(11) & 1.279(8) & 0.147(2) & 3d-5r-1 & 4.67 & 0.528(6) & 1.524(16) & 1.484(7) & 0.259(4) & 0.392(4) & 2.126(13) & 2.608(13) & 0.120(2) & 2d-5r-1 & 4.55 & 0.463(12) & 1.392(25) & 1.309(13) & 0.234(5) & 0.273(1) & 2.315(6) & --- & --- \\
4d-5r-2 & 4.68 & 0.611(7) & 1.416(12) & 1.598(6) & 0.384(5) & 0.354(5) & 2.052(10) & 2.904(95) & 0.771(127) & 3d-5r-2 & 4.67 & 0.596(5) & 1.478(13) & 1.646(6) & 0.374(5) & 0.450(4) & 2.503(19) & 1.522(5) & 0.060(1) & 2d-5r-2 & 4.70 & 0.358(13) & 1.141(27) & 1.206(18) & 0.200(6) & 0.219(0) & 2.122(2) & --- & --- \\
4d-5r-3 & 4.64 & 0.524(9) & 1.234(13) & 1.808(5) & 0.430(5) & 0.422(4) & 1.928(11) & 1.379(9) & 0.116(2) & 3d-5r-3 & 4.72 & 0.572(5) & 1.477(14) & 1.600(10) & 0.281(7) & 0.303(4) & 1.860(10) & 1.079(20) & 0.024(1) & 2d-5r-3 & 4.59 & 0.447(16) & 1.162(30) & 1.227(16) & 0.295(9) & 0.368(5) & 2.530(21) & 0.953(14) & 0.078(1) \\
4d-5r-4 & 4.63 & 0.562(6) & 1.451(12) & 2.519(45) & 0.848(75) & 0.363(5) & 1.934(11) & 2.399(35) & 0.325(21) & 3d-5r-4 & 4.64 & 0.483(8) & 1.499(19) & 1.366(11) & 0.257(6) & 0.412(3) & 2.026(12) & 0.942(41) & 0.107(10) & 2d-5r-4 & 4.55 & 0.499(12) & 1.499(33) & 1.213(14) & 0.189(5) & 0.315(2) & 2.467(9) & 0.504(7) & 0.001(0) \\
4d-5r-5 & 4.57 & 0.637(8) & 1.341(13) & 1.425(7) & 0.402(6) & 0.428(5) & 2.013(11) & 2.793(108) & 1.197(237) & 3d-5r-5 & 4.65 & 0.534(6) & 1.692(21) & 1.271(6) & 0.151(2) & 0.311(3) & 1.911(9) & 0.985(45) & 0.078(8) & 2d-5r-5 & 4.49 & 0.616(9) & 1.622(25) & 1.488(12) & 0.329(8) & 0.257(4) & 2.115(12) & 1.064(43) & 0.080(8) \\
4d-5r-6 & 4.62 & 0.722(6) & 1.492(12) & 1.710(9) & 0.471(11) & 0.425(4) & 2.189(13) & 1.040(13) & 0.140(3) & 3d-5r-6 & 4.61 & 0.603(6) & 1.615(17) & 1.291(26) & 0.319(19) & 0.355(4) & 2.251(16) & 1.276(69) & 0.009(1) & 2d-5r-6 & 4.76 & 0.383(13) & 1.195(28) & 1.129(19) & 0.212(7) & 0.215(0) & 2.215(2) & --- & --- \\
4d-5r-7 & 4.69 & 0.532(8) & 1.247(12) & 1.653(6) & 0.369(5) & 0.531(4) & 2.307(15) & 3.049(89) & 0.504(73) & 3d-5r-7 & 4.70 & 0.357(8) & 1.193(9) & 1.365(10) & 0.291(5) & 0.338(4) & 2.132(13) & 1.076(17) & 0.041(1) & 2d-5r-7 & 4.58 & 0.536(13) & 1.514(35) & 1.215(14) & 0.207(5) & 0.303(3) & 2.274(11) & 1.977(255) & 0.496(156) \\
4d-5r-8 & 4.69 & 0.638(9) & 1.338(16) & 1.596(6) & 0.391(5) & 0.384(4) & 1.720(9) & 3.420(96) & 0.354(36) & 3d-5r-8 & 4.45 & 0.607(6) & 1.776(20) & 1.471(21) & 0.296(16) & 0.321(4) & 2.196(13) & 1.094(17) & 0.036(1) & 2d-5r-8 & 4.74 & 0.577(10) & 1.418(23) & 1.331(11) & 0.317(6) & 0.287(4) & 2.089(19) & 1.023(51) & 0.171(21) \\
\midrule
4d mean & 4.67 & 0.607(66) & 1.375(100) & 1.855(428) & 0.548(270) & 0.414(55) & 2.025(177) & 2.283(919) & 0.444(376) \\
3d mean & 4.64 & 0.535(84) & 1.532(175) & 1.437(138) & 0.278(64) & 0.360(53) & 2.126(203) & 1.323(551) & 0.059(40) \\
2d mean & 4.62 & 0.485(90) & 1.368(181) & 1.265(110) & 0.248(57) & 0.280(51) & 2.266(165) & 1.104(537) & 0.165(195) \\
\midrule
4d-3r-1 & 2.71 & 0.637(7) & 0.840(1) & 1.415(7) & 0.747(10) & 0.303(6) & 1.638(2) & 1.454(9) & 0.219(3) & 3d-3r-1 & 2.74 & 0.793(9) & 1.056(7) & 1.421(20) & 0.805(31) & 0.480(8) & 1.728(15) & 1.405(8) & 0.232(3) & 2d-3r-1 & 2.85 & --- & --- & 1.430(19) & 0.642(24) & 0.443(9) & 2.101(21) & 1.021(52) & 0.248(24) \\
4d-3r-2 & 2.74 & 0.769(18) & 0.972(7) & 1.332(16) & 0.767(20) & 0.338(4) & 1.476(3) & 0.975(19) & 0.260(6) & 3d-3r-2 & 2.73 & 0.800(7) & 1.213(7) & 1.446(6) & 0.604(7) & 0.387(5) & 1.749(7) & 1.246(12) & 0.229(4) & 2d-3r-2 & 2.70 & --- & --- & 1.381(24) & 0.733(33) & 0.363(5) & 1.888(10) & 1.027(19) & 0.092(2) \\
4d-3r-3 & 2.76 & 0.813(8) & 0.951(5) & 1.482(18) & 0.835(30) & 0.330(6) & 1.399(3) & 1.109(17) & 0.292(7) & 3d-3r-3 & 2.81 & 0.549(17) & 0.896(11) & 1.482(7) & 0.682(9) & 0.438(6) & 1.785(14) & 1.945(9) & 0.156(3) & 2d-3r-3 & 2.78 & --- & --- & 1.508(27) & 0.803(48) & 0.389(5) & 2.019(12) & 2.223(6) & 0.159(1) \\
4d-3r-4 & 2.84 & 0.649(7) & 0.969(2) & 1.425(8) & 0.673(9) & 0.328(4) & 1.643(3) & 1.136(16) & 0.174(4) & 3d-3r-4 & 2.82 & 0.739(8) & 1.126(8) & 1.473(10) & 0.682(14) & 0.405(5) & 1.938(8) & 1.279(10) & 0.183(2) & 2d-3r-4 & 2.79 & --- & --- & 1.331(23) & 0.649(29) & 0.334(4) & 1.956(8) & 1.595(21) & 0.046(2) \\
4d-3r-5 & 2.85 & 0.654(15) & 0.929(4) & 1.405(17) & 0.887(23) & 0.504(8) & 1.742(8) & 1.210(10) & 0.304(4) & 3d-3r-5 & 2.73 & 0.652(11) & 1.095(6) & 1.366(10) & 0.635(10) & 0.284(4) & 1.747(2) & 1.280(12) & 0.188(3) & 2d-3r-5 & 2.77 & --- & --- & 1.454(19) & 0.723(27) & 0.292(3) & 2.062(8) & 0.927(71) & 0.144(19) \\
4d-3r-6 & 2.73 & 0.773(8) & 0.920(4) & 1.422(7) & 0.735(10) & 0.406(9) & 1.636(2) & 1.291(9) & 0.364(4) & 3d-3r-6 & 2.78 & 0.666(9) & 1.065(5) & 1.417(9) & 0.688(10) & 0.493(5) & 1.934(11) & 1.643(6) & 0.180(2) & 2d-3r-6 & 2.80 & --- & --- & 1.372(34) & 0.524(37) & 0.304(7) & 2.195(16) & 2.336(281) & 0.442(154) \\
4d-3r-7 & 2.80 & 0.787(9) & 0.972(5) & 1.409(11) & 0.769(17) & 0.582(7) & 1.597(11) & 1.438(8) & 0.318(5) & 3d-3r-7 & 2.73 & 0.697(9) & 1.118(5) & 1.393(15) & 0.710(17) & 0.490(5) & 1.668(7) & 2.617(57) & 0.407(30) & 2d-3r-7 & 2.80 & --- & --- & 1.482(31) & 0.748(51) & 0.341(5) & 1.935(11) & 1.138(28) & 0.161(8) \\
 &  &  &  &  &  &  &  &  &  & 3d-3r-8 & 2.73 & 0.718(8) & 1.109(7) & 1.443(11) & 0.657(14) & 0.539(5) & 1.863(13) & 1.090(22) & 0.202(8) & 2d-3r-8 & 2.79 & --- & --- & 1.428(23) & 0.549(26) & 0.420(5) & 2.127(11) & 2.562(184) & 1.236(378) \\
\midrule
4d mean & 2.78 & 0.726(76) & 0.936(47) & 1.413(44) & 0.773(69) & 0.399(106) & 1.590(115) & 1.230(176) & 0.276(64) \\
3d mean & 2.76 & 0.702(82) & 1.085(90) & 1.430(39) & 0.683(60) & 0.439(80) & 1.801(99) & 1.563(503) & 0.222(79) \\
2d mean & 2.79 & --- & --- & 1.423(59) & 0.671(98) & 0.361(54) & 2.035(105) & 1.604(675) & 0.316(391) \\
\midrule
4d mean & 4.47 & 0.627(102) & 1.269(277) & 1.725(444) & 0.596(282) & 0.444(100) & 2.140(656) & 1.764(827) & 0.325(273) \\
3d mean & 4.57 & 0.580(112) & 1.462(317) & 1.529(449) & 0.558(839) & 0.387(81) & 2.199(448) & 1.306(490) & 0.111(98) \\
2d mean & 4.54 & 0.487(104) & 1.579(412) & 1.370(300) & 0.391(268) & 0.318(82) & 2.263(385) & 1.296(547) & 0.193(277) \\
\bottomrule
\end{tabular}%
}
\end{sidewaystable}
\subsection{Trajectory parameter statistics and near-field details}
\label{app:traj-details}

\subsubsection{Far-field parameters and their systematics}
The $rd$-normalized far-field parameters of the nine conditions are examined at each jet spacing to check whether the $rd$ normalization still absorbs the velocity ratio (Figs.~6(j),(k) of the main text): for the rear jet the $B_f$ dependence on the measured velocity ratio within a spacing is weak and mixed (variations of at most $0.1$), whereas the front jet keeps a clear residual dependence ($B_f$ increasing as $r$ decreases), so the $rd$ absorption is only approximate for the front jet. The velocity-ratio dependence of the penetration is carried by $A_f$ in both jets: the condition means are fitted by the linear law $A_f=a\,r+b$, the single jet giving $A_s(r)=0.1433\,r+1.2384$ and the twin jets spanning $a=0.168$--$0.213$ with $b=0.42$--$0.58$ (front jet) and $a=0.132$--$0.323$ with $b=0.635$--$1.666$ (rear jet) across $L=4d$, $3d$, $2d$. The fitted lines form a hierarchy across the nine conditions: the rear-jet line lies above the front-jet line at all $r$, the front-jet lines lie below the single-jet line, and the rear-jet lines lie above it at $L=3d$ and $2d$ (at $L=4d$, $r\approx3$ the rear-jet condition mean $A_f=1.59$ falls marginally below the single-jet value $1.61$, the three-record mean $A_f=1.617$ of table ST2).

As $L$ decreases the slopes flatten and the rear-jet intercept rises ($0.635\rightarrow1.666$) while the front-jet intercept stays within $0.4$--$0.6$: this is the channel through which the spacing modulates the normalized penetration, because at smaller spacing the rear jet is more completely shielded, so it penetrates deeper while its bending shape approaches the point-momentum-source limit. The rear-jet far-field exponent $B_f$ decreases with decreasing spacing and spans the interval between the classical limits $B=1/2$ (line momentum source) and $B=1/3$ (point momentum source) of the single-nozzle JICF~\citep{broadwell_structure_1984,smith_mixing_1998}: the tightly spaced pair approaches a point-like combined source, while the wider pair leaves the rear jet closer to a two-dimensional deflected sheet -- a qualitative analogy between the two classical limits, not a model fit.

\subsubsection{Parameter interpretation}
The two parameters have distinct physical meanings: $A_f$ is the penetration scale (the normalized height reached at $|x-x_0|=rd$) and $B_f$ sets the bending shape, a smaller $B_f$ corresponding to a more sharply bent trajectory; the rear jet therefore penetrates deeper and is bent more sharply than the front jet in eight of the nine conditions, both consequences of the aerodynamic shielding: it develops in the weakened crossflow behind the front jet, which preserves its vertical momentum longer and lets it penetrate further before its sharper far-field bend. The front jet is not passive: its $A_f$ lies below the single-jet line and at the high and low velocity ratios approaches it as $L$ decreases (by about $14\%$ from $L=4d$ to $2d$ at $r\approx7$), while its $B_f$ remains above the single-jet value, so the interaction is mutual even though the front jet is usually assumed to be insensitive to the rear jet~\citep{gutmark_dynamics_2011}. The rear-jet $A_f$ responds more strongly to the spacing change; the resulting ratio $A_f^{\rm rear}/A_f^{\rm front}\approx1.4$--$2.0$ means that the rear jet penetrates roughly $40$--$100\%$ deeper in the $rd$-normalized coordinates.

\textbf{Turning angle.} The turning angle $\Delta\theta$ of Sec.~III A of the main text is defined as the angle between the initial segment direction (arc length $0$--$3$\,mm) and the late-segment direction (the last $30\%$ of the arc length) of the extracted trajectory; the single-jet group means are $56.6^\circ$, $48.6^\circ$ and $31.9^\circ$ at $r\approx3$, 5 and 7. It is a different quantity from the far-field deflection $\varphi$ of table ST1, which is the angle between the late-segment tangent and the wall normal and is evaluated over the last $40\%$ of the trajectory.

\subsubsection{Near-field fits: reporting and reliability}
The near field is analyzed in the same format as the far field (panel (b) of Fig.~5 of the main text and Figs.~6(l),(m) of the main text), with the near-field exponent $B_n$ fitted on the straight near-field segment in the $x$-on-$y$ orientation and the origin set to the measured jet-exit point, the same convention as for the single jet. The fits are reported for $129$ of the $136$ jet trajectories: the seven excluded cases are rear jets whose quasi-vertical near-field segment gives an ill-conditioned fit, and the $L=2d$, $r\approx3$ front jets are reported here but have no far-field entries, their trajectory being swallowed by the crossflow before a self-similar far field develops. All reported fits lie well above the gating thresholds (front $R^2=0.99$--$1.00$, rear $0.88$--$1.00$), so no marginal fit enters the statistics. The near-field coefficient $A_n$ carries a much larger relative uncertainty, because it is an extrapolation to $y=rd$ of a fit made on a short quasi-vertical segment; it is therefore not used in the scaling analysis, and the record-to-record scatter of one condition exceeds the plotted range of Fig.~6(l) of the main text, where the excess is marked by an arrow. The single jet is quasi-vertical in the near field, so its $A_n$ is not defined (the extrapolation is ill-conditioned), and its near field is characterized by $B_n$ and the near-field transition height $t_{y,n}$; its $B_n$ passes the $|B_n|\le4$ gate only for the $r\approx3$ and 5 groups, which is why the main-text reference in Fig.~6(m) covers those two groups.

\subsubsection{Near-field results}
The near-field transition height $t_{y,n}$ is the height at which the straight near-field segment ends and the bend-over begins, i.e.\ the length of the quasi-vertical rise; for the single jet it increases with $r$ (group means $4.8$--$9.2$\,mm from $r\approx3$ to $r\approx7$, individual records $3.9$--$11.8$\,mm). In the twin configuration the front jet bends over up to roughly $40\%$ earlier than the single jet at the same velocity ratio, the shortening being largest at the smallest spacing of the lowest ratio and at the larger spacings of the higher ratios; at $L=3d$, $r\approx7$ the detection scatters widely because three cases bend smoothly from the exit without a distinct vertical segment. The rear jet stays vertical longer at small spacing, consistent with the shielding that weakens the effective crossflow it encounters; at $L=3d$ and $4d$ its values scatter widely because it develops more independently and its near field is not always a clean two-segment shape, and at $L=2d$ it merges with the front jet shortly after the nozzle exit, so its extracted near-field trajectory represents the developing merged plume rather than an independent rear jet.

The $rd$-normalized coefficient $A_n$ carries the velocity-ratio dependence of the initial deflection: for both jets it tends to increase as $r$ decreases (by up to a factor of roughly $6$ across the groups), and at the lowest velocity ratio the front jet's $A_n$ systematically exceeds the rear jet's, i.e.\ the front jet bends over earlier and more strongly in the developing region. The near-field exponent $B_n$ is stable for the front jet and much smaller than the single-jet value: in the $x$-on-$y$ convention the front jet's streamwise offset grows from the nozzle exit (an early bend), whereas the single jet remains at the nozzle until the top of the near-field segment (a quasi-vertical rise); the rear-jet $B_n$ scatters more because at $L=2d$ its near field represents the developing merged plume.

\textbf{Physical-space penetration and deflection.} The penetration depth $H$ and the far-field deflection $\varphi$ of every condition, read directly from the extracted trajectories without any fitted parameter, are collected in table~\ref{tab:st9-penetration} together with the number of records that are still rising at the last resolved station (for which $H$ is a lower bound set by the measurement window).

\subsubsection{Velocity-normalization robustness}
The conclusions of the trajectory section of the main text are robust to the velocity-normalization choice: using the measured per-jet exit velocity instead of the documented bulk value shifts only the absolute penetration scale (by at most $21\%$), with the ordering of all comparisons unchanged.

\subsection{Trajectory extraction and transition detection}
\label{app:algorithm}

\SetAlgoLined\SetInd{2em}{1em}

The algorithm proceeds in five stages (i)--(v), described below together with the corresponding step numbers of Algorithm~\ref{alg:track}: (i) trajectory extraction, (ii) far-field transition detection, (iii) near-field-transition detection, (iv) truncation and near-field gating, and (v) power-law fitting. The natural-coordinate transformation of Algorithm~\ref{alg:nat} is summarized separately under (vi).

\textbf{(i) Trajectory extraction (Steps 1--9).} The trajectory extraction is deliberately independent of the power-law fitting (Steps 20--22). Step 1 initializes two empty lists $F$ and $R$; Steps 2--7 then track the two jets independently: for each jet and row (Step 2) the speed-profile peaks are detected within a search window of half-width $d$ centered on the previous point (on the nominal nozzle position for the first rows; Step 3), refined by a parabola fit (Step 4), and the refined peak nearest to the previous point is kept (Step 5); each candidate is validated (speed jump $\le30\%$, monotone $x$; Step 6) and appended to the track (Step 7). Steps 8--9 refine each point perpendicular to the trajectory (maximum displacement 1\,mm) and smooth by an arc-length B-spline. For the single-nozzle data the same procedure applies, with the exit origin anchored to the tracked point closest to the nominal nozzle position among the first six rows (within $0.5$\,mm). The measured velocity ratio of each case uses the free-stream velocity, the median $x$-velocity in the clean upstream region $x\in[-15,-10]$\,mm, $y\in[25,y_{\max}]$\,mm; within each nominal ratio group the $L=4d$ batch runs up to 4\% lower than the $L=2d$/$3d$ batches (e.g.,\ $1.54$ versus $1.60$\,m/s at $r\approx7$), so its measured ratios sit marginally above the others, and comparisons are made within the same batch wherever the ratio is held fixed.

\textbf{(ii) Far-field transition detection (Steps 11--13).} Following the classical picture~\citep{pratte_profiles_1967}, the trajectory is divided into a quasi-vertical near-field developing region and a far-field self-similar region following a single power law~\citep{broadwell_structure_1984,smith_mixing_1998}, the transition being determined per case. Steps 11--12 scan all candidate splits, fitting the near and far segments in their natural orientations and scoring each by combined straightness ($R^2$) and exponent stability; Step 13 selects the earliest split whose far field is a converged power law (far-field $R^2\ge0.9$, stability $\ge0.85$, exponent changing by $\le5\%$ when the first 15\% of the far-field points are dropped), with the balanced split as a fallback. For the rear jets the far field must start where the trajectory already develops downstream ($\mathrm{d}x/\mathrm{d}y\ge0.05$); case 3d-7r-4 (rear) uses a relaxed tolerance of 12\% because its far-field exponent stabilizes very late. The origin $x_0$ is the measured jet-exit point (the first trajectory point), the same convention as in the fitting stage below and for the single jet; all thresholds are common to all cases except this single tolerance.

\textbf{(iii) Near-field transition detection (Steps 14--18).} Because the near field contains a curved bend-over part, a near-field transition is detected inside it with the same straight-line logic: Steps 14--17 scan all candidate points $k_n$ before the far-field transition, fitting and scoring each; Step 18 selects the best near-field transition and discards the bend-over segment, so the near-field power law is fitted only on the straight segment from the nozzle exit to the near-field transition.

\textbf{(iv) Truncation and near-field gating (Steps 10 and 19).} Step 10 truncates the trajectory near the measurement-domain boundary: for the rear jet, the first point within 1.2\,mm of the right edge is retained as the final point; for low-velocity-ratio rear jets reaching the boundary while $y$ still grows, the cut is placed at the end of the bend-over phase, where $\mathrm{d}x/\mathrm{d}y$ reaches its maximum, removing the spurious near-vertical segment along the boundary. Step 19 gates the near field: the near-field power law is reported only when the fit on $0\ldots k_n$ satisfies $R^2\ge0.5$ (rear: $\ge0.8$), uses at least six valid points and gives $|B_n|\le4$; otherwise $B_n$ is not reported. All transitions and parameters are determined per case; the only cross-case input is the relaxed tolerance of case 3d-7r-4.

\textbf{(v) Power-law fitting (Steps 20--22).} After determining the two transition points, Steps 20--22 fit the near-field segment $0\ldots k_n$ and the far-field segment separately to the logarithmic power law, yielding $(B_n,A_n)$ and $(B_f,A_f)$ of the trajectory power law of the main text ($y_0=0$, $d=4$\,mm, $r$ the measured velocity ratio); the logarithmic fit returns the exponent $B$ and the intercept $c=\ln A$. The near-field segment is fitted in the $x$-on-$y$ orientation ($x-x_0=A y^{B_n}$) and the far-field segment in the $y$-on-$x$ orientation; the origin $x_0$ is the measured jet-exit point (the first trajectory point) for both single- and twin-jet cases, so the near-field parameters are directly comparable. In total, 68 twin-nozzle cases and the ten retained single-jet records (three single-nozzle conditions) are included in the scaling analysis.

\textbf{Velocity-normalization sensitivity.} Because $A_f$ is normalized by $(rd)^{\,B-1}$, its absolute value inherits the reference velocity: re-normalizing with the measured per-jet exit velocity ($u_j=14$\,m/s, the Limitations section of the main text) instead of the documented $V_j=10$\,m/s scales every $A_f$ by $1.4^{\,B-1}$ with $B=B_f<1$, i.e.\ by condition-mean factors of $0.84$--$0.91$ (front jet), $0.79$--$0.85$ (rear jet) and $0.81$--$0.82$ (single jet); the absolute penetration scale therefore shifts by at most 21\%, while the ordering used by the conclusions is preserved (rear/front ratio $1.3$--$2.0$ instead of $1.4$--$2.0$; condition-mean $A_f$ ratios $0.52$--$1.15$ instead of $0.46$--$1.11$). The documented $V_j=10$\,m/s convention is retained.

\clearpage

\section{The effective velocity ratio (Sec.~III B of the main text)}
\label{app:fit-robust}

The estimator protocol is specified here operationally, and its robustness tests are collected in figures S4 and S5.
The effective velocity ratio of Sec.~III B of the main text [Eq.~(4) of the main text] is measured by re-scaling each front-jet trajectory by the trial ratio $r'$ in the single-jet coordinates $\tilde u=(x-x_{\rm exit})/(r'd)$, $\tilde v=y/(r'd)$ and matching it against a reference library built from the 10 retained single-jet records (the $r\approx7$ outlier is not used): the three velocity-ratio groups ($r\approx2.70$, $4.50$ and $6.05$) provide calibrated reference curves, a reference at any intermediate ratio is obtained by linear interpolation in $r'$ between the group means, and below or above the calibrated range the nearest group curve is scaled by the single-jet penetration law $A_s(r')$ (the shape of the nearest group is retained). The effective ratio $r_{\mathrm{eff}}$ minimizes the sum of squared residuals between the measured trajectory and the reference curve over the matching window $\tilde u\in[0.05,\,2.30]$ (the resolvable near-to-mid field), the minimization being a golden-section search over $r_{\mathrm{eff}}\in[0.25r,\,1.45r]$ with a tolerance of $10^{-3}r$. The estimator returns a value for all 68 twin cases, including the 12 whose near-field transition detection fails, and recovers the single-jet records themselves with $r_{\mathrm{eff}}\approx r$, so the reference library is self-consistent (Fig.~7 of the main text): the closure is $8$ of $11$ records within $5\%$ and all $11$ within $10\%$ when each tested record is removed from the library that builds the reference family (leave-one-out; the in-sample closure is $10$ of $11$ within $5\%$, which is optimistic because a record then contributes to its own reference mean). The far-field trajectory coefficient gives a third route to the same ratio: evaluated in the ratio form $r_{\mathrm{eff},A}/r=A_f^{\rm front}/A_s(r)$ against the single-jet law $A_s(r)=0.1433\,r+1.2384$ (evaluated at the measured $r$, with the highest-ratio conditions slightly beyond the single-jet calibration range, where the law is extrapolated), the per-case front-jet coefficients of the 60 resolvable cases reproduce the condition ranking of the curve matching at condition-level rank correlation $0.89$, but at a uniform $19$--$31\%$ upward offset; the highest-ratio twin conditions lie slightly beyond the single-jet calibration range and use the extrapolated law -- the far-field coefficient retains a memory of the interaction that the whole-trajectory matching removes, so it ranks the conditions but does not calibrate them (panels (b) and (c) of figure S4 of this supplement); the $r\approx5$ conditions at $L=2d$ and $3d$, the conditions retaining the largest decay-rate excess ($35$--$42\%$), are also where the front-jet coefficient sits furthest above the single-jet line. The three estimators are compared case by case over their common subset in figure S5 of this supplement. One calibration caveat concerns the $r\approx3$ groups: their absolute ratios $r_{\mathrm{eff}}\approx1.13$, $1.39$ and $1.27$ (at $L=2d$, $3d$, $4d$) lie below the lowest calibrated group ($r\approx2.70$), so their reference curves are obtained by extending the linear law $A_f(r)$ below the calibrated domain; the single-jet self-recovery test of Fig.~7(a) of the main text covers only $r\gtrsim2.7$, so the absolute $r_{\mathrm{eff}}$ of these groups carries an extrapolation uncertainty that the test does not quantify -- the law is therefore stated, and fitted, in the normalized form $r_{\mathrm{eff}}/r$, and the window-robustness test (figure S4(b) of this supplement) shows that the $r\approx3$ groups are also the ones whose extracted ratio depends most on the matching window.
\textbf{Per-case statistics.} The dilution is a property of the condition means; the per-case scatter is $\pm0.05$--$0.12$ within conditions (68 cases), larger than the condition-to-condition spread about the fit ($\pm0.03$) and dominated by batch-to-batch variations rather than by the interaction trend, so the condition means are the appropriate calibration and testing level (the same conclusion as the earlier per-case analysis of the transition-height estimator, whose scatter $\pm0.1$--$0.3$ was dominated by detection variability).
\textbf{Window robustness.} The matching window is the single calibration choice of the estimator, and its effect is a systematic one that we state explicitly rather than average away. The primary window $u\in[0.05,\,2.30]$ covers the near field and the early bending region of all ratio groups; shrinking it to the narrow near-field window $u\in[0.05,\,0.50]$, which excludes the bending region altogether, extracts deeper ratios at $r\approx3$ ($0.27$--$0.30$ against $0.41$--$0.51$) while leaving the $r\approx5$ and 7 groups nearly unchanged, so the ordering $3r<5r<7r$ is preserved within every spacing and the fit constant moves from $c=3.31$ ($R^2=0.80$) to $c=4.45$ ($R^2=0.75$): the form of the law is a property of the flow, but the constant carries a window systematic of order $+34\%$, an order of magnitude larger than the case-resampling confidence interval of the primary fit ($[3.18,\,3.43]$ at 95\%); that interval quantifies the statistical uncertainty at a fixed window, not the calibration freedom of the estimator (Fig.~16(a) of the main text). Resampling the reference library and the matching window together with the cases (400 draws per layer) widens it to a total 95\% interval of $[3.03,\ 4.81]$, consistent with the $[3.31,\ 4.86]$ span of eighteen estimator variants (relative- and logarithmic-residual matching and a kernel-smoothed reference family); all variants preserve the hyperbolic form ($R^2=0.75$--$0.86$) and the condition ordering (Spearman $\ge0.82$), while relative- or logarithmic-residual matching reduces the window shift from $+34\%$ to $+18$--$24\%$ (to $+9\%$ combined with a kernel-smoothed reference family) at the price of a $+16$--$24\%$ level shift. The model comparison must be read with the same caveat: in the primary window the affine two-parameter alternative $a\,r+b$, which would allow a residual linear drift in $r$, fits the nine condition means with $R^2=0.74$ and leave-one-out error $0.055$, against $0.042$ for the one-parameter hyperbola, and the momentum-dilution square-root variant $(r_{\mathrm{eff}}/r)^2=r/(r+c)$ gives $c=9.25$ with $R^2=0.79$ and a leave-one-out error that degrades by 29\% across leave-one-group-out folds; in the narrow window, however, the affine form fits marginally better than the hyperbola ($R^2=0.83$ against $0.75$; with $n=9$ condition means the two are indistinguishable), so the preference for the hyperbola rests on the primary window, on parsimony and on the dilution-bookkeeping reading of the main text, not on winning every window (the comparison of the fitted forms is listed above).
\begin{figure}
\centerline{\includegraphics[width=0.995\textwidth]{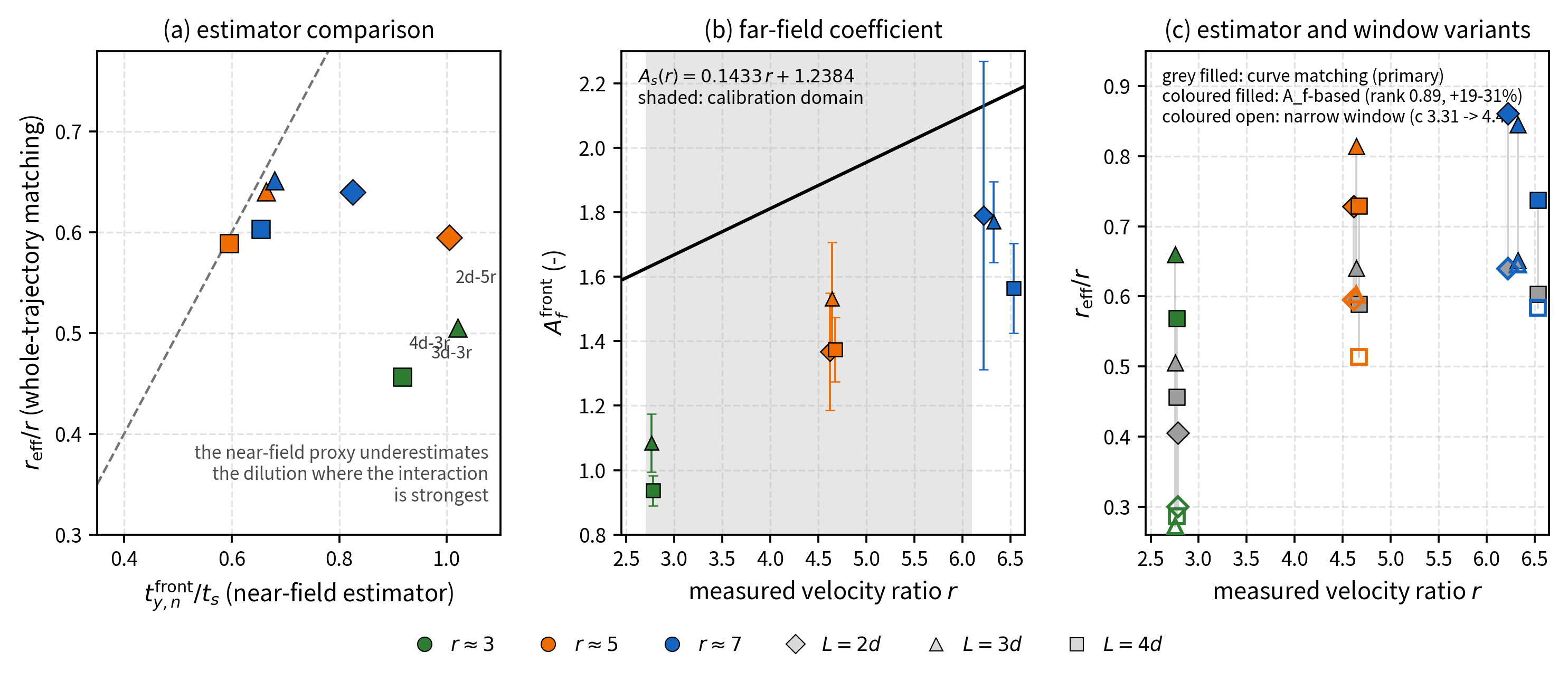}}
\caption{\textbf{Figure S4.} Robustness of the effective-ratio estimators at the condition level. (a) The transition-height ratio $t_{y,n}^{\rm front}/t_s$ against the curve-matching ratio $r_{\mathrm{eff}}/r$ (dashed: 1:1); the $L=2d$, $r\approx3$ condition has no detectable near-field transition and is omitted. (b) The front-jet far-field coefficient $A_f^{\rm front}$ (condition means $\pm$ one SD) against the measured ratio, with the single-jet law $A_s(r)=0.1433\,r+1.2384$ and its calibration domain (shaded). (c) Curve matching with the primary window (grey filled), the far-field-coefficient route (coloured filled) and curve matching with the narrow window (coloured open), all in the $r_{\mathrm{eff}}/r$ form; colour: ratio group, shape: spacing.}\alttext{Three-panel robustness figure: the near-field proxy against the whole-trajectory ratio, the far-field coefficient against the single-jet law, and all estimator and window variants in the ratio form.}
\end{figure}

\begin{figure}
\centerline{\includegraphics[width=\textwidth]{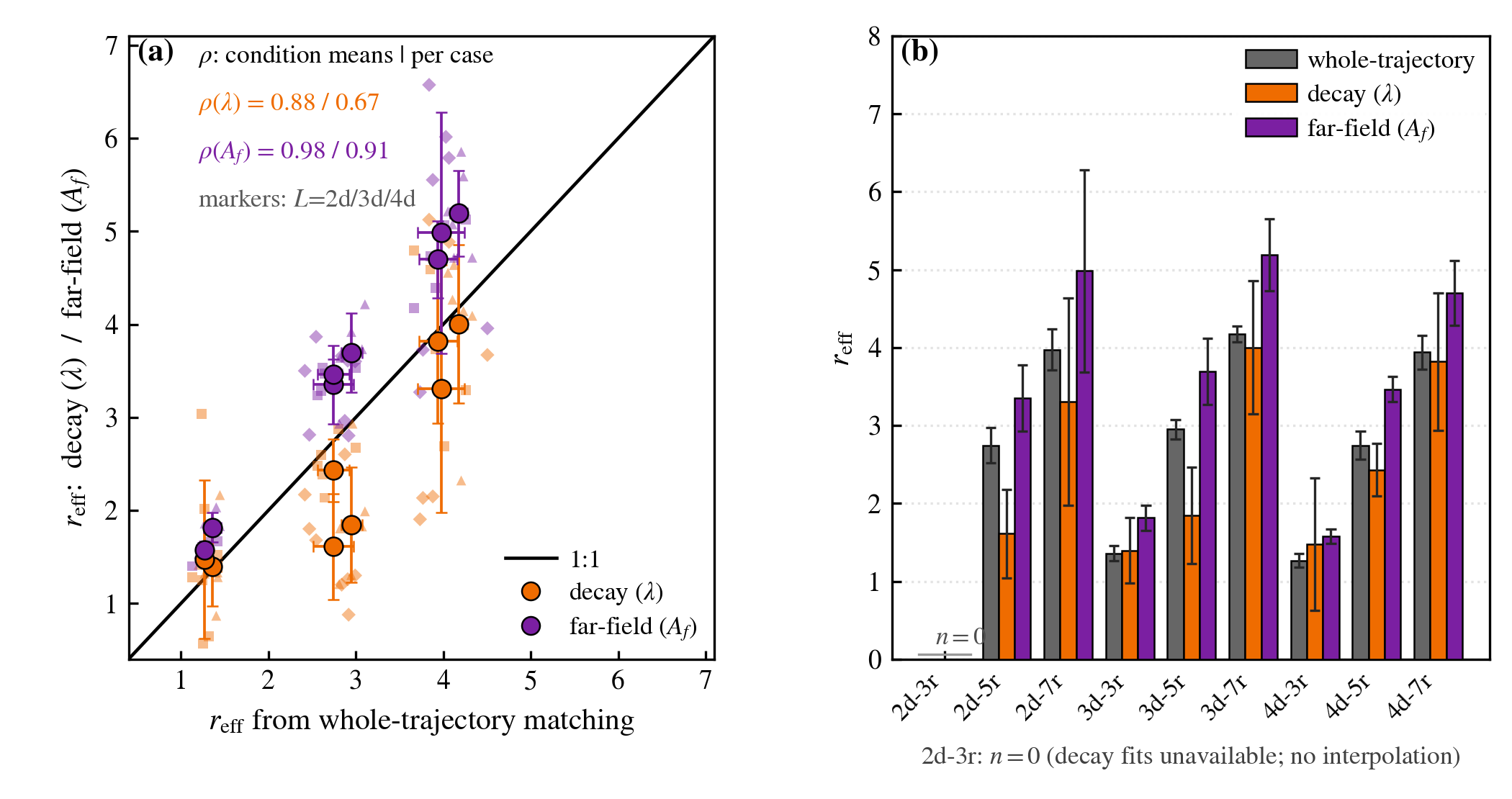}}
\caption{\textbf{Figure S5.} The three estimators of the effective ratio over their common subset (52 of the 68 twin cases; the $L=2d$, $r\approx3$ condition has no far-field coefficient). (a) Per-case estimates from the decay inversion $r_{\mathrm{eff},\lambda}$ (orange) and the far-field coefficient $r_{\mathrm{eff},A}$ (purple) against the curve-matching $r_{\mathrm{eff}}$ (dashed: 1:1). (b) Condition means of the three estimators ($\pm$ one SD; marker shape: spacing). Spearman rank correlations with the curve-matching $r_{\mathrm{eff}}$ on the absolute values: condition level $\rho(\lambda)=0.88$, $\rho(A)=0.98$ ($n=8$); case level $\rho(\lambda)=0.67$, $\rho(A)=0.91$ ($n=52$).}\alttext{Two-panel comparison of the three effective-ratio estimators over their common subset of 52 cases, per case and as condition means.}
\end{figure}

\clearpage

\section{Velocity decay in natural coordinates (Sec.~III C of the main text)}

{\fussy\emergencystretch=0pt The natural-coordinate transformation applied to the velocity field is summarized in Algorithm~\ref{alg:nat}; it produces the straightened fields, the centerline decay and the cross-track width shown in Figs.~9--11 of the main text, and the per-condition decay detail of Fig.~10 of the main text is given first below, with the condition-mean parameters collected in table ST3.\par}
\begin{algorithm}[H]
\algtitle{Natural-coordinate transformation of the velocity field}{alg:nat}
\KwIn{trajectory points $\{(x_i,y_i)\}_{i=1}^{N}$; velocity field $V(x,y)$; arc-length grid $\{s_j\}$; normal grid $\{n_k\}$}
\KwOut{normalized field $\tilde V(s_j,n_k)$ and width $\sigma(s_j)$}
compute the arc length $s_i=s_{i-1}+\sqrt{(x_i-x_{i-1})^2+(y_i-y_{i-1})^2}$, $s_1=0$\;
interpolate the trajectory onto the grid $\{s_j\}$: $x(s_j)$, $y(s_j)$\;
compute the tangent $\mathbf t=(dx/ds,dy/ds)$ and the normal $\mathbf n=(-dy/ds,dx/ds)$, both normalized\;
\For{each $s_j$}{
  \For{each $n_k$}{
    $\mathbf p=(x(s_j),y(s_j))+n_k\,\mathbf n(s_j)$\;
    $\tilde V(s_j,n_k)\gets$ bilinear interpolation of $V(\mathbf p)$ on the PIV grid\;
  }
}
normalize: $\tilde y=(\tilde V-U_\infty)/(V_0-U_\infty)$ with $V_0$ the jet-exit centerline value\;
average $\tilde y$ over the cases of the same condition at each $(s_j,n_k)$\;
\For{each $s_j$}{
  find the peak of $\tilde y(s_j,n_k)$ and the half-maximum crossings on both sides\;
  compute $\sigma_L(s_j)={\rm HWHM}_L/\sqrt{2\ln2}$ and $\sigma_R(s_j)={\rm HWHM}_R/\sqrt{2\ln2}$\;
  record $\sigma(s_j)=(\sigma_L+\sigma_R)/2$ and the asymmetry $(\sigma_R-\sigma_L)/(\sigma_R+\sigma_L)$\;
}
\Return $\tilde y(s_j,n_k)$, $\sigma(s_j)$
\end{algorithm}

\subsubsection{Excess-velocity decay: per-condition detail}
The condition-mean decay curves of Sec.~III C of the main text (Fig.~10) are collected here with their interpretation. Every condition follows the same sequence: a potential-core plateau, a deflection region in which the centerline excess velocity begins to fall, and a far-field segment in which the decay is well described by the Weibull form $\exp[-(s/\eta)^{\beta}]$ with the shape $\beta$ and the scale $\eta$ quoted in the panels. The rate ratio $\lambda/\lambda_{\rm single}$ -- the condition-mean exponential rate divided by the single-jet value interpolated at the same measured velocity ratio -- separates the two jets: the front jet lies between $1.3$ and $1.8$ times the single-jet rate in every condition, the largest values occurring at the low velocity ratios and at the smallest spacing, whereas the shielded rear jet stays between $0.8$ and $1.1$ times that rate, i.e.\ it decays at the single-jet rate within the scatter.

The front jet loses its plateau almost immediately -- its excess velocity leaves the unity plateau already at the first sampled station ($s\approx0.5$\,mm) and falls by $17$--$19\%$ within the first $2$\,mm at $r\approx3$, $5$--$8\%$ at $r\approx5$ and $1$--$5\%$ at $r\approx7$ -- so it has essentially no potential core, whereas the rear jet keeps the plateau over the first few millimeters ($-3$--$0\%$ change at $r\approx5$ and 7, $+3$--$6\%$ even at $r\approx3$) and the single jet accelerates slightly ($+2$--$7\%$) before decaying. The front-jet rate grows as the spacing decreases at $r\approx5$ ($\lambda=0.099$, $0.119$ and $0.130$\,mm$^{-1}$ at $L=4d$, $3d$ and $2d$) and is non-monotonic at $r\approx3$ ($0.148$, $0.138$ and $0.164$\,mm$^{-1}$), while the rear-jet rate stays within $0.045$--$0.096$\,mm$^{-1}$ in every condition. The condition means of all four quantities -- the Weibull shape $\beta$, the decay scale $\eta$, the exponential rate $\lambda$ and the rate ratio $\lambda/\lambda_{\rm single}$ -- are collected in table ST3, together with the single-jet reference of the three velocity-ratio groups.

\clearpage

\begin{table}
\centering
\small
\caption{\textbf{Table ST3.} Condition-mean parameters of the excess-velocity decay of Fig.~10 of the main text, for the front and the rear jet of the nine conditions: the Weibull shape $\beta$ and decay scale $\eta$, the exponential rate $\lambda$ and its ratio to the single-jet value interpolated at the same measured velocity ratio (reference rows at the foot of the table). Every entry is the mean over the cases of the condition of the per-case fits to the natural-coordinate centerline decay, started at the near-field transition -- these are the values shown in the $\beta$, $\eta$ and $\lambda/\lambda_{\rm single}$ panels of Fig.~10 of the main text; Table~IV of the main text quotes instead the condition-mean rates fitted past the far-field transition, which are systematically smaller. The single-jet reference rows are evaluated at the group-mean ratios ($r=6.06$, $4.50$, $2.70$), whereas the $\lambda/\lambda_{\rm single}$ column interpolates the single-jet rate at the measured ratio of each condition ($0.0709$--$0.0714$\,mm$^{-1}$ across the $r\approx5$ group), so the column ratios do not divide exactly by the reference-row values. The case-to-case scatter is drawn in the same panels of Fig.~10.}\alttext{Table of the condition-mean Weibull and exponential decay parameters of the front and rear jets with the single-jet reference rows.}
\label{tab:st7-decay}
\begin{tabular}{llcccccccc}
\toprule
& & \multicolumn{4}{c}{front jet} & \multicolumn{4}{c}{rear jet} \\
\cmidrule(lr){3-6}\cmidrule(lr){7-10}
$r$ & $L$ & $\beta$ & $\eta$ (mm) & $\lambda$ (mm$^{-1}$) & $\lambda/\lambda_{\rm s}$ & $\beta$ & $\eta$ (mm) & $\lambda$ (mm$^{-1}$) & $\lambda/\lambda_{\rm s}$ \\
\midrule
6.53 & $4d$ & 1.47 & 16.7 & 0.078 & 1.36 & 2.60 & 28.5 & 0.052 & 0.91 \\
4.67 & $4d$ & 1.36 & 12.8 & 0.099 & 1.40 & 1.84 & 22.8 & 0.067 & 0.95 \\
2.78 & $4d$ & 1.26 & 9.1 & 0.148 & 1.64 & 1.00 & 9.7 & 0.083 & 0.91 \\
6.32 & $3d$ & 1.62 & 17.9 & 0.076 & 1.33 & 3.27 & 31.1 & 0.054 & 0.94 \\
4.64 & $3d$ & 1.69 & 12.9 & 0.119 & 1.67 & 1.83 & 24.1 & 0.060 & 0.84 \\
2.76 & $3d$ & 1.28 & 9.3 & 0.138 & 1.53 & 1.54 & 14.3 & 0.093 & 1.03 \\
6.22 & $2d$ & 1.86 & 16.5 & 0.088 & 1.54 & 2.05 & 29.7 & 0.045 & 0.78 \\
4.62 & $2d$ & 1.67 & 11.9 & 0.130 & 1.82 & 2.23 & 25.2 & 0.069 & 0.96 \\
2.79 & $2d$ & 1.50 & 7.5 & 0.164 & 1.81 & 1.51 & 13.7 & 0.096 & 1.06 \\
\midrule
\multicolumn{10}{l}{\emph{single-jet reference (measured ratio of each group)}}\\
6.06 & --- & 2.98 & 27.6 & 0.057 & --- & & & & \\
4.50 & --- & 2.37 & 22.8 & 0.073 & --- & & & & \\
2.70 & --- & 1.77 & 17.1 & 0.091 & --- & & & & \\
\bottomrule
\end{tabular}
\end{table}
Fig.~11(a)--(i) and Fig.~8(i) of the main text show a V-shaped evolution shared by all conditions: the jet first narrows under the crossflow compression, reaches a minimum near the near-field transition, and then grows slowly downstream, so the centerline decay is controlled by mixing, while the cross-section evolves through compression and deformation rather than self-similar spreading. The overall width $\bar\sigma$ (the $\sigma(s)$ mean over the usable stations) is nearly universal, $0.75$--$1.45$\,mm ($0.19$--$0.36d$), with the growth rate after the minimum small ($k=d\sigma/ds\approx0.02$--$0.05$\,mm/mm at $L=4d$ and $3d$) and the relative width asymmetry $\langle|\Delta\sigma|/(\sigma_L+\sigma_R)\rangle$ modest, $0.02$--$0.13$ when averaged over the jet core, i.e.\ over the stations where the profile peak exceeds half of its maximum (Fig.~11(l) of the main text).

The only systematic exceptions are the $L=2d$ front jets, which stay narrow and continue to contract ($\bar\sigma\approx0.75$--$0.91$\,mm, $k\approx-0.04$ to $-0.01$\,mm/mm) because they merge with the rear jet immediately, and the slightly smaller growth rate of the $L=2d$ rear jet ($k\approx0.02$--$0.03$\,mm/mm). The width is therefore not a sensitive discriminator; the single jet shows the same V-shaped evolution ($\bar\sigma\approx1.2$--$1.4$\,mm; Fig.~8(i) of the main text), so the contraction--spreading sequence is an intrinsic feature of the deflected jet rather than a twin-jet effect.

At fixed spacing, the soft RZ of both jets increases with $r$ (front: $26.1$--$32.8$\,mm$^2$ at $r\approx3$, $80.4$--$97.7$\,mm$^2$ at $r\approx5$, $97.4$--$121.8$\,mm$^2$ at $r\approx7$; rear: $29.3$--$32.2$, $55.7$--$101.3$ and $75.7$--$119.9$\,mm$^2$), so the velocity-ratio dependence is stronger than the spacing dependence. At $L=2d$ the front and rear zones merge into the inter-jet recirculation, attributed to the front jet; the rear-side soft area still exceeds the front-side condition mean at every velocity ratio ($29.3$ versus $26.1$, $101.3$ versus $80.4$, $119.9$ versus $97.4$\,mm$^2$ at $r\approx3$, 5 and 7), reflecting the intermittently reversing wake downstream of the rear jet; this merged recirculation sustains the merged plume, while the rear jet rises in the shielded region behind it.

\subsection{Natural-coordinate transformation and width/decay definitions}
\textbf{(vi) Natural-coordinate transformation and width/decay definitions (Algorithm~\ref{alg:nat}).} The transformation of Algorithm~\ref{alg:nat} converts the time-averaged field of each case into the natural coordinates $(s,n)$ of its extracted trajectory: the trajectory is interpolated onto a uniform arc-length grid, the local tangent and normal vectors are evaluated, and the velocity magnitude is interpolated onto the regular $(s,n)$ grid, each case being normalized by its jet-exit excess velocity $(V_0-U_\infty)$. Because every case is straightened along the arc length of its own trajectory and scaled by its own exit excess velocity, the fields of cases with different trajectories share one frame: the cases of a condition can therefore be averaged cell by cell, and the independent single-jet records give the reference on the same grid. The cross-track profile at each $s$ is characterized by its half-maximum width: the peak and the half-maximum crossings on the two sides give the equivalent Gaussian widths $\sigma_L(s)={\rm HWHM}_L/\sqrt{2\ln2}$ and $\sigma_R(s)={\rm HWHM}_R/\sqrt{2\ln2}$; the mean $\sigma(s)=(\sigma_L+\sigma_R)/2$ records the width and the ratio $(\sigma_R-\sigma_L)/(\sigma_R+\sigma_L)$ its asymmetry, and this yields the trajectory-aligned centerline decay $\hat y(s)=(V-U_\infty)/(V_0-U_\infty)$ at $n=0$ and the cross-track width $\sigma(s)$.

\clearpage

\clearpage

\section{Decay confirms the dilution (Sec.~III D of the main text)}

Figure S6 resolves the collapse of the decay rate condition by condition. Read at the measured ratio (open circles), every front jet decays faster than the single jet at the same ratio; moved to its effective ratio (filled circles), the condition mean lands on the single-jet calibration $\lambda_s=0.162\,r^{-0.563}$ to within $\pm8\%$ for four of the nine conditions and within $\pm18\%$ for seven (median $8\%$), the two conditions retaining the largest excess being $L=2d$ and $3d$ at $r\approx5$. The three $r\approx3$ conditions move below the lowest calibrated group ($r\approx2.7$), where the calibration is extrapolated.
\begin{figure}
\centerline{\includegraphics[width=\textwidth]{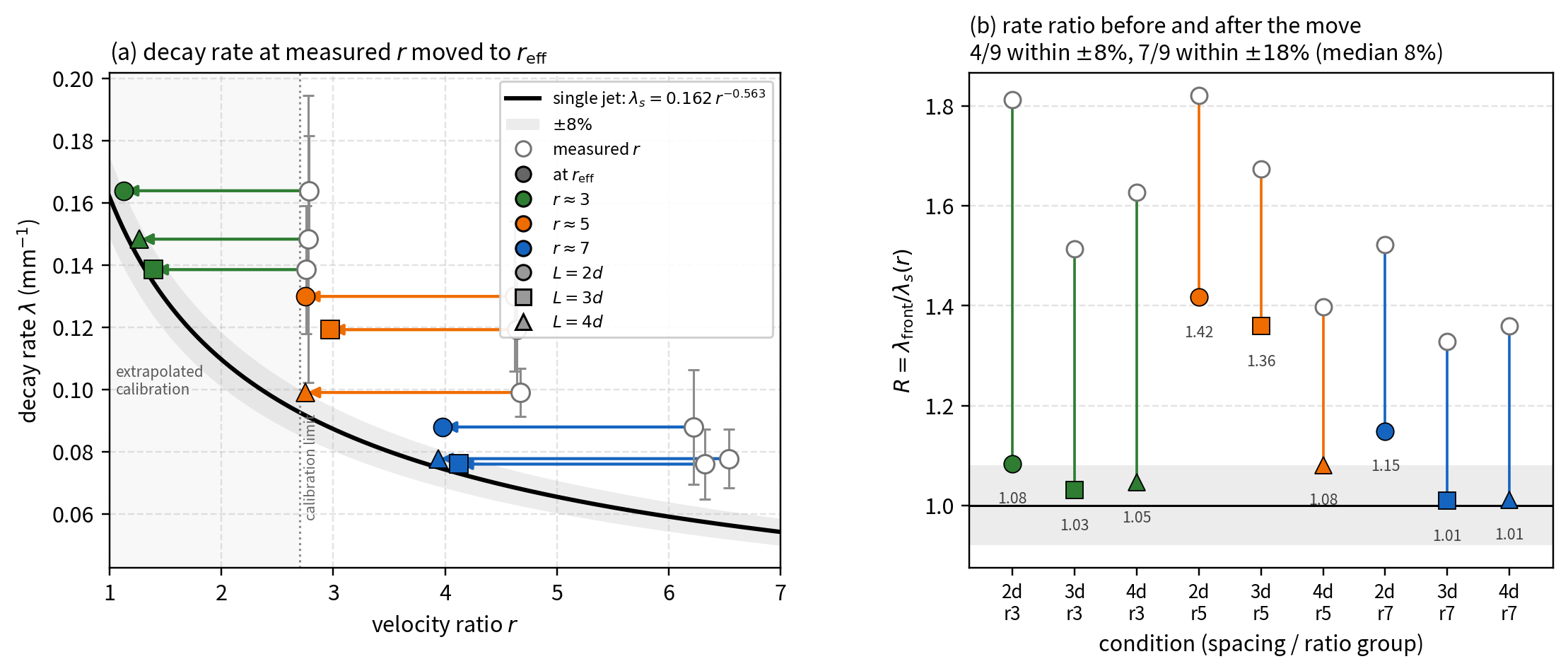}}
\caption{\textbf{Figure S6.} The decay-rate collapse at the effective ratio (detail of Fig.~12(b) of the main text). (a) Condition-mean front-jet decay rate against the measured ratio, with the single-jet calibration $\lambda_s=0.162\,r^{-0.563}$ ($\pm8\%$ band); arrows move each condition to $r_{\mathrm{eff}}$, where filled markers show the same rate. (b) The rate ratio $R=\lambda_{\rm front}/\lambda_s$ at the measured ratio (open) and at $r_{\mathrm{eff}}$ (filled); the band is $\pm8\%$.}\alttext{Two-panel detail of the decay-rate collapse at the effective velocity ratio: the condition-mean decay rate versus measured ratio with arrows to the effective ratio, and the rate ratio before and after the move.}
\label{fig:si-lam-reff}
\end{figure}

\clearpage

\section{Reverse-flow region (Sec.~III E of the main text)}

\textbf{Reverse-flow threshold sensitivity.} The soft reverse-flow metrics of the main text (the frame-resolved probability $P$ and the soft area $A_{RZ}$, strength $S$ and mean reverse speed $\bar{u}_{\rm rev}$ of Eqs.~(6)--(9)) are recomputed at the alternative thresholds $u_{\rm thr}=0.025$ and $0.1$\,m/s (same definitions as the $0.05$\,m/s baseline, differing only in the threshold). table ST4 lists the percentage change of the condition-mean soft area $A_{RZ}$ and strength $S$ relative to the $0.05$\,m/s baseline: $A_{RZ}$ moves by $-10.2\%$ to $+5.4\%$ (front side $-6.6\%$ to $+3.5\%$) and $S$ by at most $1.5\%$. The conclusions are unchanged at every threshold: the front side dominates in $S$ in all nine conditions and in $A_{RZ}$ in five of the six conditions with $L\ge3d$ (the $L=2d$ rear-side excess persists at both alternative thresholds; at $L=4d$, $r\approx3$ the front-side area falls marginally below the rear-side value at every threshold), both metrics increase monotonically with the spacing at fixed velocity ratio, and the twin/single comparison of the recirculation-comparison figure of the main text keeps its ratios within $0.85$--$1.18$ (area) and $0.90$--$1.66$ (strength) across the three thresholds (these are soft reverse-flow ratios; the largest-connected-component ratios of the main text are $1.40$--$1.55$ for the area at $r\approx7$).
\begin{table}
\centering
\begin{tabular}{lcccccccc}
\hline
 & \multicolumn{4}{c}{$u_{\rm thr}=0.025$\,m/s} & \multicolumn{4}{c}{$u_{\rm thr}=0.1$\,m/s} \\
condition & $\Delta A_{RZ}^{f}$ & $\Delta A_{RZ}^{r}$ & $\Delta S^{f}$ & $\Delta S^{r}$ & $\Delta A_{RZ}^{f}$ & $\Delta A_{RZ}^{r}$ & $\Delta S^{f}$ & $\Delta S^{r}$ \\
\hline
$L=2d$, $r\approx3$ & $+2.1$ & $+2.9$ & $+0.1$ & $+0.1$ & $-4.1$ & $-5.6$ & $-0.3$ & $-0.5$ \\
$L=2d$, $r\approx5$ & $+1.9$ & $+3.8$ & $+0.1$ & $+0.2$ & $-3.7$ & $-7.4$ & $-0.3$ & $-0.9$ \\
$L=2d$, $r\approx7$ & $+3.0$ & $+4.2$ & $+0.1$ & $+0.3$ & $-5.9$ & $-8.1$ & $-0.5$ & $-1.1$ \\
$L=3d$, $r\approx3$ & $+2.1$ & $+2.6$ & $+0.1$ & $+0.1$ & $-4.2$ & $-4.9$ & $-0.3$ & $-0.4$ \\
$L=3d$, $r\approx5$ & $+2.4$ & $+4.6$ & $+0.1$ & $+0.3$ & $-4.6$ & $-8.8$ & $-0.4$ & $-1.0$ \\
$L=3d$, $r\approx7$ & $+3.5$ & $+5.4$ & $+0.2$ & $+0.4$ & $-6.6$ & $-10.2$ & $-0.7$ & $-1.5$ \\
$L=4d$, $r\approx3$ & $+2.1$ & $+2.3$ & $+0.1$ & $+0.1$ & $-4.1$ & $-4.5$ & $-0.3$ & $-0.3$ \\
$L=4d$, $r\approx5$ & $+2.3$ & $+4.0$ & $+0.1$ & $+0.2$ & $-4.4$ & $-7.6$ & $-0.3$ & $-0.7$ \\
$L=4d$, $r\approx7$ & $+2.7$ & $+5.1$ & $+0.1$ & $+0.3$ & $-5.2$ & $-9.6$ & $-0.4$ & $-1.2$ \\
\hline
\end{tabular}
\caption{\textbf{Table ST4.} Reverse-flow threshold sensitivity: percentage change of the condition-mean soft reverse-flow area $A_{RZ}$ and strength $S$ at $u_{\rm thr}=0.025$ and $0.1$\,m/s, each taken relative to the $u_{\rm thr}=0.05$\,m/s baseline (a positive entry therefore means that the alternative threshold gives the larger value: the looser $0.025$\,m/s threshold enlarges the area, the stricter $0.1$\,m/s threshold shrinks it), on the front (f) and rear (r) side of each of the nine conditions (Table~I of the main text); each entry is the mean over the cases of the condition.}\alttext{Table of the percentage change of the soft reverse-flow area and strength at the alternative velocity thresholds.}
\label{tab:rz-thr-sens}
\end{table}
The hard-threshold statistics behind the twin/single ratios of Fig.~14(a) of the main text are collected in figure S7: the reverse-flow area and the pixel-summed speed over it, taken from the time-averaged fields ($u<-u_{\rm thr}$, largest connected component) and normalized by the single-jet baseline of the same ratio. They are the absolute counterparts of the soft, frame-resolved metrics of Fig.~13 of the main text and preserve their ordering: the front side dominates the recirculation, and both statistics increase with the spacing at the highest ratio, where the blockage excess is clearest.
\begin{figure}
\centerline{\includegraphics[width=\textwidth]{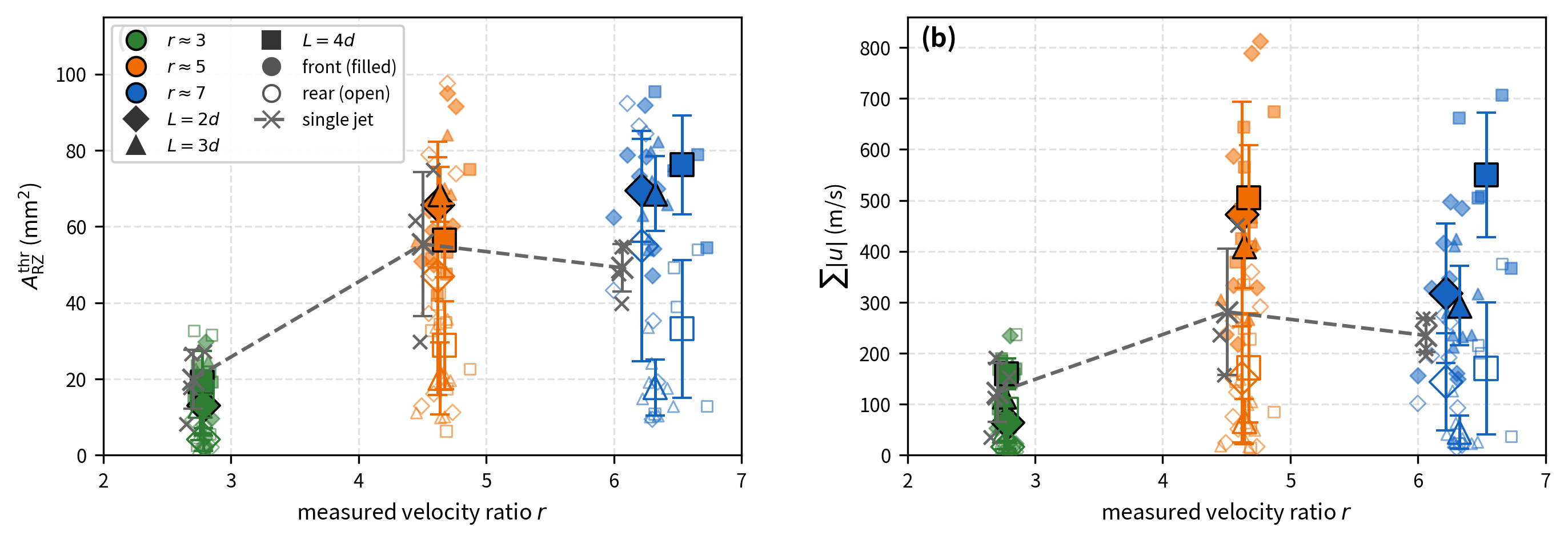}}
\caption{\textbf{Figure S7.} Hard-threshold reverse-flow statistics of the nine conditions (detail of Fig.~14(a) of the main text): (a) reverse-flow area $A_{\rm RZ}^{\rm thr}$ and (b) pixel-summed speed $\sum|u|$ of the time-averaged fields versus measured ratio, with the single-jet baseline of the same ratio. Per-case values (small symbols), condition means $\pm$ SD (large symbols); filled/open: front/rear, colour: ratio group, shape: spacing.}\alttext{Two-panel figure of the hard-threshold reverse-flow area and pixel-summed speed versus measured velocity ratio, compared with the single-jet baseline.}
\end{figure}

\clearpage

\section{Low-pressure attraction (Sec.~III F of the main text)}

\textbf{Pressure boundary condition.} The Poisson reconstruction uses a zero Dirichlet boundary condition on the domain boundary, so the absolute pressure level is arbitrary and only pressure differences within the field are meaningful. All pressure statistics of the main text are sampled in the interior of the field (the gap, the trajectory stations and the inter-jet region), far from the boundary, so the reported differences are insensitive to the boundary value. The case-to-case scatter of the field range ($190\pm36$\,Pa) includes this reconstruction uncertainty.
\textbf{Pressure-gradient boundary sensitivity.} The Poisson reconstruction uses $P=0$ on the domain boundary; pressure \emph{gradients} near the wall are the most exposed to the boundary condition, while the interior pressure differences of the main text are insensitive to it (see the next paragraph). The momentum-balance comparison of the companion paper samples the gap-side zone above $y=2$\,mm and is therefore only weakly affected, but is intended as a magnitude comparison, not a boundary-condition-free budget.
\textbf{Local-reference pressure check.} Because the reconstruction fixes $P=0$ on the domain boundary, and the Poisson equation is elliptic, a gap-side excess computed against that global reference could in principle be an artifact of the boundary condition. The check is therefore repeated against a \emph{local} reference: for each case and each height $y$, the mean pressure over the gap zone between the two nozzles ($x=10$ to $10+L$\,mm, with $L$ the center-to-center spacing) is compared with the mean pressure over the clean upstream inflow at the same height ($x=-14$ to $-5$\,mm), which removes any global offset. Over the near-field heights $y=2$--$16$\,mm the gap-side excess is positive in all 68 cases, with a case-mean of $+12.0$\,Pa and a range of $+6.6$ to $+27.4$\,Pa (per-condition means $+8.4$ to $+18.8$\,Pa). The elevation of the inter-jet pressure relative to the surrounding flow at the same height is therefore independent of the global reference level. The local excess is smaller than the value obtained against the global zero reference (up to $+34$\,Pa at $L=2d$), because part of the latter reflects the overall pressure level of the field rather than the blockage; the sign and the ordering across conditions are unchanged.
The reconstruction chain was validated on a synthetic field and checked for averaging convergence; the two diagnostics are collected in figures S8 and S9 and are described in the next two paragraphs.
\textbf{Synthetic-field validation.} The reconstruction chain was verified on synthetic velocity fields that reproduce the experimental grid ($130\times111$ vectors, $0.345$\,mm spacing) and the experimental velocity scale: two jets, each modeled by a counter-rotating Lamb--Oseen vortex pair (core radius $2.5$\,mm, $|\Gamma|=0.20$\,m$^2$/s, half-separation $3$\,mm), are placed $2d$ apart in a $3$\,m/s crossflow with a stagnation-type return flow in the gap, and the pair height oscillates at $500$\,Hz to supply large-scale unsteadiness; per-vector Gaussian noise of $2.5\%$ of the local velocity magnitude is added to each of $400$ instantaneous frames. The reference pressure is obtained from the exact velocity field: the Poisson source is evaluated analytically and the five-point Dirichlet problem is solved exactly by a sine-transform spectral solver, Richardson-extrapolated onto the experimental grid and cross-checked against the closed-form Lamb--Oseen pressure, so the reference's own error is below $0.15$\,Pa. The synthetic field carries the magnitudes of the paper's diagnostics: a field range of $186$\,Pa (experimental $190\pm36$\,Pa), a gap-zone excess of $+24.5$\,Pa over the upstream reference (paper: $+6.6$ to $+27.4$\,Pa, case mean $+12.0$\,Pa) and an inter-jet pressure contrast of $16.6$\,Pa (paper: the order of $10$\,Pa, the $p_{\rm int}=22\pm13$\,Pa of the main text). The chain reproduces them with errors of $0.19$, $0.24$ and $0.04$\,Pa (global RMSE $0.22$\,Pa, gap-zone RMSE $0.39$\,Pa, $0.1\%$ of the range). Varying the Reynolds-stress smoothing over $\sigma=0.5$, $1$, $2$ grid points changes the RMSE by less than $0.01$\,Pa, because random noise is strongly attenuated by the elliptic inversion ($1/k^2$ spectral weighting); halving the resolution ($0.69$\,mm) raises the global/gap RMSE to $0.57/1.08$\,Pa and the inter-jet difference error to $0.70$\,Pa; a smooth $1\%$ of $u_{\rm rms}$ bias in the mean field costs $0.27/0.36$\,Pa; and window-filtering the mean field with the $1.38$\,mm PIV interrogation window ($24$\,px at $75\%$ overlap), the dominant PIV systematic error, costs a global/gap RMSE of $1.84/3.20$\,Pa with the key differences still recovered within $0.44$\,Pa. The reconstruction error of the pressure differences is therefore of order $0.2$--$0.7$\,Pa under realistic noise, bias and resolution perturbations -- $1$--$4\%$ of the $12$--$34$\,Pa differences on which the pressure conclusions rest (figure S8).
\textbf{Averaging-convergence diagnostic.} The $400$-frame records ($40$\,kHz, $10$\,ms) were checked for convergence of the quantities entering the Poisson source, in the gap-side band ($x=10$ to $10+L$\,mm, $y=2$--$16$\,mm) of two representative cases ($2d$-$3r$-$1$ and $4d$-$7r$-$1$). The mean field converges well: the sampling error of the gap-side mean streamwise velocity is $2.6\%$ and $4.4\%$. The zone-mean Reynolds stress $\langle u'v'\rangle$ converges slowly -- its sampling error is $54\%$ and $61\%$ -- because the zone average mixes contributions of opposite sign; the frame series has an integral time scale of $0.68$ and $0.43$\,ms, so the $10$\,ms record contains $15$--$23$ correlation times, equivalent to $7$--$12$ independent large-scale samples. The diagnostic quantity of the paper, however, is the reconstructed pressure, not the zone-mean stress: reconstructed with the full chain from the first and the second $200$ frames, the two pressure fields differ by an rms of $10.1$ and $6.7$\,Pa in the gap zone ($14.6$ and $6.9$\,Pa over the whole field), so the $400$-frame average carries a gap-zone sampling error of $5.0$ and $3.4$\,Pa, while the turbulent source term as a whole contributes only $1.4$ and $1.0$\,Pa rms to the gap-zone pressure. The $15$--$34$\,Pa differences of the main text therefore exceed the single-record sampling error by a factor of $3$--$10$; averaged over the $7$--$8$ records of a condition, the error shrinks by another factor of $\approx2.8$ to $1$--$2$\,Pa, well below the reported differences. Only the lower end of the gap-side excess range ($+6.6$\,Pa) is comparable to the single-record error, and that statistic is quoted as a range, not as a condition-mean result.
\begin{figure}
\centerline{\includegraphics[width=\textwidth]{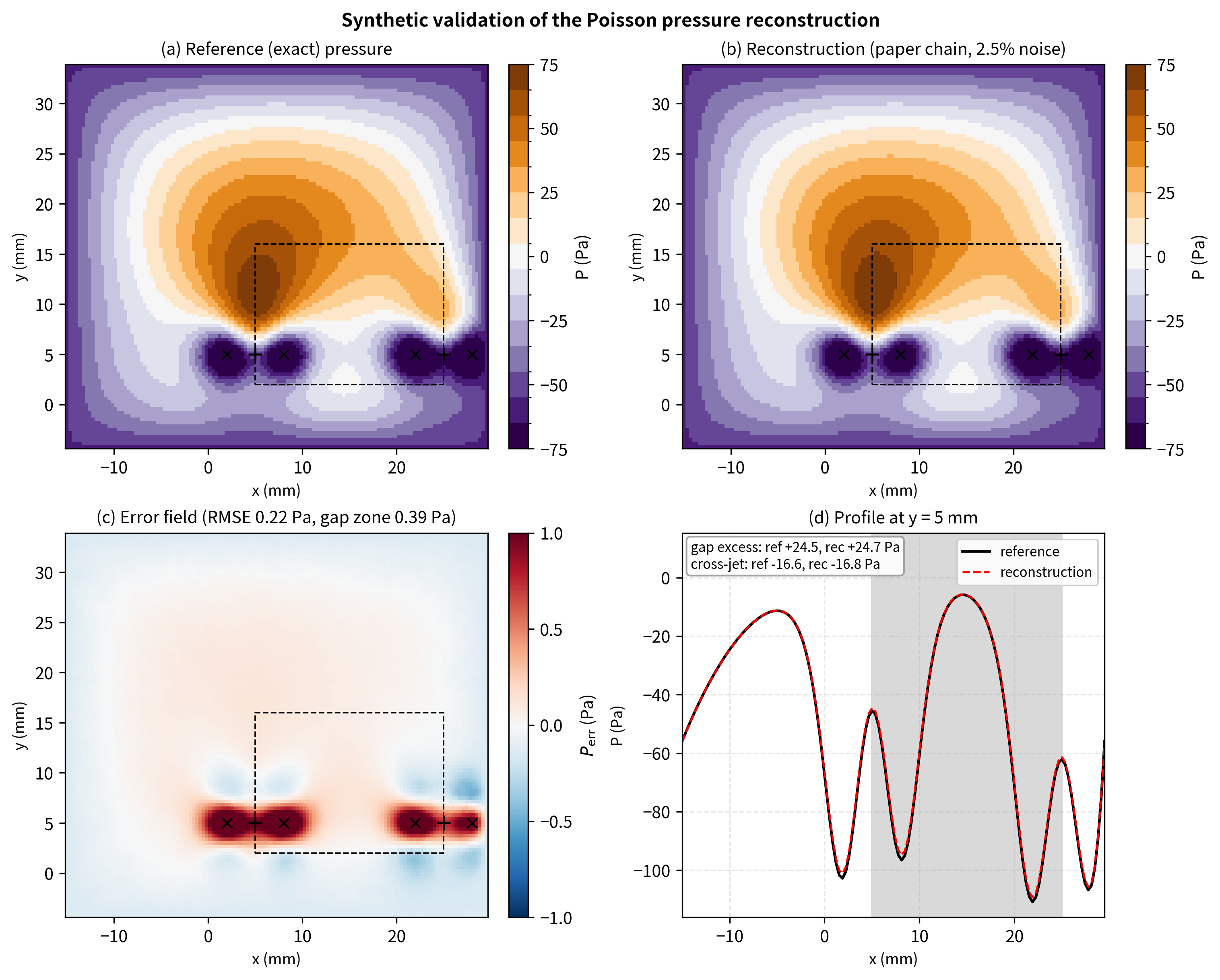}}
\caption{\textbf{Figure S8.} Synthetic-field validation of the pressure reconstruction: (a) exact reference pressure, (b) reconstructed field, (c) error field and (d) horizontal profile at $y=5$\,mm (grey band: the gap zone).}\alttext{Four-panel synthetic validation of the pressure reconstruction: the exact reference field, the reconstructed field, the error field and a profile comparison.}
\label{fig:si-pressure-synthetic}
\end{figure}

\begin{figure}
\centerline{\includegraphics[width=\textwidth]{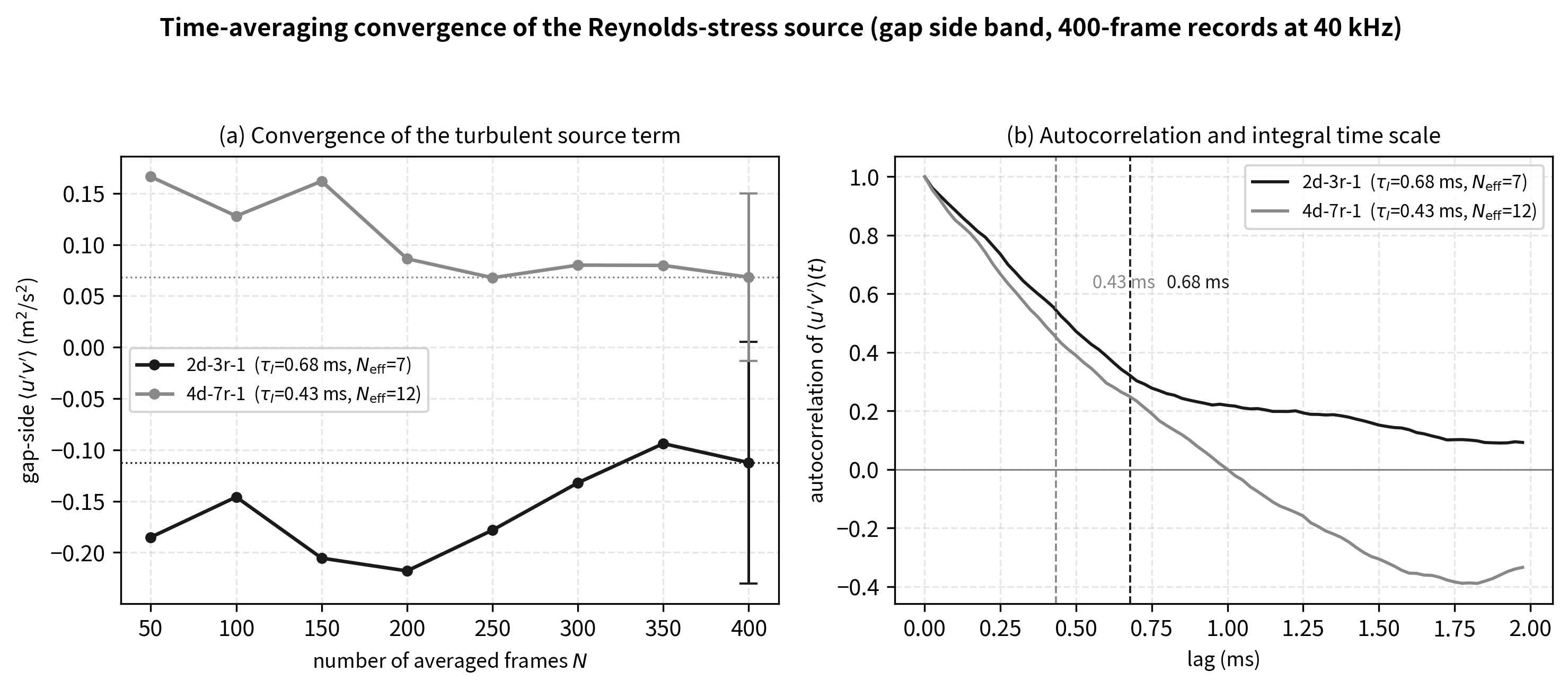}}
\caption{\textbf{Figure S9.} Averaging convergence of the Reynolds-stress source behind the reconstructed pressure, in two representative cases ($2d$-$3r$-$1$ and $4d$-$7r$-$1$): (a) running zone-mean Reynolds stress versus the number of averaged frames; (b) autocorrelation of the frame series, with the integral time scale marked.}\alttext{Two-panel convergence diagnostic: the running zone-mean Reynolds stress versus the number of averaged frames, and its frame-series autocorrelation.}
\label{fig:si-pressure-convergence}
\end{figure}

\clearpage

\section{The scaling law (Sec.~III G of the main text)}

\textbf{Why the near-field transition-height estimator fails at $r\approx3$.} The transition detection of Algorithm~1 (Steps 14--18) identifies the end of the near-field straight segment, which is confined to the region outside the potential core; at $r\approx3$ the strongly bent front jet leaves the straight segment inside that region, so the detection either fails (12 cases) or latches onto an already-distorted segment, returning near-unity ratios where the whole-trajectory matching gives $0.41$--$0.51$ (figure S4(a) of this supplement: the two estimators agree at $r\approx5$ and 7 within the case scatter, $0.59$--$0.68$ against $0.60$--$0.68$, but the transition-height values cluster at $0.92$--$1.02$ at $r\approx3$ and at $1.00$ for $L=2d$, $r\approx5$, against $0.60$ from the curve matching). The scaling correlation built on this estimator -- the power law $r_{\mathrm{eff}}/r=2.90\,\Pi^{-1/2}$ in the combined scale $\Pi=rL/d$ -- was fitted over the eight conditions the estimator could resolve, and its apparent quality ($R^2=0.84$) is an artifact: the estimator returns near-unity values exactly at low $\Pi$ and reduced values at high $\Pi$, which is the trend the power law was fitted to. Retested against the curve-matching ratios over the same eight condition means, the power law collapses ($R^2=-4.3$, its residuals carrying a $+0.87$ correlation with $\Pi$, a systematic under-prediction at large $\Pi$); over all nine conditions it fails outright ($R^2=-4.8$, residual correlation $+0.86$). The power law is therefore an extraction artifact of the transition-height estimator and is superseded by the velocity-dilution law of the main text; its parameter values ($\ell_g/d=2.90$, onset bound $\Pi\approx8.4$) are retained here only as the record of the superseded analysis.

An earlier formulation of the scaling, retained here as the record of the superseded analysis, condensed the exchange into the momentum-area model
\begin{equation*}
J_{\mathrm{eff}}\,\ell_M L = J\,(\ell_g)^2,
\qquad
\frac{J_{\mathrm{eff}}}{J}=\frac{(\ell_g)^2}{\ell_M L}=\frac{(\ell_g/d)^2}{\Pi},
\qquad \Pi\equiv r\,\frac{L}{d}=\frac{\ell_M L}{d^2},
\end{equation*}
with $\ell_M=rd$ the momentum length, $\ell_g$ the saturation gap length and $\Pi d^2$ the interaction area in nozzle-area units. The geometric reading sets $\ell_g$ by the wall-to-wall clearance of the widest configuration ($\approx L-d=3d$ at $L=4d$), where the enclosed recirculation is complete and strongest. Under the equal-density condition $J_{\rm eff}/J=(r_{\rm eff}/r)^2$, the model takes the power-law form $r_{\mathrm{eff}}/r=(\ell_g/d)\,\Pi^{-1/2}$. Fitted to the transition-height ratios $t_{y,n}^{\rm front}/t_s$ over the eight conditions within the range of validity $\Pi\ge(\ell_g/d)^2\approx8.4$ (95\% CI $[7.4,\ 9.6]$; the $L=2d$, $r\approx3$ condition lies below it and was excluded), the single constant calibrates to $\ell_g/d=2.90$ (fit-only 95\% CI $[2.72,\ 3.10]$; $[2.5,\ 3.6]$ once the single-jet baseline scatter is propagated), the model explaining 84\% of the condition-mean variance against 82\% for the fixed geometric value $3d$; the free-exponent fit over the same eight means returns $b=-0.48\pm0.10$, consistent with the structural $-1/2$, and the far-field coefficient in the ratio form refits the family at $\ell_g/d=2.74$ (95\% CI $[2.28,\ 3.26]$). The agreement of $\ell_g\approx3d$ with the widest-gap clearance was reported only as a post-hoc consistency check, and the model only as a candidate empirical scaling, not an independently validated law.

This formulation is superseded because it rests on the fragile transition-height estimator: retested against the curve-matching ratios of Section~\ref{app:fit-robust}, the power law collapses, with the free exponent becoming $+0.25$ against the structural $-1/2$ (the full diagnosis is given below and in Sec.~III G of the main text). The momentum-area model, read in momentum units as $J_{\rm eff}/J=[r/(r+c)]^2$, is retained in the main text as the momentum formulation of the same exchange, with the hyperbolic form as the law supported by the calibration-free estimator.

\clearpage

\section{The near-field capacity (Sec.~III H of the main text)}

\textbf{The near-field height carrier.} The spacing effect is carried by the exit-relative height ratio $y_t/y_s$ at small offsets $\Delta x=1$--$5$\,mm from the exit (Sec.~III H of the main text). At every offset up to $\Delta x=3$\,mm the ratio is strictly ordered in both controls -- all nine spacing pairs ($2d>3d>4d$, i.e.\ the ratio falls as the spacing grows) and all nine ratio pairs ($3r<5r<7r$, i.e.\ it rises with the velocity ratio) -- while at $\Delta x\ge4$\,mm the spacing ordering breaks as the $L=2d$ front jet merges into the rear jet; the window $\Delta x\le3$\,mm ($0.1$--$0.3\,d$) is the clean near-field window (Fig.~16(c) of the main text). Reading each spacing through the hyperbolic form, $y_t/y_s=r/(r+c_h)$, gives the per-spacing capacities $c_h=2.50$, $2.63$ and $3.30$ at $L=2d$, $3d$ and $4d$: the wider gap hosts the larger near-field entrainment capacity, consistent with the reversed-flow core being strongest at the widest spacing (Sec.~III E of the main text). These capacities are station-dependent: at $\Delta x=1$\,mm the same reading gives $c_h=3.76$, $4.15$ and $4.95$ (the same ordering, uniformly larger), at $\Delta x\ge4$\,mm the ordering breaks as the $L=2d$ trajectories merge, and $\Delta x=3$\,mm is the representative station quoted in the main text; the case-level scatter of the ratio propagates to $\pm0.2$--$0.35$ in the per-spacing mean $c_h$, so the $2d$--$3d$ difference ($0.13$) is not significant on its own and the robust statement is the wider-gap direction ($4d$ exceeding both narrower spacings by $0.7$--$0.8$, beyond the propagated scatter). The near-field capacity $c_h$ and the whole-trajectory capacity $c$ of the dilution law are distinct quantities -- $c_h$ measures the excess deflection accumulated over the first $3$\,mm ($0.1$--$0.3\,d$), where the reversed core acts, whereas $c$ measures the dilution integrated over the whole resolvable trajectory, where the gap-side exchange acts -- and the numerical overlap of $c=3.31$ with the $c_h$ range ($2.50$--$3.30$) is a coincidence of the two measures, not a mutual confirmation: no single value is expected to serve both paths.

\section{Processing and figure-generation pipeline (all sections)}
\label{si-code}

This section documents how the processed data, the figures and the tables of this supplement are produced, and it
gives the pseudocode of the remaining numerical stages. The corresponding Python~3 programs are attached to the
submission as a code archive; all paths in the archive point to the processed PIV data set described in
Sec.~1 and to the intermediate caches written by the stages below, so that the pipeline can be re-run without
touching the raw camera files. The pipeline is a sequence of independent stages: each figure or table is the
output of one script, and the data flow between stages is through the NumPy/JSON caches listed in table ST6.

\begin{algorithm}[H]
\algtitle{Batch averaging and single-jet reference library (Secs.~I and III A)}{alg:library}
\KwIn{processed average fields $\{V(x,y)\}$ of the records, one jet-exit position $x_0$ per record}
\KwOut{single-jet library $t_s(r), A_s(r), \lambda_s(r), B_s$; batch table}
exclude the outlier record of the $r\approx7$ group (one record, judged on the far-field fit quality)\;
for each record: re-plot $x\leftarrow x-x_0$ and store $(B_n,A_n,B_f,A_f,t_{y,n},\lambda,H)$ from Algorithm~\ref{alg:track}\;
group the records by measured velocity ratio ($r\approx2.8,4.6,6.5$) and form the group means $\pm$ SD\;
fit the penetration law $A_s(r)=a r+b$ to the group means; interpolate $t_s(r)$ and $\lambda_s(r)$ between the group means\;
pool the ten retained records: $B_f=0.393\pm0.017$, $A_f=1.85\pm0.22$ (the classical single-nozzle scaling check)\;
\end{algorithm}

\begin{algorithm}[H]
\algtitle{Condition-mean fields in physical and natural coordinates (Secs.~III C and III E)}{alg:meanfields}
\KwIn{retained cases of a condition; per-case trajectory and instantaneous fields}
\KwOut{condition-mean fields $\bar V(x,y)$ and $\bar{\tilde V}(s,n)$; per-case statistics}
for each case: average the instantaneous fields over the 400 frames of the record\;
for the natural coordinates: straighten each case onto its own trajectory with Algorithm~\ref{alg:nat}\;
normalize $\tilde y=(\tilde V-U_\infty)/(V_0-U_\infty)$ only where the main text states it\;
average cell by cell over the cases of the condition (identical $(s,n)$ grid for all cases)\;
store the per-case and condition statistics in the caches used by the statistics tables\;
\end{algorithm}

\begin{algorithm}[H]
\algtitle{Curve-matching estimator of the effective velocity ratio (Sec.~III B)}{alg:curve-match}
\KwIn{single-jet reference family (natural-coordinate fields of the 11 single records); front-jet trajectory of a twin case}
\KwOut{$r_{\mathrm{eff}}/r$ per case; hyperbolic fit $r_{\mathrm{eff}}/r=r/(r+c)$; bootstrap CI}
build the single-jet family $\tilde V_{\mathrm{single}}(s,n;r)$ on the $(s,n)$ grid from the reference records\;
for each twin case: take the front-jet trajectory in single-jet scaling coordinates $(u,v)$ over the primary window
$\tilde u\in[0.05,\,2.30]$\;
scan the reference ratio $r_{\mathrm{ref}}$ over the calibrated range and keep the one minimizing the RMS distance
between the twin trajectory and the single-jet family trajectory at $r_{\mathrm{ref}}$\;
$r_{\mathrm{eff}}\leftarrow r_{\mathrm{ref}}$; form the ratio $r_{\mathrm{eff}}/r$ and repeat over the 68 cases\;
fit $r_{\mathrm{eff}}/r=r/(r+c)$ at the condition level (one constant $c$); bootstrap the cases ($10^4$ resamples)
for the 95\% CI; repeat with the narrow window $\tilde u\in[0.05,\,0.50]$ for the window-robustness figure\;
\end{algorithm}

\begin{algorithm}[H]
\algtitle{Reverse-flow probability and pressure reconstruction (Secs.~III E and III F)}{alg:rz-pressure}
\KwIn{400 instantaneous fields per record; time-averaged field $\bar V$}
\KwOut{reverse-flow probability field $P(x,y)$; condition-mean pressure field $\bar P(x,y)$}
for each frame: mark the reverse flow $u<-u_{\rm thr}$ ($u_{\rm thr}=0.05$\,m/s); keep the largest connected component\;
$P(x,y)\leftarrow$ fraction of frames in which the pixel is reverse-flow (frame-resolved probability)\;
pressure: mean-flow source $-\rho[(U_x)^2+2U_yV_x+(V_y)^2]$ from the time-averaged field\;
Reynolds-stress source from the 400 frames (Gaussian smoothing, $\sigma=1$ grid point), second derivatives by finite differences\;
solve the five-point Poisson equation with Dirichlet $P=0$ boundary (sparse direct solver); subtract the mean of the
reference region ($y>8$\,mm) to fix the offset\;
validate against the synthetic Lamb--Oseen field (figure S8) and the averaging-convergence diagnostic (figure S9)\;
\end{algorithm}

\begin{table}
\centering
\small
\caption{\textbf{Table ST6.} Figure-to-code map of the supplement and of the main-text figures that this supplement reproduces. The scripts are attached as the code archive; intermediate caches are written by the stages of Algorithms~\ref{alg:library}--\ref{alg:rz-pressure} and re-used by the figure scripts.}
\label{tab:si-code-map}
\resizebox{\textwidth}{!}{%
\begin{tabular}{ll}
\toprule
Figure / table & Generating script (code archive) \\
\midrule
S1--S3, ST5 & \texttt{trajectory\_analysis/plot\_v28\_examples.py}; \texttt{velocity\_analysis/plot\_natural\_appendix.py} \\
main Fig.~3 & \texttt{trajectory\_analysis/plot\_single\_jet\_v28.py}, \texttt{plot\_v28\_examples.py} \\
main Fig.~4 & \texttt{trajectory\_analysis/plot\_avg\_traj\_vs\_single.py} \\
main Figs.~5--6 & \texttt{trajectory\_analysis/make\_traj\_figures.py}, \texttt{make\_fig5\_merged.py} \\
main Figs.~2, 8--9 & \texttt{velocity\_analysis/make\_fig\_inst\_velocity.py}, \texttt{plot\_natural\_examples.py}, \texttt{make\_fig\_nat\_mean\_nine.py} \\
main Figs.~10--12 & \texttt{velocity\_analysis/make\_fig\_nat\_stats.py}, \texttt{mechanism\_analysis/plot\_bend\_vs\_decay.py} \\
main Figs.~13--14 & \texttt{rz\_analysis/analyze\_rz.py}, \texttt{mechanism\_analysis/make\_fig\_rz\_trends\_combined.py} \\
main Figs.~7, 16 & \texttt{reff\_upgrade\_20261014/*}; \texttt{mechanism\_analysis/make\_fig\_reff\_scaling\_combined.py} \\
main Fig.~15; S8, S9 & \texttt{pressure\_analysis/analyze\_pressure\_tke\_nine.py}; \texttt{pressure\_validation\_20261015/*} \\
S4--S6 & \texttt{reff\_upgrade\_20261014/*}; SI scripts in the archive \\
S7, ST3, ST4 & \texttt{mechanism\_analysis/make\_fig\_rz\_trends\_combined.py}; \texttt{velocity\_analysis/make\_fig\_nat\_stats.py} \\
\bottomrule
\end{tabular}}
\end{table}

{\setlength{\emergencystretch}{2em}\noindent\textbf{Code archive.} The attached archive contains the analysis, statistics and figure-generation
programs in their run order, together with the shared modules (\texttt{plot\_style.py} for the fixed colour
tables, \texttt{analysis\_common.py} for the data loaders, \texttt{paths.py} for the data roots and
\texttt{measured\_ratio.py} for the velocity-ratio convention). The programs are plain Python~3 (NumPy, SciPy,
Matplotlib; Matplotlib $\ge$ 3.5), read the processed PIV data from the documented data root (set through the
environment variable \texttt{PIV\_DATA\_ROOT}) and write the intermediate caches to \texttt{output\_analyses/};
the figure scripts then write the EPS/PNG/PDF files used in the manuscript. The archive is self-contained in the
sense that no stage requires the raw camera frames, only the processed fields and the trajectory JSON files.}

\bibliographystyle{aipauth4-1-titles}
\bibliography{twin-jet-piv,liutex}